\documentclass[12pt]{article}
\pdfoutput=1 
   
\usepackage{hyperref}  
\usepackage{amsmath}   
\usepackage{amssymb} 
\usepackage{graphicx}
\usepackage{url}

\usepackage{float,wrapfig,color} 

\allowdisplaybreaks
\def\eq#1{(\ref{#1})}
\def\s[#1\s]{\begin{align}\begin{split}#1\end{split}\end{align}}
\def\[#1\]{\begin{align}#1\end{align}}
\def\tphi{{\tilde \phi}}

\begin{document}

\begin{titlepage}

\title{
\hfill\parbox{4cm}{ \normalsize YITP-19-59}\\ 
\vspace{1cm} 
Numerical and analytical analyses of a matrix model with non-pairwise
contracted indices}

\author{
Naoki Sasakura\footnote{sasakura@yukawa.kyoto-u.ac.jp} and 
Shingo Takeuchi\footnote{shingo.portable@gmail.com}
\\
$^*${\small{\it Yukawa Institute for Theoretical Physics, Kyoto University,}}
\\ {\small{\it  Kitashirakawa, Sakyo-ku, Kyoto 606-8502, Japan}}
\\
$^\dagger${\small{\it Phenikaa Institute for Advanced Study and Faculty of Basic Science, }}
\\ {\small{\it Phenikaa University, Hanoi 100000, Vietnam}}
}

\date{\today}

\maketitle

\begin{abstract}
We study a matrix model that has $\phi_a^i\ (a=1,2,\ldots,N,\ i=1,2,\ldots,R)$ as its dynamical
variable, whose lower indices are pairwise contracted, but upper ones are not always done so.
This matrix model has a motivation from a tensor model for quantum gravity,
and is also related to the physics of glasses, because it has the same form as what appears 
in the replica trick of the spherical $p$-spin model for spin glasses, though the parameter range of our interest is different. To study the dynamics, which in general depends on $N$ and $R$,
we perform Monte Carlo simulations and compare
with some analytical computations in the leading and the next-leading orders.
A transition region has been found around $R\sim N^2/2$, which matches a relation 
required by the consistency of the tensor model.
The simulation and the analytical computations agree well outside the transition region, but not  
in this region, implying that some relevant configurations are not properly included by the analytical
computations.
With a motivation coming from the tensor model, 
we also study the persistent homology of the configurations generated in
the simulations, and have observed its gradual change from
$S^1$ to higher dimensional cycles with the increase of $R$ around the transition region. 
\end{abstract}

\end{titlepage}

\section{Introduction}
\label{sec:introduction}
Quantization of gravity is one of the major fundamental problems in theoretical physics.
The quantization of general relativity by the standard perturbative methods 
of quantum field theory
fails due to non-renormalizable divergences.  Various approaches have been proposed
and being studied to solve the fundamental problem, depending on views of authors.
In one approach, general relativity (with higher derivative terms) is directly quantized
as quantum field theory with the modern technique of the functional renormalization 
group \cite{Reuter:2019byg}.
In other approaches, fundamental discreteness is introduced to represent spacetimes,
which include (causal) dynamical triangulations \cite{Loll:2019rdj}, 
loop quantum gravity \cite{Rovelli:2014ssa}, causal sets \cite{Surya:2019ndm}, 
quantum graphity \cite{Konopka:2006hu}, matrix models \cite{Wigner,thooft-planar,Brezin:1990rb,Douglas:1989ve,Gross:1989vs}, 
tensor models \cite{Ambjorn:1990ge,Sasakura:1990fs,Godfrey:1990dt,Gurau:2009tw}, and so on. In these discretized approaches, an important 
criterion for success is whether macroscopic spacetimes are generated, 
or in other words,
whether there exist appropriate continuum limits that recover the usual continuum picture 
of spacetime with dynamics described by general relativity as low-energy effective theory.

The criterion above can in principle be checked by studying the properties of 
a wave function of each theory. If the wave function has a peak
at a configuration that can well be described by a macroscopic spacetime picture,
then the model can be considered to be potentially successful.
An indirect motivation for the present paper is to understand the properties of 
the wave function \cite{Narain:2014cya} that is an exact solution to a tensor model 
in the Hamilton formalism \cite{Sasakura:2011sq,Sasakura:2012fb} 
(See \ref{Chap:Introtoctm} for the tensor model.). 
It has been argued and shown for some simple cases that the wave function has peaks 
at the tensor values that are invariant under Lie groups \cite{Obster:2017dhx}.
By using the correspondence developed in \cite{Kawano:2018pip} between tensor values and 
spaces with geometries, 
this would imply that the spacetimes symmetric under Lie groups are favored quantum
mechanically. 
However, the main difficulty in arguing this is that only little part of the peak structure 
(often called landscape in such contexts) of the wave function is known,
not enough to discuss ``probabilities of spacetimes."

To simplify the problem keeping the main structure from the tensor model as much as possible, 
one of the authors of the present paper and his collaborators 
considered the following two simplifications in the former papers. 
One is that they considered a toy wave function
rather than the actual wave function of the tensor model  \cite{Obster:2017pdq}. 
The actual wave function is 
expressed by a certain power, say $R$-th, 
of a function expressed by an integration over $N+1$ variables, 
but, in the toy wave function, the function is simplified to 
the one expressed by an integration over $N$ variables 
by fixing a certain integration variable to a constant. 
While this substantially simplifies the analysis, the toy wave function keeps
the most crucial property mentioned above 
that there appear peaks at the tensor values that are symmetric under 
Lie groups  as the actual wave function of the tensor model does
  \cite{Obster:2017pdq}.

Though this toy wave function is simpler than the actual one, it is still difficult to 
perform thorough analyses,
because the dimension of the argument (a symmetric tensor with three indices) of the wave function
is very large with the order of $\sim N^3/6$.
Therefore, as an additional simplification, the authors  in \cite{Lionni:2019rty}
considered a model that can be obtained by integrating over the argument of the toy 
wave function.
This gives a dynamical system of a matrix, say $\phi_a^i\ (a=1,2,\ldots,N,\ i=1,2,\cdots,R)$,
rectangular in general, where the lower indices are pairwise contracted, but the upper ones are not 
always done so.
While this model does not fall into the known solvable models such as 
the rectangular matrix models \cite{Anderson:1990nw,Anderson:1991ku,Myers:1992dq} or the vector models
\cite{Nishigaki:1990sk,DiVecchia:1991vu}, 
 it has the same form as what appears when the replica trick is applied to 
the spherical $p$-spin model for spin glasses \cite{pspin,pedestrians},
 where  $R$ plays the role of the replica number.
Here, though the form is exactly so, the concerned ranges of the dynamical 
variables and the parameters
are different between our model and the spin glass model, 
and it seems reasonable to reanalyze the matrix model with fresh eyes:
(i) While the replica number $R$ is taken to the vanishing limit in the replica trick, it takes 
a finite value $R\sim N^2/2$ in the correspondence to the tensor model, and should rather be taken 
to infinity in the thermodynamic limit $N\rightarrow \infty$; 
(ii) A coupling constant\footnote{More specifically, $\lambda$ in \eq{eq:parfun}.
This sign difference originates from the pure imaginary value of the tensor coupling in \eq{eq:wavefun},
which is real in the spherical $p$-spin model.} 
takes the opposite sign in our model compared to the spin glass model; 
(iii) There is a spherical constraint on the dynamical variable in the spherical $p$-spin model, but there is none in our model.
 
In this paper, we numerically study the matrix model by the Monte Carlo simulation 
with the standard Metropolis update method.
This contrasts with the perturbative analytical computations performed in the previous 
paper \cite{Lionni:2019rty}. 
We also perform some additional analytical computations to compare with the numerical results.
We have obtained the following main results:
\begin{itemize}
\item
The expectation values of some observables are computed by the numerical simulations, 
and it is observed that there exists a transition region around $R\sim N^2/2$.  
Intriguingly, the location is in good coincidence with  
$R=(N+2)(N+3)/2$ that is required by the consistency of the tensor model (Namely, the hermiticity of the Hamiltonian constraints. See \ref{Chap:Introtoctm} for more details.) 
\cite{Obster:2017dhx,Narain:2014cya,Sasakura:2013wza}.  
Presently, this coincidence is mysterious, since there are no apparent connections
between the transition and the consistency. 
The observables seem to continuously but substantially change their behavior at the transition region,  but it has not been determined whether this transition is a phase transition or a crossover
in the thermodynamic limit $N\rightarrow \infty$.
The method for the Monte Carlo simulations performed in this paper is 
not powerful enough for 
the determination because of an issue explained below.

\item
The expectation values of some observables are analytically computed in the leading order, 
mostly based on the treatment in the previous paper \cite{Lionni:2019rty},
and are compared with the numerical results.
Good agreement between them is obtained outside the vicinity of the transition region,
while there exist deviations in the transition region.
The deviations are such that they soften the transition to make it look more like a crossover. 
A next-leading order computation has also been performed, but this does not well correct the deviations.

\item
The tensor model suggests the presence of topological characteristics for the dominant configurations
of the matrix $\phi_{a}^i$ (See Section~\ref{sec:persistent} and \ref{Chap:Introtoctm}.).
Therefore, we have studied topological characters of the configurations 
that are generated in the simulations by using the modern technique called persistent homology \cite{carlsson_topology_2009} in the topological data analysis. 
This technique extracts homology groups possessed by a data, 
which is a value of the matrix $\phi_{a}^i$ in our case. 
The dominant topology gradually changes from $S^1$ to higher-dimensional cycles
with the increase of $R$ in the vicinity of the transition region.

\item
The Monte Carlo simulation becomes substantially difficult in
the region with $R\gtrsim N^2/2$ and large $\lambda/k^3$,
where $\lambda$ and $k$ are the parameters of our model \eq{eq:parfun}.
In the region, the step sizes of the Metropolis updates chosen for reasonable acceptance rates  
become too small to reach thermal equilliburiums in $\sim 10^{10}$ Metropolis updates.

\item
A characterization of the transition can be done by the sizes of the matrix elements, 
which take relatively large values at the region with $R\gtrsim N^2/2$ and large $\lambda/k^3$, but 
otherwise fluctuate around small values.
In the former case, our model may behave like the spherical $p$-spin model,
since the matrix elements are effectively constrained to non-zero sizes,
which would approximately realize the spherical constraint in the spherical $p$-spin model.
This may partly explain the bad performance of the Monte Carlo simulation 
in the region, seemingly reflecting the high viscous nature of glassy fluids.

\end{itemize}

This paper is organized as follows.  In Section~\ref{sec:model},
we define the model and summarize the previous results \cite{Lionni:2019rty}
that are relevant to the present paper.
In Section~\ref{sec:observables},  some observables 
are introduced and the analytical expressions of their expectation values are obtained in a leading order. 
In Section~\ref{sec:fextend}, the details of the computation in the leading order 
are given. The result of the next-leading order is also presented, 
while the details of the computation are given in \ref{app:fnext}.
In Section~\ref{sec:saddle}, we perform a saddle point analysis of the expectation values of 
the observables in the leading order.  This describes the transition as a continuous 
phase transition in the large $N$ limit, where the first derivatives of the expectation values of the observables 
with respect to $R$ are discontinuous.  
In Section~\ref{sec:mc},
we compare the Monte Carlo and the analytical results. They agree well outside the transition region.
In the transition region, however, there exist deviations,
which smoothen the transition to make it look more like a crossover.
In Section~\ref{sec:persistent}, we analyze the homology structure
of the configurations generated by the simulations. The preference changes from 
$S^1$ to higher-dimensional cycles with the increase of $R$ in the vicinity of the transition region.
The last section is devoted to a summary and future prospects.
In \ref{Chap:Introtoctm}, the motivation for the matrix model is explained from the viewpoint of the 
tensor model.
 In \ref{app:Req2}, an instructive computation of the partition function for 
 $R=2$ is given.
 In \ref{app:derivationofconnect}, \ref{app:fnext}, and \ref{sec:estimation}, 
 some equations used in the main text are explicitly derived.
 In \ref{app:persistent}, a brief introduction to persistent homology is given.

 %%%%%%%%%%%%%%%%%%%%%%%%%%%555
 %%%%%%%%%%%%%%%%%%%%%%%%
 %%%%%%%%%%%%%%%%%%%%%%%
 
\section{The model}  
\label{sec:model}
The partition function of our matrix model is given by
\[
Z_{N,R}(\lambda,k):=
\int_{{\mathbb R}^{NR}} d\phi \ \exp\left(-\lambda U(\phi)- k \hbox{Tr}(\phi^t \phi)\right),
\label{eq:parfun}
\]
where $\lambda$ and $k$ are the coupling constants assumed to be positive,
$\phi$ is a (generally rectangular) real matrix, $\phi_a^i\ (a=1,2,\ldots,N,\  i=1,2,\ldots,R$),
and $d\phi:=\prod_{a,i=1}^{N,R}d\phi_a^i$. The integration is over 
 the $NR$-dimensional real space denoted by ${\mathbb R}^{NR}$.  
 The coupling terms are defined by
\s[
&U(\phi):=\sum_{i,j=1}^R \left( \phi_a^i \phi_a^j \right)^3,\\
&{\rm Tr}(\phi^t \phi):=\sum_{i=1}^R \phi_a^i \phi_a^i, 
\label{eq:defofUtr} 
\s]
where the repeated lower indices are assumed to be summed over. We use this standard 
convention for the lower indices throughout this paper, unless otherwise stated.
On the other hand, we do not use this convention for the upper indices: 
A sum over them must always be written explicitly\footnote{One can avoid this unusual convention
by introducing some external variables. An example
is $U(\phi,C)\equiv  C_{i_1i_2i_3} C_{j_1j_2j_3} \phi_{a_1}^{i_1} \phi_{a_1}^{j_1} \phi_{a_2}^{i_2} \phi_{a_2}^{j_2} \phi_{a_3}^{i_3} \phi_{a_3}^{j_3}$, which reproduces $U(\phi)$ by putting $C_{i_1i_2i_3}=\delta_{i_1i_2}\delta_{i_1i_3}$.}.  
The background motivation for the matrix model is explained in \ref{Chap:Introtoctm} from the viewpoint 
of the tensor model.

In \eq{eq:defofUtr}, the lower indices are contracted pairwise, while the upper indices
are not necessarily so.
Therefore the model has the symmetry of the $O(N)$ transformation on the lower indices, 
but only the permutation symmetry $S_R$ of relabeling $\{1,2,\ldots,R\}$ on the upper indices. 
These symmetries are not enough to diagonalize $\phi_a^i$ in general, and 
therefore this model cannot be solved in a similar way as the usual matrix model.

In the previous paper \cite{Lionni:2019rty},
we considered an expression which can just be obtained by separating
the radial and angular part of the integration in \eq{eq:parfun}:
By the change of variable, $\phi_a^i=r \tilde \phi_a^i$, with $r^2=\sum_{i=1}^R \phi_a^i \phi_a^i$
and $\tilde \phi$ representing the angular coordinates, we obtain
\[
Z_{N,R}(\lambda,k)=V_{NR-1} \int_0^\infty dr\, r^{NR-1} f_{N,R}(\lambda\, r^6) \, e^{-k \, r^2},
\label{eq:zwithf}
\]
where
\[
f_{N,R}(t):=\frac{1}{V_{NR-1}}\int_{S^{NR-1}} d \tilde \phi \, e^{-t\, U(\tilde \phi)}
\]
with $V_{NR-1}=\int_{S^{NR-1}} d\tilde \phi$, the volume of the $NR-1$-dimensional unit sphere. 

This rather trivial change of expression is actually very useful, 
because $f_{N,R}(t)$ can be shown to be 
an entire function of $t$ and therefore has a Taylor expansion in $t$ with the infinite 
convergence radius around $t=0$ (actually around arbitrary $t\neq\infty$).
Therefore, in principle, the dynamics can be solved by obtaining the entire perturbative 
series of $f_{N,R}(t)$. Note that the corresponding perturbative expansion
of $Z_{N,R}(\lambda,k)$ in $\lambda$ around $\lambda=0$, 
often obtained by perturbative methods,  is merely an asymptotic series,
because $Z_{N,R}(\lambda,k)$ has an essential singularity at $\lambda=0$. 
The $f_{N,R}(t)$ has also the property that it is a decreasing positive function of $t$ with 
$f_{N,R}(0)=1$ for real $t$. This property provides a good criterion for assessing
the validity of an approximation of $f_{N,R}(t)$.
In the previous paper \cite{Lionni:2019rty}, 
$f_{N,R}(t)$ in the leading order of $1/R$ has been determined 
by a Feynman diagrammatic method with the result,
\s[
&f^{1/R,{\rm leading}}_{N,R}(t)=\left(1+\frac{12 t }{N^3 R^2}\right)^{-\frac{ N(N-1)(N+4)}{12}}
\left(1+\frac{6(N+4) t }{N^3 R^2}\right)^{-\frac{N}{2}}.
\label{eq:repf}
\s]
In particular, \eq{eq:repf} indeed satisfies the properties above: It is a decreasing function for real $t$ 
with $f^{1/R,{\rm leading}}_{N,R}(0)=1$, and 
is almost an entire function, since the locations of the singularities are far away from the relevant region 
$t\geq 0$ for large $N,R$.  

Since there are two parameters $N,R$, which can be taken large, the range of validity of \eq{eq:repf},
which was derived in the leading order of $1/R$, is not obvious.  
However, in later sections, we will find that \eq{eq:repf}\footnote{More precisely, because 
of the difference of our strategy of computations taken in this paper, the expression \eq{eq:fextendlamd0}
is slightly different from \eq{eq:repf} obtained in \cite{Lionni:2019rty}. 
However, the difference is negligible for large $N,R$, and is not essential.} 
will give results which agree well\footnote{
In fact, the expression \eq{eq:repf} cannot be correct for small $R$ such as $R=2$.
This is explicitly shown in \ref{app:Req2}. However, the difference shown there in the asymptotic region $t\sim\infty$ is not relevant for the thermodynamic properties, since for small $R$, the dominant 
contributions come from $t \sim 0$, as can be explicitly observed 
in the Monte Carlo simulations in Section~\ref{sec:mc}.}
with those of the numerical simulations
except in the transition region around $R\sim N^2/2$.

\section{Observables}
 \label{sec:observables}

The purpose of this section is to introduce some observables, say 
$\cal O(\phi)$, and discuss the expectation values defined by
\[
\langle {\cal O(\phi)}\rangle:=
\frac{1}{Z_{N,R}(\lambda,k)}
\int_{{\mathbb R}^{NR}} d\phi \ {\cal O}(\phi)\ e^{-\lambda U(\phi)- k {\rm Tr}(\phi^t \phi)}.
\label{eq:evO}
\]
The observables must respect the symmetry $O(N)\times S_R$ mentioned in Section~\ref{sec:model}. 
 Among various possibilities, we consider $\hbox{Tr}( \phi^t \phi)$ and  $U(\phi)$ 
 in \eq{eq:defofUtr}, and also  
 \[
 U_d(\phi):=\sum_{i=1}^R \left( \phi_a^i \phi_a^i \right)^3.
 \]
 The last one is the diagonal part of the sum in $U(\phi)$ in \eq{eq:defofUtr}.
 Since these observables are some parts contained in the exponent of \eq{eq:parfun},
 they can be implemented in the numerical simulations with little additional computational costs.

 To compute the expectation values of these observables,  it is convenient to extend \eq{eq:parfun} by introducing the coupling constant $\lambda_d$ conjugate to $U_d(\phi)$
 as  
 \[
 Z_{N,R}(\lambda,\lambda_d,k):=\int_{{\mathbb R}^{NR}} d\phi \ e^{-\lambda U(\phi)-\lambda_d U_d(\phi)- k {\rm Tr}(\phi^t \phi)}.
\label{eq:parfundiag}
 \]
 Then the expectation values can respectively be expressed as 
 \[
 \begin{split}
&\langle \hbox{Tr}(\phi^t \phi) \rangle 
=\frac{\partial}{\partial k} F_{N,R}(\lambda,\lambda_d=0,k), \\
&\langle U(\phi) \rangle =\frac{\partial}{\partial \lambda} F_{N,R}(\lambda,\lambda_d=0,k), \\
&\langle U_d(\phi) \rangle =\left.\frac{\partial}{\partial \lambda_d} F_{N,R}(\lambda,\lambda_d,k)\right |_{\lambda_d=0},
\end{split}
\label{eq:obsbyderf}
 \]
where $F_{N,R}(\lambda,\lambda_d,k):=-\log Z_{N,R}(\lambda,\lambda_d,k)$,
which is the free energy of the model.
Here we have put $\lambda_d=0$ at last, since our interest is in 
\eq{eq:parfun} corresponding to $\lambda_d=0$ of \eq{eq:parfundiag}.
 
To compute the partition function \eq{eq:parfundiag}, it is convenient to  
first separate the radial and the angular part as in \eq{eq:zwithf}: 
\[
Z_{N,R}(\lambda,\lambda_d,k)=
V_{NR-1} \int_0^\infty dr\ r^{NR-1} f_{N,R,\lambda,\lambda_d}(r^6)\, e^{-k r^2},
\label{eq:parextbyf}
\]
where
\[
f_{N,R,\lambda,\lambda_d}(t):=\frac{1}{V_{NR-1}}
\int_{S^{NR-1}}d\tilde \phi \ e^{-\lambda\, t\, U(\tilde \phi)-\lambda_d\, t \, U_d(\tilde \phi)}.
\label{eq:flamd}
\]
The $f_{N,R,\lambda,\lambda_d}(t)$ has the similar properties as $f_{N,R}(t)$ explained
in Section~\ref{sec:model}: 
It is an entire function of $t$; For $\lambda>0, \lambda_d\geq 0$, it is positive and 
decreasing in $t$ for real $t$; $f_{N,R,\lambda,\lambda_d}(0)=1$.
With $f_{N,R,\lambda,\lambda_d}(t)$, the observables \eq{eq:obsbyderf} can be
expressed by
\s[
&\langle \hbox{Tr}(\phi^t \phi) \rangle=\frac{1}{{\cal N}_f}\int_0^\infty dr\ r^{NR+1} f_{N,R,\lambda,0}(r^6)\, e^{-k r^2},\\
&\langle  U(\phi) \rangle=-\frac{1}{{\cal N}_f}\int_0^\infty dr\ r^{NR-1}
\frac{\partial}{\partial \lambda} f_{N,R,\lambda,0}(r^6)\, e^{-k r^2}, \\
&\langle U_d(\phi) \rangle=-\frac{1}{{\cal N}_f}\int_0^\infty dr\ r^{NR-1} 
\left.
\frac{\partial}{\partial \lambda_d}f_{N,R,\lambda,\lambda_d}( r^6)\right|_{\lambda_d=0}
\, e^{-k r^2},
\label{eq:obsbyf}
\s]
where the normalization is given by
\[
{\cal N}_f=\int_0^\infty dr\ r^{NR-1} f_{N,R,\lambda,0}(r^6)\, e^{-k r^2}.
\label{eq:normalizationf}
\]

From the leading order computation, which is detailed in Section~\ref{sec:fextend},
we obtain
\[
f^{leading}_{N,R,\lambda,\lambda_d}(t)=
h_{N,R}(\lambda R t +\lambda_d t) \ h_{N,R}(\lambda_d t )^{R-1}
\label{eq:fleadwithh}
\]
with 
\[
h_{N,R}(x):=\left(1+12 \gamma_3 x \right)^{-\frac{ N(N-1)(N+4)}{12}}
\left(1+6(N+4) \gamma_{3} x \right)^{-\frac{N}{2}},
\label{eq:exph}
\]
where
\[
\gamma_n:=\frac{\Gamma\left(\frac{NR}{2}\right)}{2^n\Gamma\left(\frac{NR}{2}+n\right)}.
\label{eq:defofgamma}
\]

When $\lambda_d=0$ is taken, \eq{eq:fleadwithh} becomes
\s[
&f^{leading}_{N,R,\lambda,0}(t)=\left(1+12 \gamma_3 \lambda  R t  \right)^{-\frac{ N(N-1)(N+4)}{12}}
\left(1+6(N+4) \gamma_{3}  \lambda R t \right)^{-\frac{N}{2}}.
\label{eq:fextendlamd0}
\s]
This could be expected to agree with \eq{eq:repf},
but there is a slight difference coming from \eq{eq:defofgamma}
with $n=3$. This difference originates from 
the fact that the strategy of the computation we take in Section~\ref{sec:fextend} 
is different from the one taken previously in \cite{Lionni:2019rty}, and therefore 
the expressions of the leading orders are slightly different from each other.
However, they agree with each other in the leading order of $1/R$, since 
$\gamma_n\sim (NR)^{-n}$ for $NR \gg n$, as expected.

By putting these expressions into \eq{eq:obsbyf},  we obtain
\s[
&\langle \hbox{Tr}(\phi^t \phi) \rangle_{leading}=\frac{1}{{\cal N}_f}\int_0^\infty dr\ r^{NR+1} f_{N,R,\lambda,0}^{leading}(r^6)\, e^{-k r^2},\\
&\langle  U(\phi) \rangle_{leading}=-\frac{R}{{\cal N}_f}\int_0^\infty dr\ r^{NR+5} \frac{h_{N,R}'(\lambda R r^6)}{h_{N,R}(\lambda R r^6)} f_{N,R,\lambda,0}^{leading}( r^6)\, e^{-k r^2}, \\
&\langle U_d(\phi) \rangle_{leading}=-\frac{1}{{\cal N}_f}\int_0^\infty dr\ r^{NR+5} \left(
\frac{h_{N,R}'(\lambda R r^6)}{h_{N,R}(\lambda R r^6)} +(R-1)\frac{h_{N,R}'(0)}{h_{N,R}(0)}
\right)f_{N,R,\lambda,0}^{leading}( r^6)\, e^{-k r^2},
\label{eq:expobss}
\s]
where ${\cal N}_f$ is given by \eq{eq:normalizationf} with $f_{N,R,\lambda,0}^{leading}(t)$. 
The actual values of these integrations can be obtained numerically.

%%%%%%%%%%%%%%%%%%%%%%%%%%%%%%%%
%%%%%%%%%%%%%%%%%%%%%%%%%%%%%%555
%%%%%%%%%%%%%%%%%%%%%%%%%%%%%%%%%%

\section{Computations of $f_{N,R,\lambda,\lambda_d} (t)$ in the leading and 
the next-leading orders}
\label{sec:fextend}
In this section, we will compute $f_{N,R,\lambda,\lambda_d}(t)$ in the leading order of $t$,
and will also show the result in the next-leading order, whose detailed computations are given 
in \ref{app:fnext}.
In \cite{Lionni:2019rty}, the computation in the leading order of $1/R$ has been performed 
by using the Feynman diagrams for the $\phi_a^i$ variables.  
In this paper, however, we will take a different strategy.
This is because the new strategy makes more transparent the rather complicated
counting of combinatorics performed in \cite{Lionni:2019rty},  
and make it straightforward to include the extra coupling $\lambda_d U_d(\phi)$ and also 
to consider the next order in $t$. 
For $\lambda_d=0$,  the new strategy gives essentially 
the same result as \cite{Lionni:2019rty} in the leading order, 
as commented below \eq{eq:fextendlamd0}.

Let us define 
\[
f_{N,R,\Lambda_{ij}}(t):=\frac{1}{V_{NR-1}}\int_{S^{NR-1}} d\tilde \phi  \ 
e^{-t \, U_{\Lambda_{ij}}(\tilde \phi)},
\label{eq:fext}
\]
where 
\[
U_{\Lambda_{ij}}(\tilde \phi):=\sum_{i,j=1}^R \Lambda_{ij} \left(  \tilde \phi_a^i
\tilde \phi_a^j\right)^3
\]
with a real symmetric matrix $\Lambda_{ij}$. 
The eigenvalues of the matrix $\Lambda_{ij}$ are assumed to be non-negative for the convergence 
of the corresponding partition function.    
The $f_{N,R,\Lambda_{ij}}(t)$ also has the same nice properties as $f_{N,R}(t)$ that it is an entire function,
which has a Taylor series expansion in $t$ with the infinite convergence radius around $t=0$, 
and is a decreasing positive function of $t$ for real $t$ and $\Lambda\neq 0$ 
with $f_{N,R,\Lambda_{ij}}(0)=1$. 
If we take $\Lambda_{ij}=\lambda+\lambda_d \delta_{ij}$, \eq{eq:fext} gives $f_{N,R,\lambda,\lambda_d}(t)$ in \eq{eq:flamd}. 
By introducing a new variable $P_{abc}^i\ (a,b,c=1,2,\ldots,N,\ i=1,2,\ldots,R)$, which is 
symmetric for the lower indices, one can rewrite \eq{eq:fext} as
\[
\begin{split}
&f_{N,R,\Lambda_{ij}}(t)\\&=const. \int_{S^{NR-1}} d\tilde \phi \int_{-\infty}^\infty
\prod_{i=1}^R \prod_{a\leq b \leq c=1}^N dP_{abc}^i 
\exp \left( -\sum_{i=1}^R  P_{abc}^i P_{abc}^i + 2 I \sqrt{t} \sum_{i,j=1}^R 
\tilde \Lambda_{ij} P^i_{abc} \tilde \phi_a^j \tilde \phi_b^j \tilde \phi_c^j
\right),
\end{split}
\label{eq:fextwithP}
\]
where  $I$ denotes the imaginary unit and $\tilde \Lambda_{ij}$ is a symmetric matrix 
satisfying\footnote{In general, 
such a matrix $\tilde \Lambda$
can be obtained as $\tilde \Lambda =M^t \sqrt{D} M$ by diagonalizing the matrix $\Lambda$ as  
$\Lambda=M^t D M$ with a diagonal matrix $D$ and an orthogonal one $M$.}, 
\[
\Lambda_{ij}=\sum_{k=1}^R \tilde \Lambda_{ik} \tilde \Lambda_{kj}.
\label{eq:defoftildelam}
\]
The constant prefactor in \eq{eq:fextwithP} can be determined by $f_{N,R,\Lambda_{ij}}(0)=1$.

To compute \eq{eq:fextwithP}, let us first integrate over $\tilde \phi$. This change of the 
order of the integrations can be done, because the integration over $P^i_{abc}$ 
with the infinite integration region converges 
uniformly for any $\tilde \phi \in S^{NR-1}$. Then our task is to compute 
\[
\left  \langle e^{ 2 I \sqrt{t} P\tilde \phi^3} \right\rangle_{\tilde \phi} = \frac{1}{V_{NR-1}}
\int_{S^{NR-1}} d\tilde \phi \ e^{2 I \sqrt{t}\, P\tilde \phi^3},
\label{eq:defofexp}
\]
where we have used a short-hand notation, 
\[
P\tilde \phi^3:=
 \sum_{i,j=1}^R 
\tilde \Lambda_{ij} P^i_{abc} \tilde \phi_a^j \tilde \phi_b^j \tilde \phi_c^j,
\label{eq:formO}
\]
and $\langle \cdot \rangle_{\tilde \phi}$ denotes
the expectation value for the uniform probability distribution on the unit sphere $S^{NR-1}$.

For further computations, let us introduce the cumulants $\langle {\cal O}^n \rangle^c$ 
defined by\footnote{Cumulants are more familiar as connected correlation functions in particle physics, because they can be computed by summing over connected Feynman diagrams. However, we do not use this 
terminology here, because we will rewrite the cumulants for $\tilde \phi$
in terms of $\phi$ as in \eq{eq:connecttilde}, and one can explicitly see that 
these cumulants contain some disconnected diagrams in terms of the Feynman diagrams of $\phi$ in general,
because of the extra factor $\gamma_{m/2}$ in \eq{eq:connecttilde}. }
\[
\log \langle e^{s \cal O} \rangle=\sum_{n=1}^\infty \frac{s^n}{n!} 
\langle {\cal O}^n \rangle^c
\label{eq:defofcum}
\]
with arbitrary $s$.
Then \eq{eq:fextwithP} can be rewritten as 
\[
f_{N,R,\Lambda_{ij}}(t)=const.  \int_{-\infty}^\infty
\prod_{i=1}^R \prod_{a\leq b \leq c=1}^N dP_{abc}^i  \ 
e^{-S_{eff}(P)}
\label{eq:ffromseff}
\]
with
\[
S_{eff}(P)=\sum_{i=1}^R  P_{abc}^i P_{abc}^i- 
\sum_{n=1}^\infty \frac{(2 I \sqrt{t})^n}{n!} \langle ( P\tilde \phi^3)^n \rangle_{\tilde \phi}^c,
\label{eq:seff}
\]
where
$S_{eff}(P)$ can be regarded as an effective action of $P^i_{abc}$ after integrating out 
$\tilde \phi$, and it is given in terms of the perturbative expansion in $t$.
Due to the form \eq{eq:formO}, the $n$-th order cumulant gives the $n$-th order interaction term of $P^i_{abc}$, and all the terms with odd $n$ vanish because of the obvious 
invariance of the integration over $\tilde \phi$ under $\tilde \phi \rightarrow -\tilde \phi$.

Let us compute the quadratic term with $n=2$ in \eq{eq:seff}. Since $\langle P\tilde \phi^3\rangle_{\tilde \phi}=0$, we obtain
\[
\begin{split}
\langle (P\tilde \phi^3)^2\rangle_{\tilde \phi}^c&=\langle(P\tilde \phi^3)^2\rangle_{\tilde \phi}
=
 \sum_{i,j,i',j'=1}^R  \tilde \Lambda_{ij} \tilde 
 \Lambda_{i'j'}  P^i_{abc} P^{i'}_{a'b'c'} 
 \langle \tilde \phi_a^j \tilde \phi_b^j \tilde \phi_c^j \tilde 
 \phi_{a'}^{j'} \tilde \phi_{b'}^{j'} \tilde \phi_{c'}^{j'} 
 \rangle_{\tilde\phi}
 \end{split}
  \label{eq:o2}
\]
The integral over $\tilde \phi$ on $S^{NR-1}$, which is a curved compact space,
is not easy to handle,  
so we use a formula which maps this integration to the Gaussian 
integral: 
\[
\langle \tilde \phi_{a_1}^{i_1}\tilde \phi_{a_2}^{i_2}\cdots \tilde \phi_{a_m}^{i_m} 
\rangle_{\tilde \phi}=
(2\beta)^{\frac{m}{2}}\gamma_{\frac{m}{2}} \langle  \phi_{a_1}^{i_1} \phi_{a_2}^{i_2}\cdots \phi_{a_m}^{i_m} \rangle_\phi,
\label{eq:connecttilde}
\]
where $\gamma_n$ is defined in \eq{eq:defofgamma}, and
\s[
&\langle  \phi_{a_1}^{i_1} \phi_{a_2}^{i_2}\cdots \phi_{a_m}^{i_m} \rangle_\phi:=
\frac{1}{\int_{{\mathbb R}^{NR}} d\phi \ e^{-\beta {\rm Tr}\phi^t \phi}} \int_{{\mathbb R}^{NR}}
d\phi \ \phi_{a_1}^{i_1} \phi_{a_2}^{i_2}\cdots \phi_{a_m}^{i_m} \ e^{-\beta {\rm Tr} \phi^t \phi}.
\label{eq:defofexpbeta}
\s]
Here $\beta$ is a sort of dummy variable, which can be chosen freely with $\beta>0$,
and does not appear in the final expressions. In fact, as shown below, the factor $(2 \beta)^{\frac{m}{2}}$ in 
\eq{eq:connecttilde} is exactly canceled by the 
same factor from the Wick contraction \eq{eq:wick}.
The formula \eq{eq:connecttilde} was previously used in \cite{Lionni:2019rty}, and 
is proven in \ref{app:derivationofconnect} so that the present paper 
be self-contained.

\begin{figure*}
\begin{center}
\hfil
\includegraphics[width=2cm]{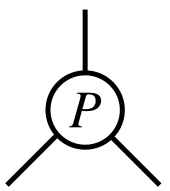}
\hfil
\includegraphics[width=8cm]{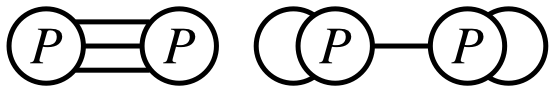}
\hfil
\end{center}
\caption{Left: The diagram for the interaction vertex $\sum_{i,j=1}^R
\tilde \Lambda_{ij} P^i_{abc} \phi_a^j \phi_b^j \phi_c^j$. Right: The diagrams obtained by evaluating \eq{eq:o2} 
through \eq{eq:connecttilde}.}
\label{fig:feynman}
\end{figure*}

Through the replacement \eq{eq:connecttilde},
\eq{eq:o2} can be computed by the standard procedure 
using the Wick theorem and Feynman diagrams. A Wick contraction is performed by
\[
\langle \phi_a^i \phi_b^j\rangle_\phi=\frac{1}{2\beta} \delta_{ab}\delta^{ij},
\label{eq:wick}
\]
which can be derived by an explicit computation of \eq{eq:defofexpbeta}.
The Feynman diagram for the 
vertex $\sum_{i,j=1}^R
\tilde \Lambda_{ij} P^i_{abc} \phi_a^j \phi_b^j \phi_c^j$ is shown in the left figure 
of Figure~\ref{fig:feynman}. Each leg is supposed to bring the two indices of $\phi_a^i$, and 
a Wick contraction connects two legs with the identification of  their indices as in \eq{eq:wick}.
A caution is that each leg on one vertex brings independent lower indices, but a common
upper index.  

Now let us apply the Wick contractions to what is obtained by replacing \eq{eq:o2} with 
\eq{eq:connecttilde}.
We find the two diagrams shown in the right figure of Figure~\ref{fig:feynman}. Their degeneracy factors are 6 and 9, respectively, by counting the numbers of the ways to connect the legs of the two vertices. 
Since $j$ and $j'$ in \eq{eq:o2} are identified by the Wick contractions, we also get 
$\sum_{j=1}^R \tilde \Lambda_{i j} \tilde \Lambda_{i'j}=\Lambda_{ii'}$ as a factor (See \eq{eq:defoftildelam}.). 
Thus, we obtain
\[
\langle (P\tilde \phi^3)^2 \rangle^c_{\tilde \phi}=\gamma_3
\sum_{i,j=1}^R  \Lambda_{ij} \left( 6 P^i_{abc} P^j_{abc}+9 P^i_{aab}P^j_{bcc}  \right),
\label{eq:cum2}
\]
where one notices that the factor $(2\beta)^3$ coming from
the replacement \eq{eq:connecttilde}
is exactly canceled by the factors of the three Wick contractions \eq{eq:wick} 
performed for the evaluation.
Putting the result \eq{eq:cum2} into \eq{eq:seff}, one obtains the effective action in
the second order of $P_{abc}^i$ as
\s[
&S_{eff}^{(2)}(P)=\sum_{i,j=1}^R 
\left( \delta_{ij} P^i_{abc} P^j_{abc}
+2 \gamma_3  
 \Lambda_{ij} t \left( 6 P^i_{abc} P^j_{abc}+9 P^i_{aab}P^j_{bcc}  \right) \right).
 \label{eq:seff2}
\s]

The computation of \eq{eq:ffromseff} has now been reduced to diagonalizing the 
quadratic expression
\eq{eq:seff2}. The upper and lower indices can independently be diagonalized, because 
\eq{eq:seff2} has the form of the direct product with respect to these indices. More 
explicitly, since $\Lambda_{ij}$ is real and symmetric, we can consider the following decomposition
into the eigenspaces: 
\[
\Lambda_{ij}=\sum_{\lambda_{ev}} \lambda_{ev}\, v^{\lambda_{ev}}_i v^{\lambda_{ev}}_j,
\label{eq:lambdadecomp}
\]
where $v^{\lambda_{ev}}$ are the orthonormal eigenvectors,
and the sum is over all the eigenvalues (with their degeneracies). By putting this and 
$\sum_{\lambda_{ev}} v^{\lambda_{ev}}_i v^{\lambda_{ev}}_j=\delta_{ij}$ into
\eq{eq:seff2}, we obtain a decomposition,
\s[
&S_{eff}^{(2)}(P)=\sum_{\lambda_{ev}} \left(
 P^{\lambda_{ev}}_{abc} P^{\lambda_{ev}}_{abc}
+2 \gamma_3  
 \lambda_{ev} t \left( 6 P^{\lambda_{ev}}_{abc} P^{\lambda_{ev}}_{abc}
 +9 P^{\lambda_{ev}}_{aab}P^{\lambda_{ev}}_{bcc}  \right) \right),
 \label{eq:seff2diaglam}
\s]
where $P^{\lambda_{ev}}_{abc}:=\sum_{i=1}^R v^{\lambda_{ev}}_iP^{i}_{abc}$.

Next let us diagonalize the lower index part in \eq{eq:seff2diaglam}, 
\[
P_{abc} P_{abc}+ 2 \gamma_3\lambda_{ev}\,t
 \left( 6 P_{abc} P_{abc}+9 P_{aab}P_{bcc}  \right),
 \label{eq:pquad}
\]
where for brevity we omit $\lambda_{ev}$ from $P^{\lambda_{ev}}_{abc}$.
Let us separate $P_{abc}$ into the 
tensor part $P^T_{abc}$ and the vector part $P^V_{abc}$, which are defined by\footnote{This decomposition
can be understood as follows. First $P_{abb}$ represents the $O(N)$-vector part of $P_{abc}$.
Its embedding to $P_{abc}$ is given by an expression proportional to \eq{eq:ptv}. 
Then the coefficient can be determined by the condition that $P^T_{abc}$ does not contain
the vector part: $P^T_{abb}=0$.}
\[
\begin{split}
P_{abc}&=P^T_{abc}+P^V_{abc}, \\
P^V_{abc}&=\frac{1}{N+2} \left( 
P_{add}\delta_{bc}+P_{bdd}\delta_{ca}+P_{cdd}\delta_{ab}
\right).
\end{split}
\label{eq:ptv}
\]
It is easy to check that $P^T_{abc}P^V_{abc}=0$ and 
$P^V_{abc} P^V_{abc}=\frac{3}{N+2}  P_{abb}P_{acc}$.
In particular, the former identity implies that $P^T_{abc}$ and $P^V_{abc}$ are independent
degrees of freedom.
Then, by using \eq{eq:ptv} and the identities above, \eq{eq:pquad} can be expressed as
\[
( 1+12\gamma_3\lambda_{ev}\,  t) P^T_{abc} P^T_{abc}+
\left( 1+6(N+4) \gamma_3 \lambda_{ev}\,  t\right) P^V_{abc} P^V_{abc}.
\label{eq:quadev}
\]
The number of independent components contained in $P^T_{abc}$ and $P^V_{abc}$ 
are $\# P^T=N(N+1)(N+2)/6-N=N(N+4)(N-1)/6$ and $\# P^V=N$, respectively. 
Therefore, by putting this diagonal form into \eq{eq:ffromseff} and integrating 
over $P^V$ and $P^T$, 
we finally obtain the expression of $f_{N,R,\Lambda_{ij}}(t)$ from the quadratic order $S_{eff}^{(2)}(P)$ 
as 
\s[
&f^{(2)}_{N,R,\Lambda_{ij}}(t)=const. \int dP e^{-S^{(2)}_{eff}(P)} \\
&\hspace{5mm}=\prod_{\lambda_{ev}} 
( 1+12\gamma_3 \lambda_{ev} t)^{-\frac{N(N+4)(N-1)}{12}}
\left( 1+6(N+4)\gamma_3 \lambda_{ev}\, t\right)^{-\frac{N}{2}} \\
&\hspace{5mm}=\prod_{\lambda_{ev}} h_{N,R}(\lambda_{ev}t),
\label{eq:expf2}
\s]
where we have determined the overall factor by requiring $f^{(2)}_{N,R,\Lambda_{ij}}(0)=1$,
the product is over all the eigenvalues of the matrix $\Lambda_{ij}$,
and $h_{N,R}(x)$ is defined in \eq{eq:exph}.

For the computation of the observables discussed in Section~\ref{sec:observables}, 
we consider $\Lambda_{ij}=\lambda+\lambda_d \delta_{ij}$. 
In this case, the matrix $\Lambda_{ij}$ has one eigenvalue $\lambda R+\lambda_d$ 
with the eigenvector $(1,1,\ldots,1)$, and the eigenvalue $\lambda_d$ 
with degeneracy $R-1$ with any of the vectors transverse to $(1,1,\ldots,1)$ as the eigenvectors. 
Therefore, from \eq{eq:expf2}, we obtain
\[
f^{(2)}_{N,R,\lambda+\lambda_d \delta_{ij}}(t)=h_{N,R}(\lambda R t +\lambda_d t) \ h_{N,R}(\lambda_d t )^{R-1}.
\]
This is the leading-order result shown in \eq{eq:fleadwithh}.

As we will see later in Section~\ref{sec:mc}, there are some deviations 
between the leading-order result above and the numerical simulations. 
To see how the situation changes by adding some corrections, 
we have also computed the next-leading order. 
The details of the computation are given in \ref{app:fnext}.
The final result is 
\[
 f^{next-leading}_{N,R,\lambda,\lambda_d}(t)
 =f^{leading}_{N,R,\lambda,\lambda_d}(t) 
 \left(1 - \langle S_{eff}^{(4)}(P) \rangle_P \right),
 \label{eq:final4}
\]
where $f^{next-leading}_{N,R,\lambda,\lambda_d}(t)$ is the sum of the leading and the next-leading orders, and 
\s[
&\langle S_{eff}^{(4)}(P) \rangle_P
\\&=-\frac{4 t^2}{4!}
\Big [
\gamma_6 
 R G_1 (x_1,y_1) -3 (\gamma_3^2-\gamma_6) 
 \big(G_2(x_2,y_2)+
 (\lambda R+\lambda_d)^2 G_3(x_3,y_3) +(R-1) \lambda_d^2 G_3(x_4,y_4)
\big)
\Big ]
\label{eq:finalexps4}
\s]
with the definitions of $G_i,x_i,y_i$ given by from \eq{eq:count4} to \eq{eq:x34}.

%%%%%%%%%%%%%%%%%%%%%%%%%%5
%%%%%%%%%%%%%%%%%%%%%%%%%%%
%%%%%%%%%%%%%%%%%%%%%%%%%%%%

\section{Saddle point analysis in the leading order}
\label{sec:saddle}
The integral expressions of the observables \eq{eq:expobss} 
in the leading order do not 
seem to have explicit expressions with known functions.
Therefore, a way to obtain their explicit values is to numerically perform the integrations. 
This will be performed in Section~\ref{sec:mc} to compare with
the Monte Carlo results.
In this section, on the other hand, we will apply the saddle point approximation to the integrals
to obtain a qualitative global picture of the phase structure of the model.

To discuss the saddle point approximation of the partition function \eq{eq:zwithf},
let us consider the minus of the logarithm of the integrand, which is given 
by\footnote{Namely, we rewrite the integral in the form, $\int dr e^{-  {\cal F}_{N,R} (\lambda, k,r)}$,
as commonly performed in the saddle point approximation.}
\s[
& {\cal F}_{N,R} (\lambda, k,r)
\equiv f_0-(NR-1)\log r - \log f_{N,R}(\lambda r^6) +k r^2.
 \label{eq:logofint}
 \s]
with $f_0=-\log V_{NR-1}$.
A saddle point $r=r_*$ of the integral is determined by
\[
\left. \frac{\partial}{\partial r} {\cal F}_{N,R} (\lambda, k,r) \right | _{r=r_*}=0
\label{eq:saddleeq}
\]
with the second derivative being positive at the point.
As $f_{N,R}$, we take \eq{eq:repf}, which is the expression in the leading order of $1/R$:
\s[
&{\cal F}^{1/R,leading}_{N,R}(\lambda,k,r):= f_0 -(NR-1) \log r+A_0 \log(1+A_1 r^6 )+B_0 \log
(1+ B_1 r^6)+k r^2
\label{eq:flead}
\s]
with
\[
\begin{split}
&A_0=\frac{N(N-1)(N+4)}{12},\ A_1=\frac{12 \lambda  }{N^3 R^2}, \\
&B_0=\frac{N}{2},\ B_1=\frac{6(N+4) \lambda}{N^3 R^2}.
\end{split}
\]

In the saddle point approximation, the free energy of the model is approximated by
\[
F^{free}_{N,R}(\lambda,k)\sim {\cal F}^{1/R,leading}_{N,R}(\lambda,k,r^*)+\hbox{unimportant terms},
\]
where the unimportant terms contain $\log V_{NR-1}$, which does not depend on $\lambda$ or $k$, 
and also some lower order terms in $N$, which will be discussed in the last paragraph of this section.

Let us first show that there exists a unique solution to the saddle point equation \eq{eq:saddleeq},
for the leading order expression \eq{eq:flead}, in the integration 
region $r\geq 0$. To see this, it is convenient to use a new parametrization of $R$ in terms of $\alpha$ as $R=R_c (1+\alpha)$ with $R_c=(N+1)(N+2)/2$ and $-1<\alpha$.
Then, by noting $N R_c=6 (A_0+B_0)$, the saddle point equation \eq{eq:saddleeq} with \eq{eq:flead} can be written as
\[
N R_c \alpha-1 +\frac{6 A_0}{1+A_1 r_*^6}+\frac{6 B_0}{1+B_1 r_*^6}=2 k r_*^2.
\label{eq:saddleinr}
\]
The lefthand side is obviously  a decreasing function of $r_*$ with a maximum at $r_*=0$, while 
the righthand side is an increasing function from zero to the infinity. 
Since the maximum on the lefthand side is 
$NR_c\alpha-1+6A_0 + 6B_0=NR-1$,
there always exists a unique solution of $r_*$ for $N,R\geq1$.  
Moreover, the solution is smooth: $r_*$ does not jump in a discrete manner, 
when the parameters are continuously changed, because the $r_*$-dependence 
of each side continuously changes.  
This turns down the possibility that the model has a discontinuous phase transition
in this treatment.  

To discuss the solution more quantitatively with approximations, 
let us restrict ourselves to the parameter range of 
our interest: $\lambda\sim O(1)$, $N\gtrsim O(10)$, and $k\lesssim O(1)$.
In addition, for the simplicity of the following discussions, let us avoid the region around 
$\alpha\sim 0$.  
By noting that $NR_c,$ and $A_0$ are of order $O(10^3)$ or larger, 
one can find that, for each case of $\alpha<0$ and $\alpha>0$, 
there are only two relevant terms among all in \eq{eq:saddleinr}. 
For $\alpha<0$,  the first and third terms on the lefthand side are relevant, and 
for $\alpha>0$, the first term on the lefthand side and the one on the righthand side are relevant.
By solving the equations under taking these relevant terms only, the solutions are respectively given 
by
\[
r_*^2\sim\left \{ 
\begin{array}{ll}
\left(\frac{1}{A_1} \left( -
\frac{6 A_0}{N R_c \alpha}-1
\right)\right)^{\frac{1}{3}},& \alpha<0,\\
\frac{N R_c \alpha}{2 k}, & \alpha>0.
\end{array}
\right.
\label{eq:rstar}
\]
The first case shows divergent behavior for $\alpha\rightarrow -0$. However, this should not
be taken as it is, since the transition should not have such an abrupt behavior as discussed above. In fact, the simplification taken above brakes down in the vicinity of $\alpha\sim0$, and the real behavior is such that $r_*^2$ smoothly 
interpolates between the two parameter regions in the vicinity of $\alpha\sim 0$. 

\begin{figure*}
\begin{center}
\includegraphics[width=5cm]{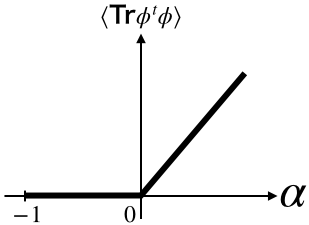}
\hfil
\includegraphics[width=5cm]{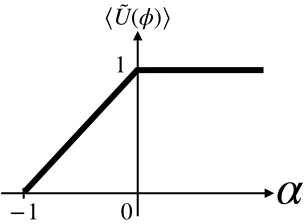}
\hfil
\includegraphics[width=5cm]{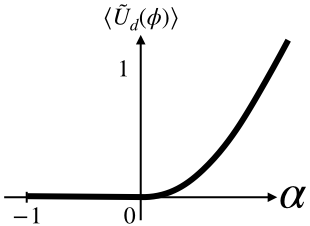}
\caption{The behavior of the observables with respect to $\alpha=(R-R_c)/R_c$
in the large $N$ limit, based on the saddle point analysis with the use of the leading order 
result of $f_{N,R}$.  
The parameters are assumed to be  $\lambda\sim O(1)$ and $k\lesssim O(1)$.
}
\label{fig:obs}
\end{center}
\end{figure*}
The $r_*^2$ in \eq{eq:rstar} has different large-$N$ behavior in the two regions: $N^\frac{7}{3}$ for $\alpha<0$ and
$N^3$ for $\alpha>0$. By normalizing  $r_*^2$ with the common 
factor $(NR_c)^{-1}$ for both the regions, we obtain
\[
\tilde r_*^2\sim\left \{ 
\begin{array}{cl}
0,& \alpha\leq 0,\\
\frac{\alpha}{2k}, & \alpha>0,
\end{array}
\right.
\label{eq:rtildestar}
\]
in the $N\rightarrow \infty$ limit, where $\tilde r_*^2:=(NR_c)^{-1} r_*^2$. 
See the leftmost figure in Figure~\ref{fig:obs}.
This characterizes the transition at $\alpha =0$ as a continuous phase transition 
with $\langle \hbox{Tr}(\phi^t \phi)\rangle /(NR_c) = 0$ 
for $\alpha\leq 0$ and $\langle \hbox{Tr}(\phi^t \phi)\rangle /(NR_c) =\frac{\alpha}{2k}$ for $\alpha>0$
in the thermodynamic limit.

The other observables can be treated in similar manners. By taking \eq{eq:expobss}, putting $r=r_*$, and taking the leading order in large $N$, one obtains
\[
\begin{split}
\langle U(\phi) \rangle_{leading}&\sim
\left\{ 
\begin{array}{cl}
\frac{N^3}{12 \lambda}(1+\alpha), &\hbox{for } \alpha \leq 0,\\
\frac{N^3}{12 \lambda}, &\hbox{for } \alpha>0,\\
\end{array}
\right.\\
\langle U_d(\phi) \rangle_{leading}&\sim
\left\{
\begin{array}{cl}
\frac{N^3(1+\alpha)}{12 \lambda (-\alpha)}, &\hbox{for } \alpha\leq 0, \\
\frac{N^5}{16 k^3} \frac{\alpha^3}{(1+\alpha)^2}, &\hbox{for } \alpha>0.
\end{array}
\right.
\end{split}
\label{eq:obsu}
\]
The divergence of $\langle U_d(\phi) \rangle_{leading}$ in $\alpha\rightarrow -0$ should not 
be taken as it is, because of the same reason mentioned above for $r_*^2$.
By normalizing $\tilde U(\phi):=(N^3/12 \lambda)^{-1} U(\phi)$
and $\tilde U_d(\phi):=(N^5/16 k^3)^{-1} U_d(\phi)$, one obtains
\[
\begin{split}
\langle \tilde U(\phi) \rangle_{leading}&\sim
\left\{ 
\begin{array}{cl}
1+\alpha, &\hbox{for } \alpha \leq 0,\\
1, &\hbox{for } \alpha>0,\\
\end{array}
\right.\\
\langle \tilde U_d(\phi) \rangle_{leading}&\sim
\left\{
\begin{array}{cl}
0, &\hbox{for } \alpha\leq 0, \\
\frac{\alpha^3}{(1+\alpha)^2}, &\hbox{for } \alpha>0.
\end{array}
\right.
\end{split}
\]
See the middle and rightmost figures in Figure~\ref{fig:obs}.
The results support the same conclusion that there is a continuous phase transition at $\alpha=0$.

Let us finally comment on the consistency of the above saddle point approximation in the large $N$ case. 
From \eq{eq:flead} and \eq{eq:rstar}, it is straightforward by some explicit computations 
to find the expansion of ${\cal F}^{1/R,leading}_{N,R}$ around $r\sim r_*$ as
\[
{\cal F}^{1/R,leading}_{N,R} \sim a_0+a_2 (r-r_*)^2 +a_3 (r-r_*)^3+\cdots,
\label{eq:fexpinr}
\]
where $a_n:=\frac{1}{n!}\frac{d}{dr^n}{\cal F}^{1/R,leading}_{N,R}(\lambda,k,r)|_{r=r_*}$
can easily be estimated as $a_0\sim O(N^3)$, and 
\s[
&a_2\sim O(N^{2/3}),\ a_3\sim O(N^{-1/2}) , \hbox{ for }\alpha<0,\\
&a_2\sim O(1),\ a_3\sim O(N^{-3/2}) , \hbox{ for }\alpha>0.
\s]
Therefore the integral is dominated by the saddle point value $a_0$,
and the Gaussian integration around the saddle 
point\footnote{$\int dr \, e^{-a_2 (r-r_*)^2}$ generates a contribution proportional to $\log a_2\lesssim O(\log N)$ to the free energy, and this is subdominant.}
 and the higher orders in $r-r^*$ are subdominant. 
In addition, the insertion of the observables ${\cal O}$ as in \eq{eq:expobss} for the computations of their
expectation values does not change the location of the saddle point in the leading order,
because the additional contribution $\log {\cal O}$ to \eq{eq:fexpinr} is $O(\log N)$, and  
cannot become comparable to the leading $O(N^3)$ terms.

 %%%%%%%%%%%%%%%%%%%%%%%%%%%%%%%   
 %%%%%%%%%%%%%%%%%%%%%%%%%%%%%%%  
 %%%%%%%%%%%%%%%%%%%%%%%%%%%%%%%     

\section{Comparison with Monte Carlo simulations} 
\label{sec:mc} 
 
%=====================
\begin{figure*}   
\begin{center}    
\includegraphics[width=47.0mm,clip]{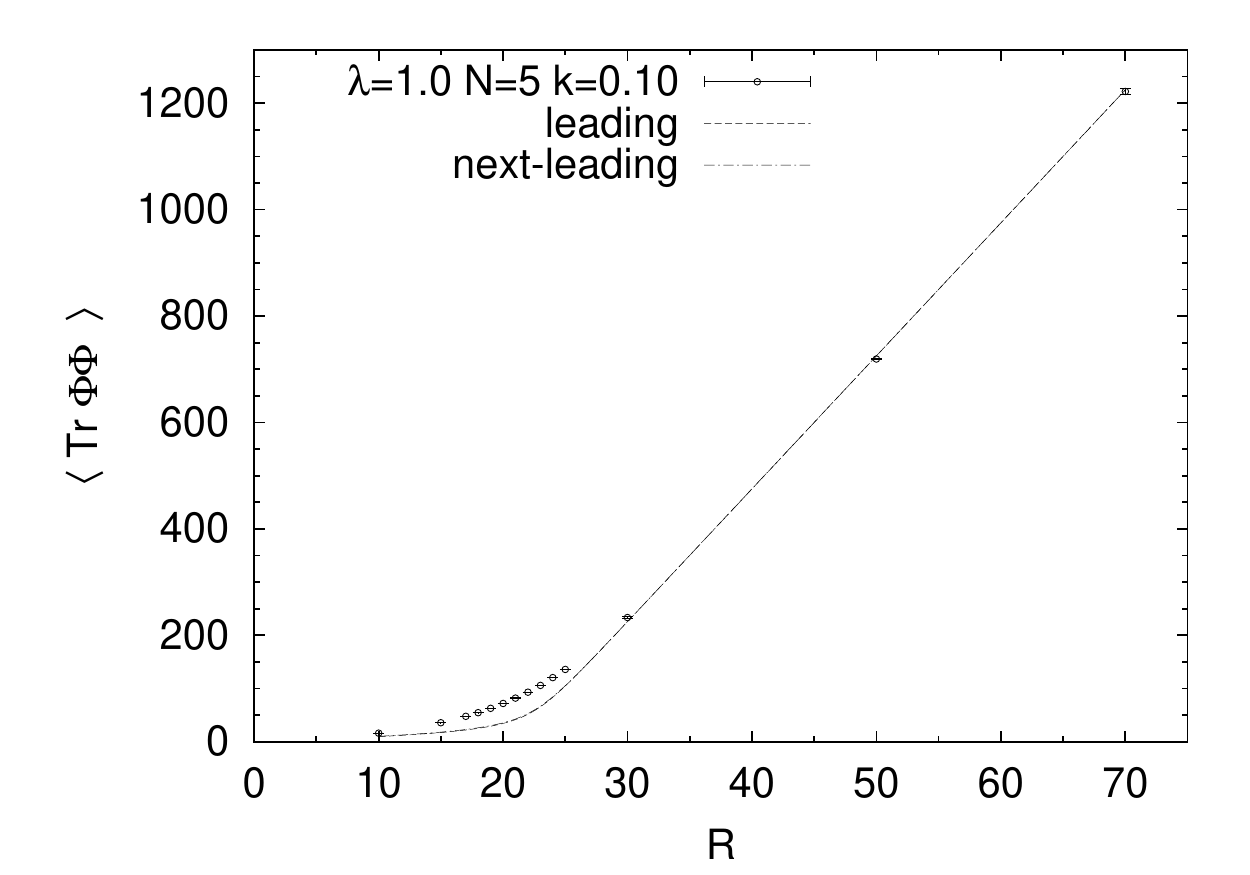}  
\includegraphics[width=47.0mm,clip]{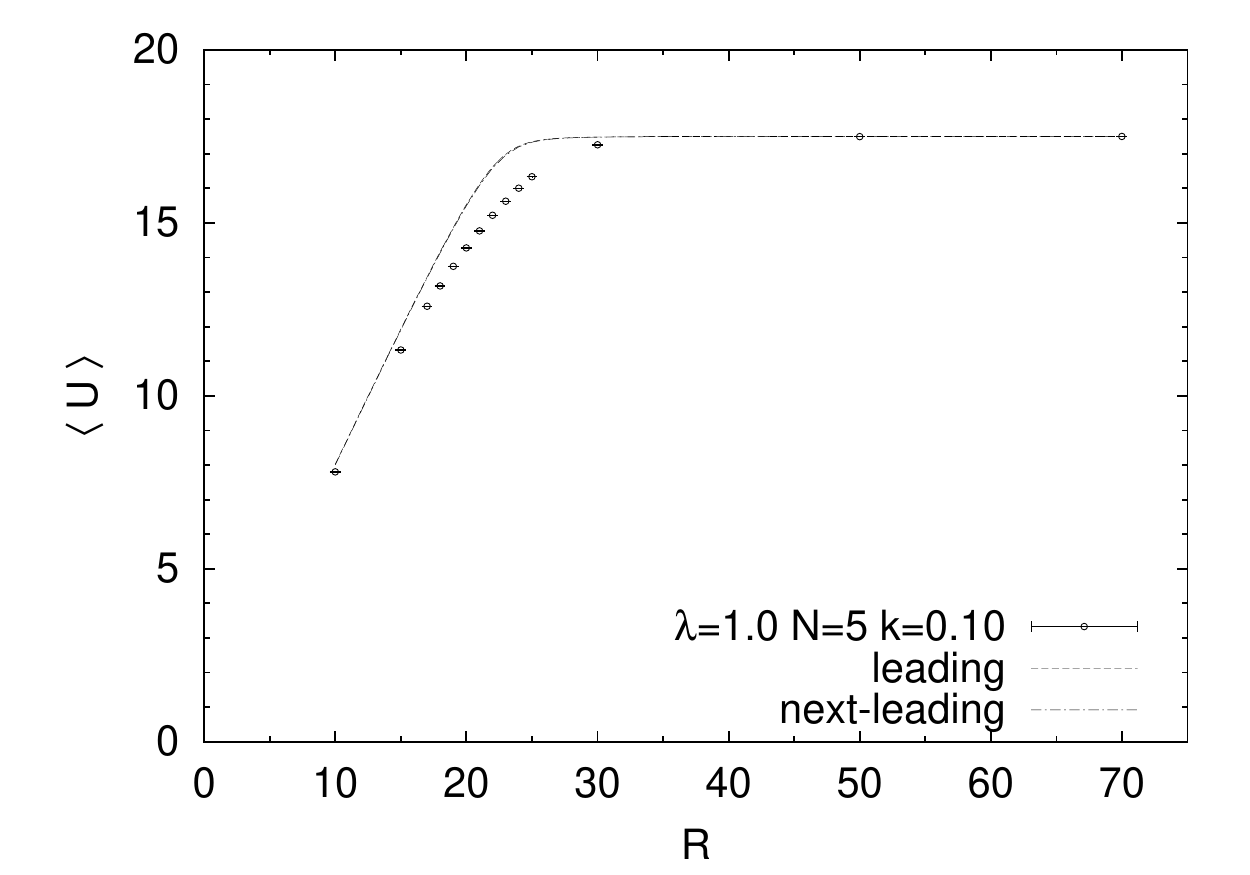}   
\includegraphics[width=47.0mm,clip]{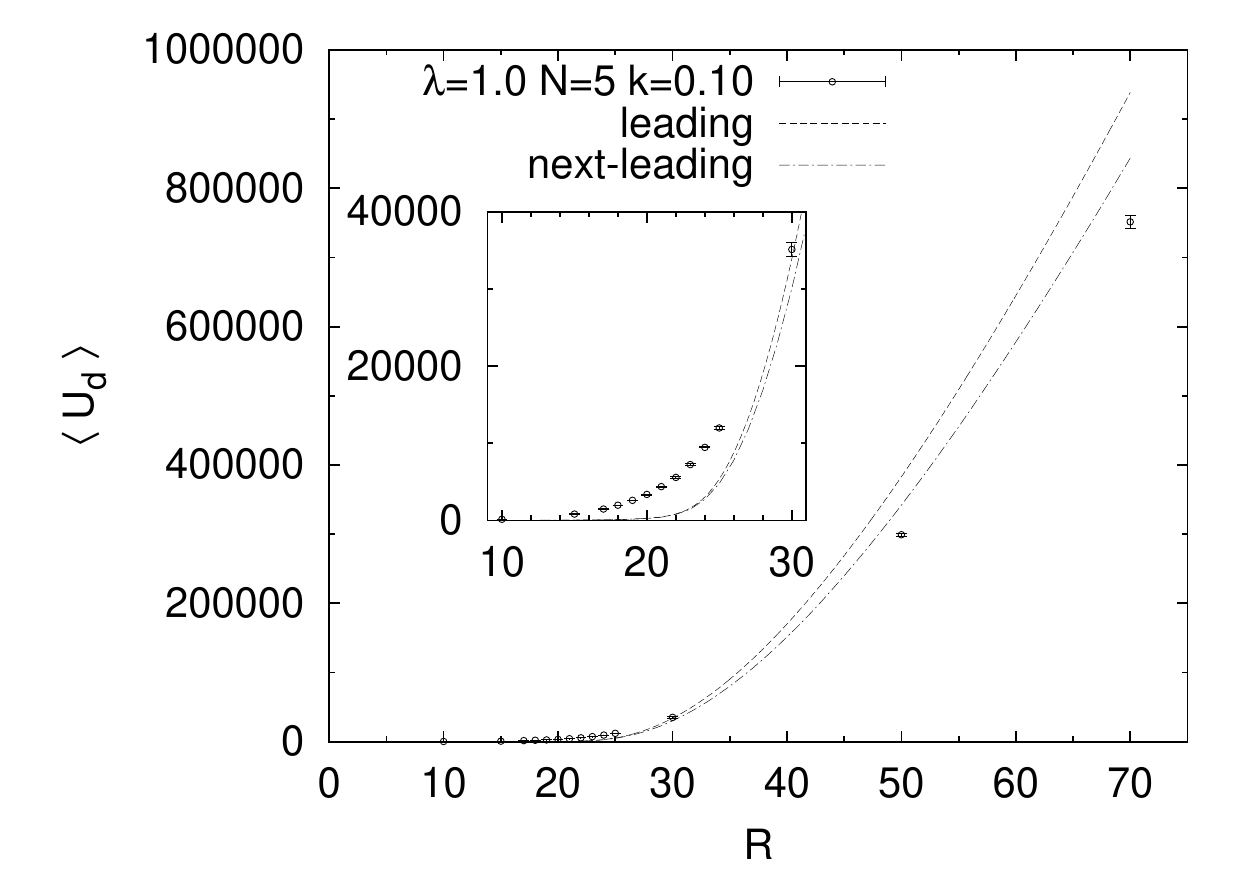}\\ 
\includegraphics[width=47.0mm,clip]{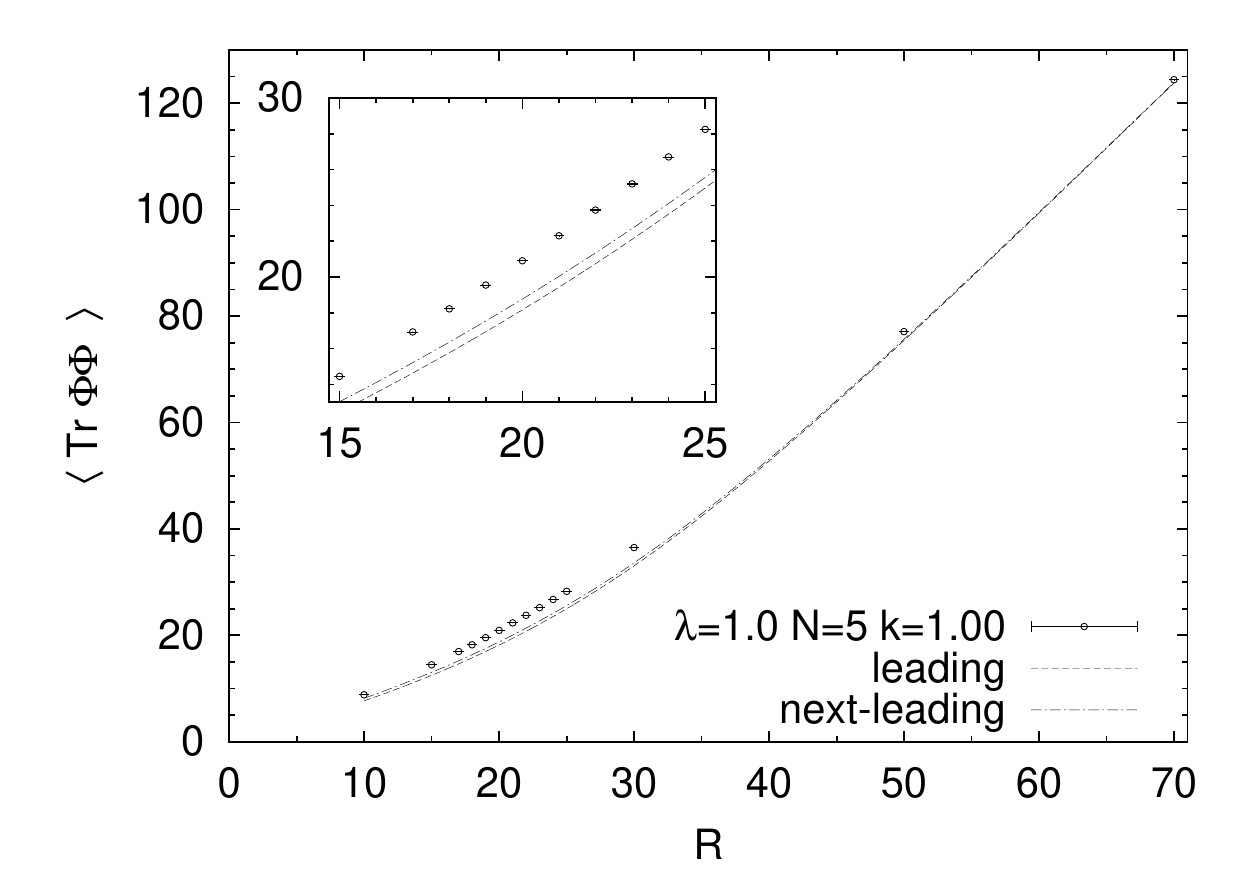} 
\includegraphics[width=47.0mm,clip]{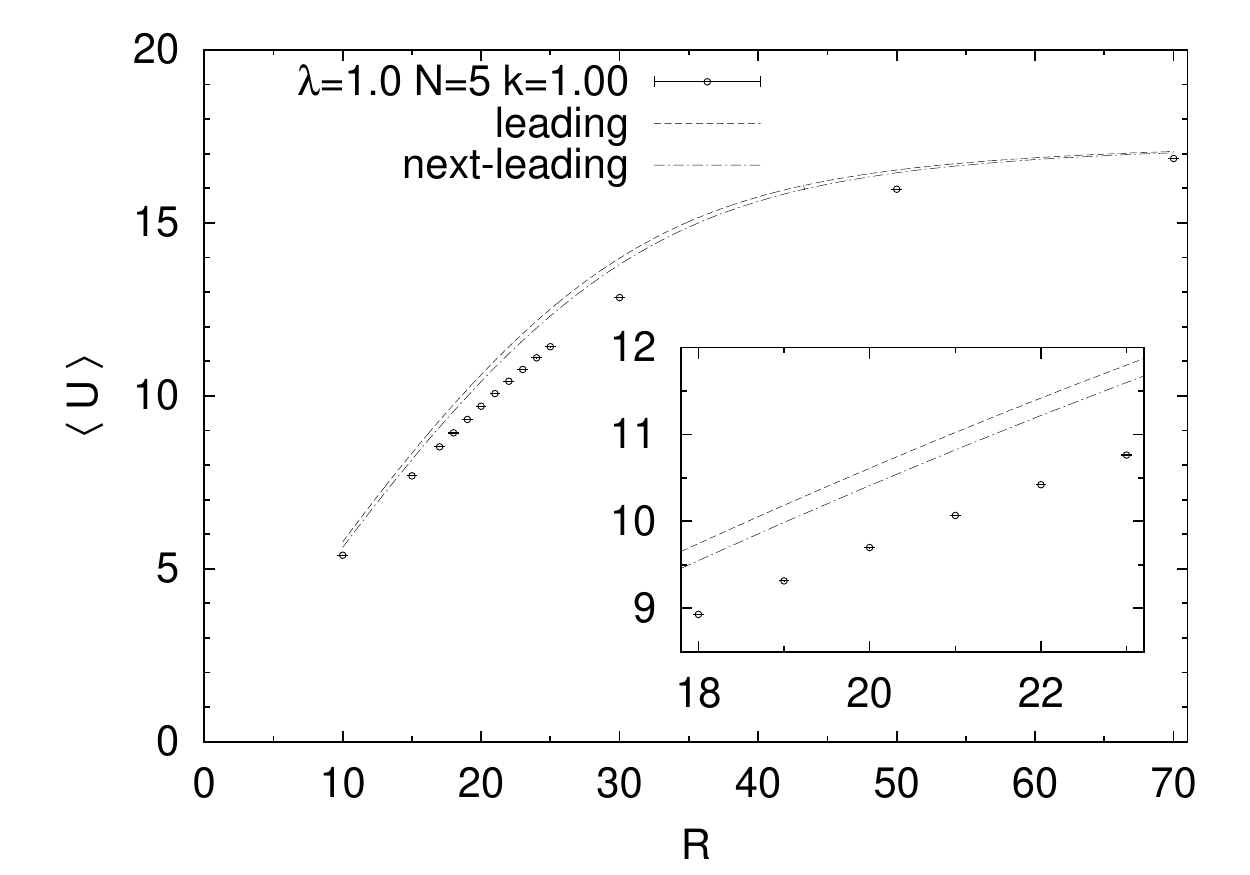} 
\includegraphics[width=47.0mm,clip]{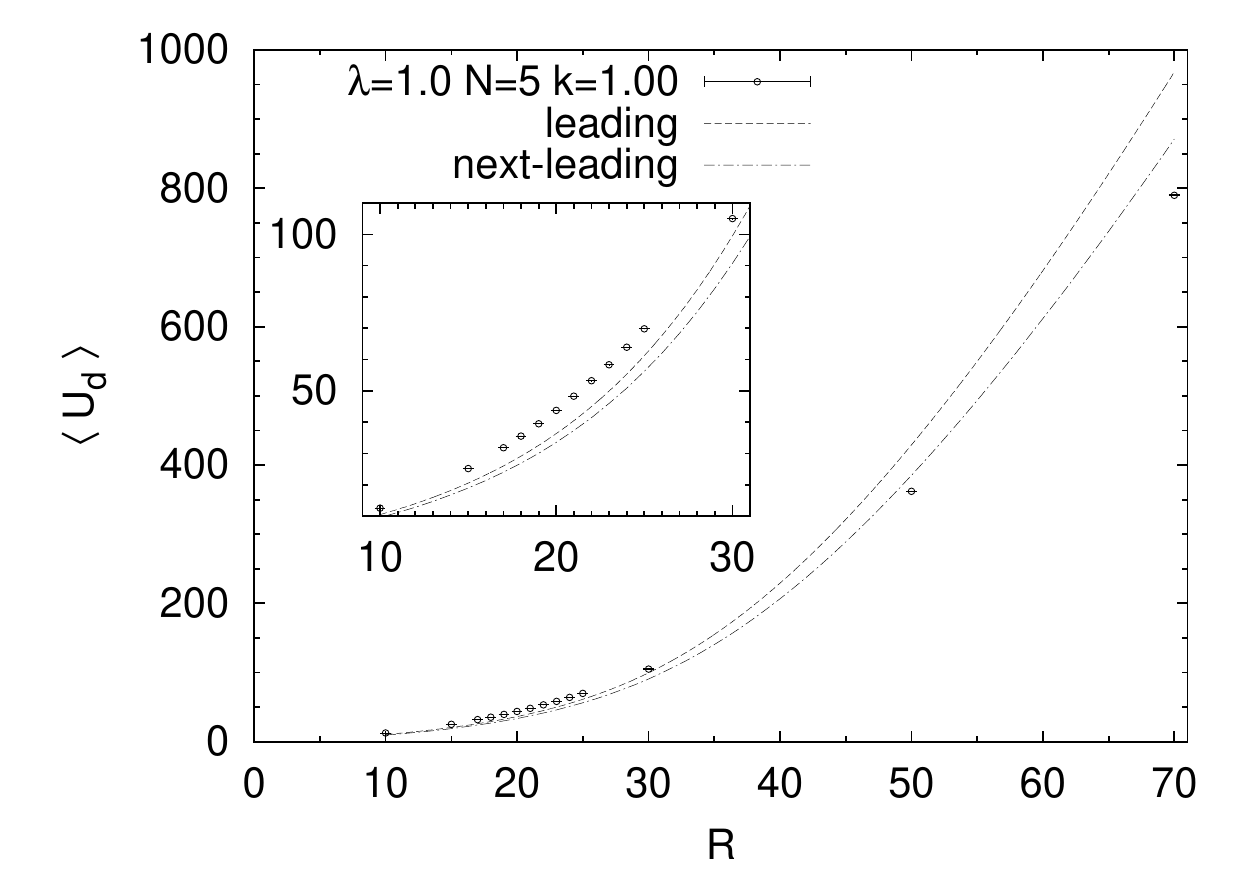}   
\end{center}   
\caption{The numerical results of e.v. of the observables discussed in Section~\ref{sec:observables} for 
$\lambda=1$, $N=5$ and 
$k=0.1$ (top three) 
and 
$k=1.0$ (bottom three) against $R$. 
Plotted points represent the Monte Carlo results and `leading' and `next-leading' mean 
the evaluations based on Eq.(12) with perturbatively evaluated $f_{N,R,\lambda,\lambda_d}(t)$ 
in the leading and next-leading orders, respectively. 
There are some small windows within the figures, where one can see which of the leading and next-leading lines
exists above/below the other, and can also see more clearly the results for $\langle U_d \rangle$ 
in the transition region around $R=R_c$.  
For the clarity of the small $R$ regions, which are unclear in some of the figures, we 
provide Figure \ref{FigN05a}.
}    
%---
\label{FigN05}  
\end{figure*}      
%=====================
\begin{figure*}     
\begin{center}    
\includegraphics[width=47.0mm,clip]{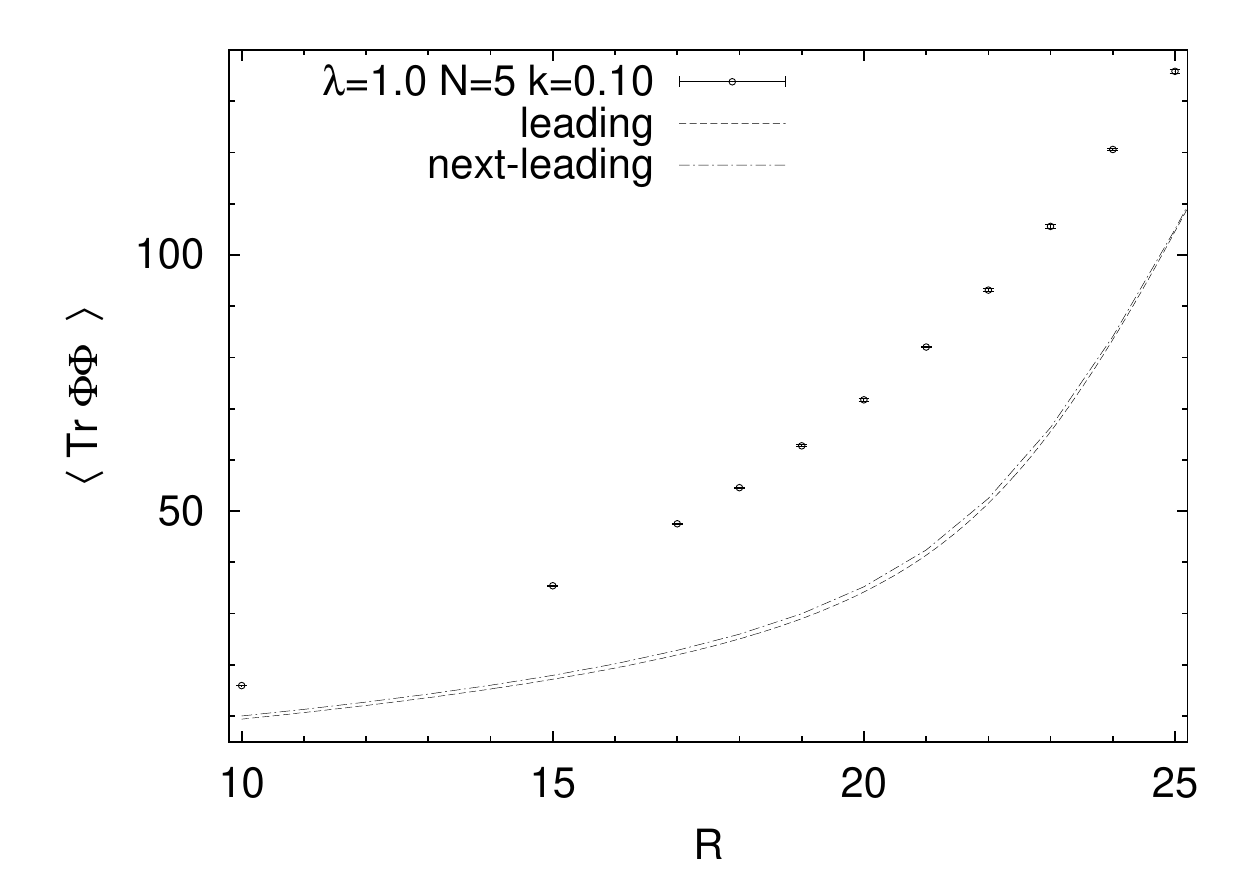} 
\includegraphics[width=47.0mm,clip]{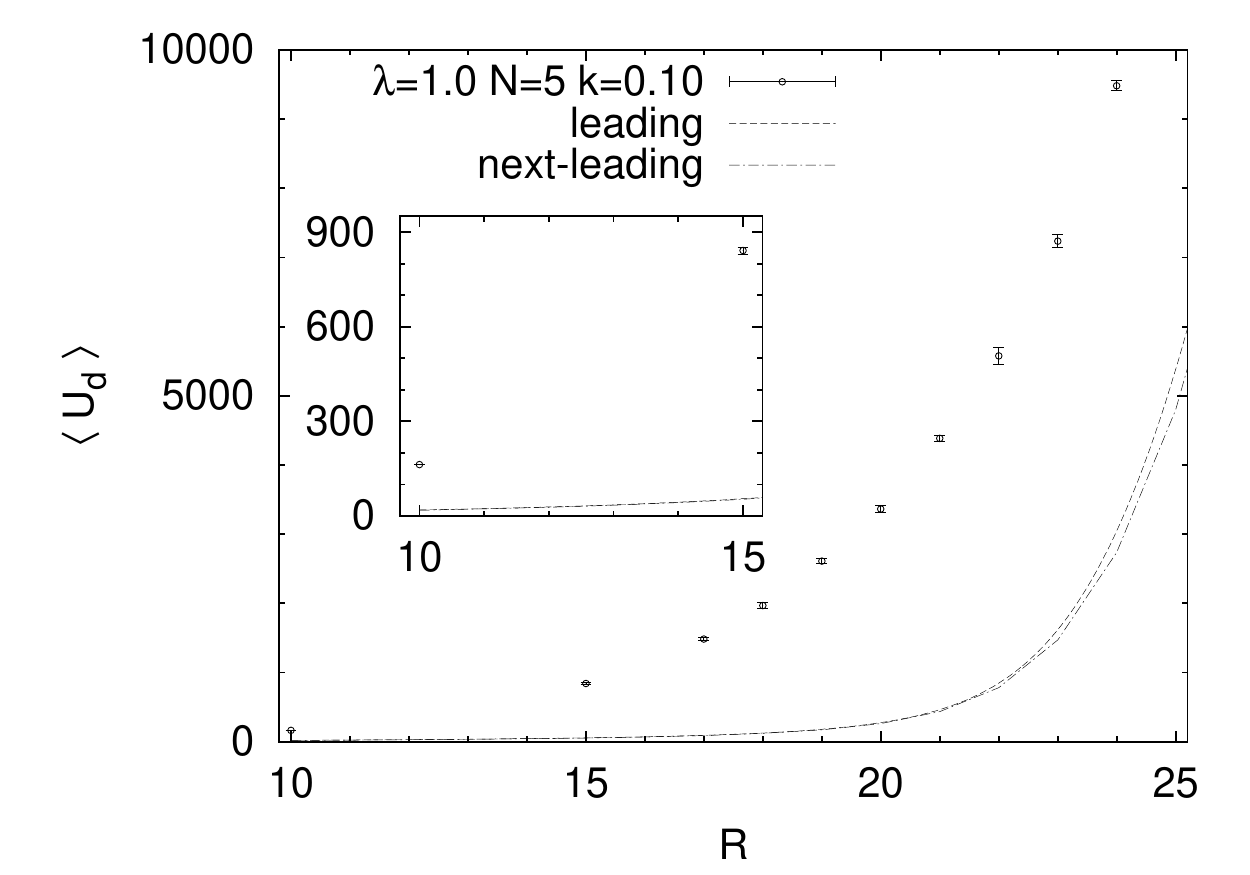}\\ 
\includegraphics[width=47.0mm,clip]{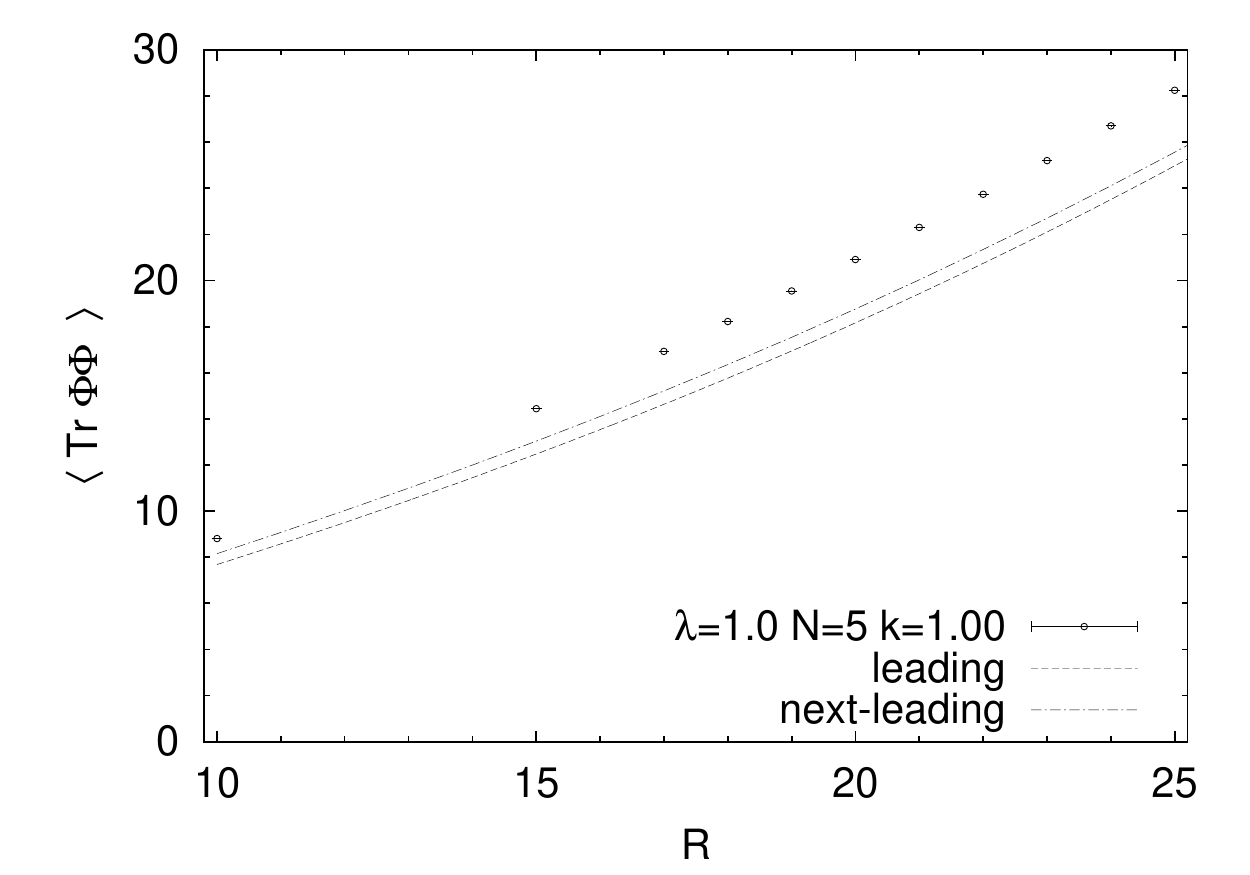} 
\includegraphics[width=47.0mm,clip]{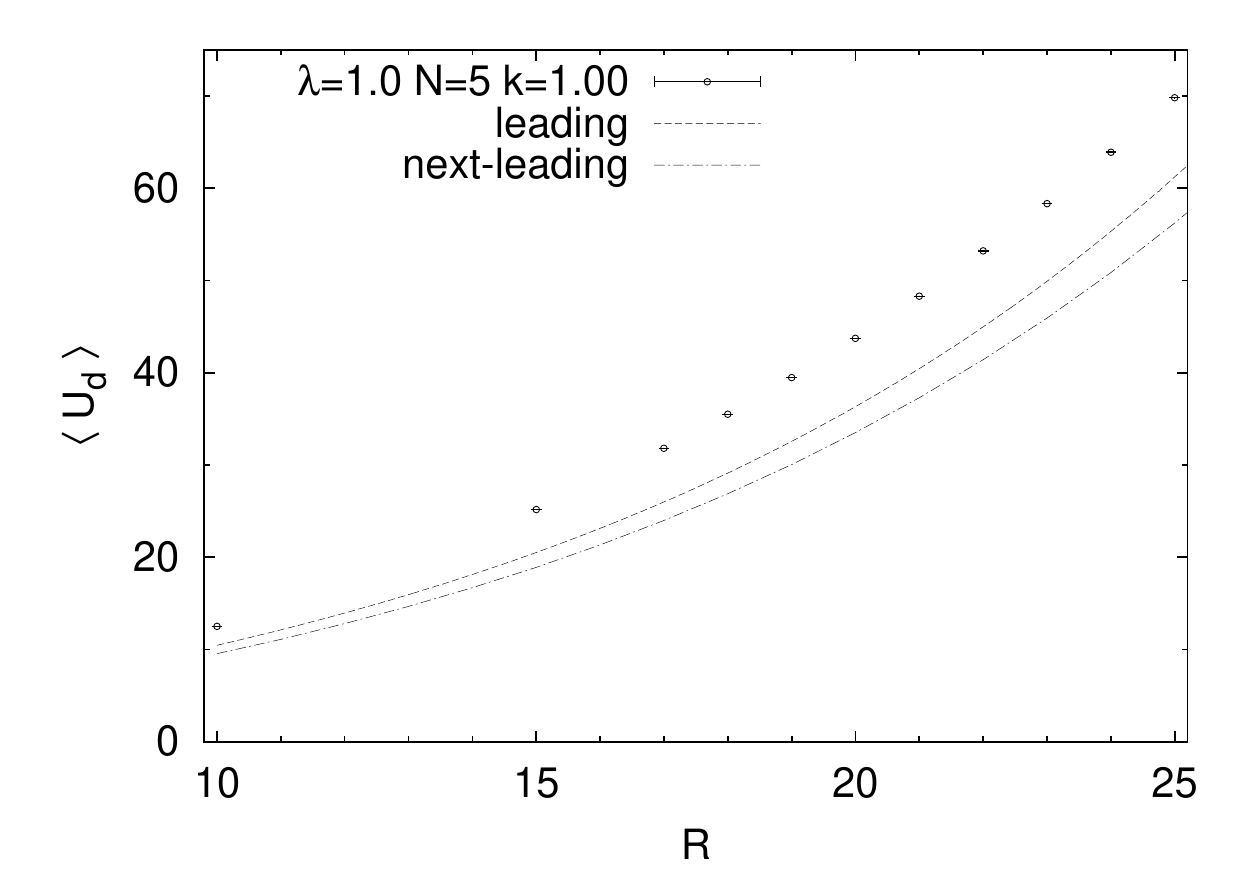}  
\end{center}   
\caption{ 
Magnification of the small $R$ regions that are unclear in Figure \ref{FigN05}.
}   
%--- 
\label{FigN05a}  
\end{figure*}    
%=====================
\begin{figure*} 
\begin{center} 
\includegraphics[width=47.0mm,clip]{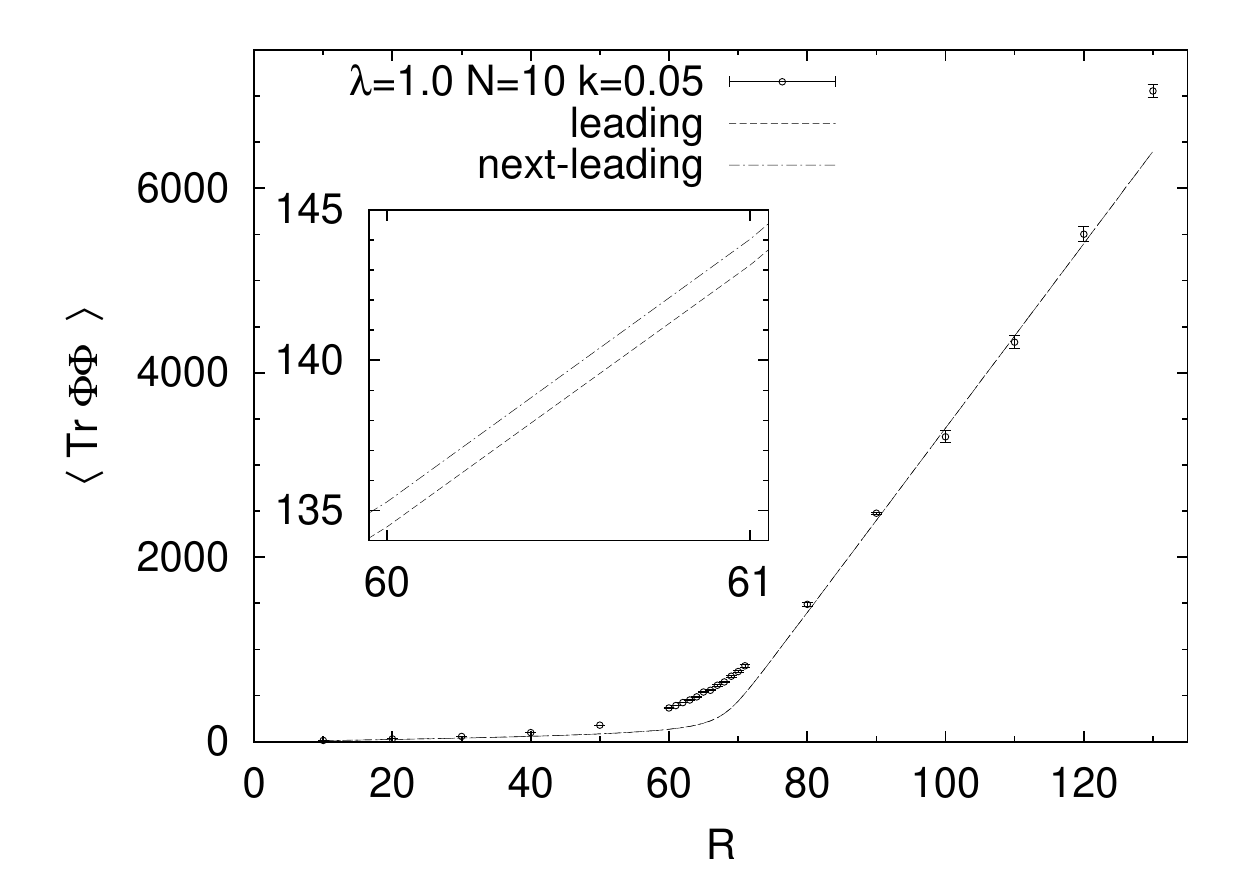}
\includegraphics[width=47.0mm,clip]{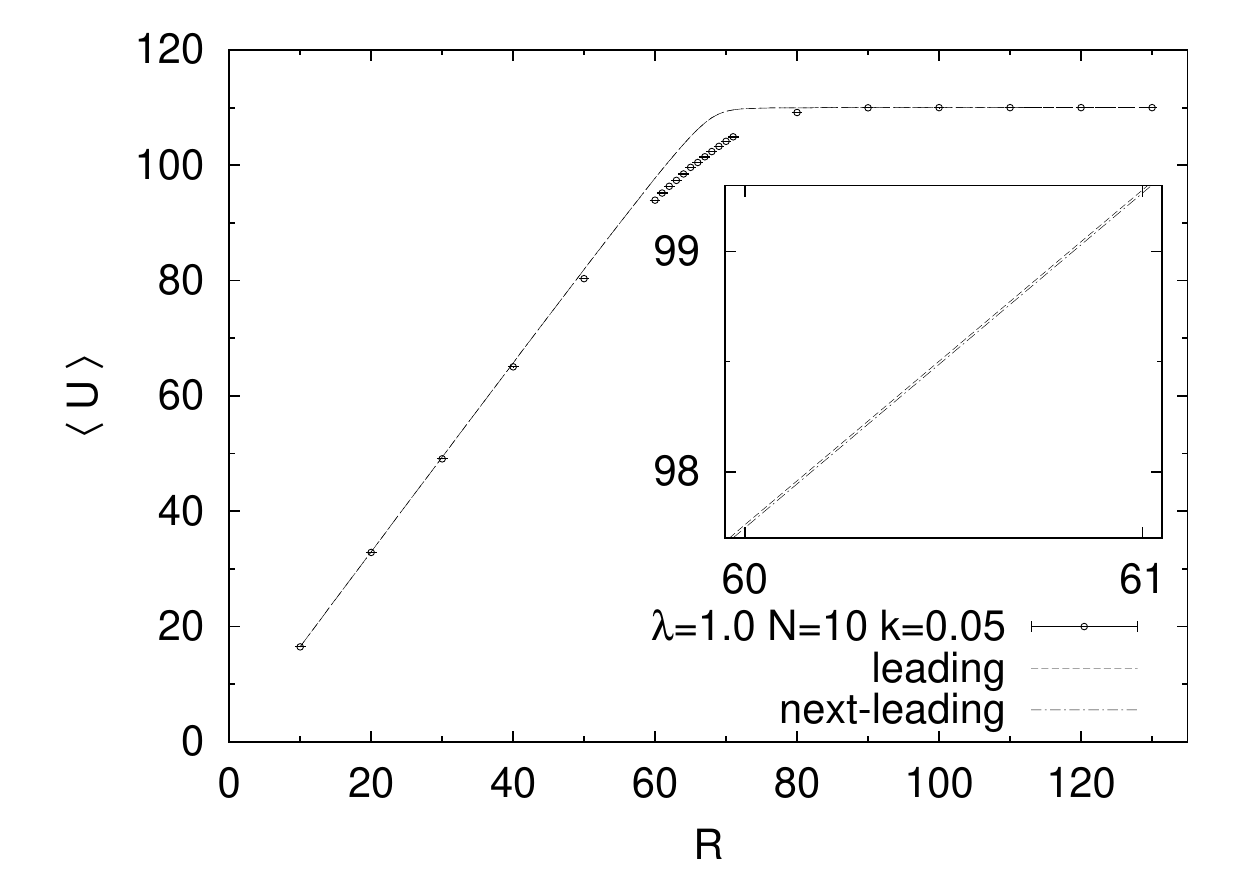}
\includegraphics[width=47.0mm,clip]{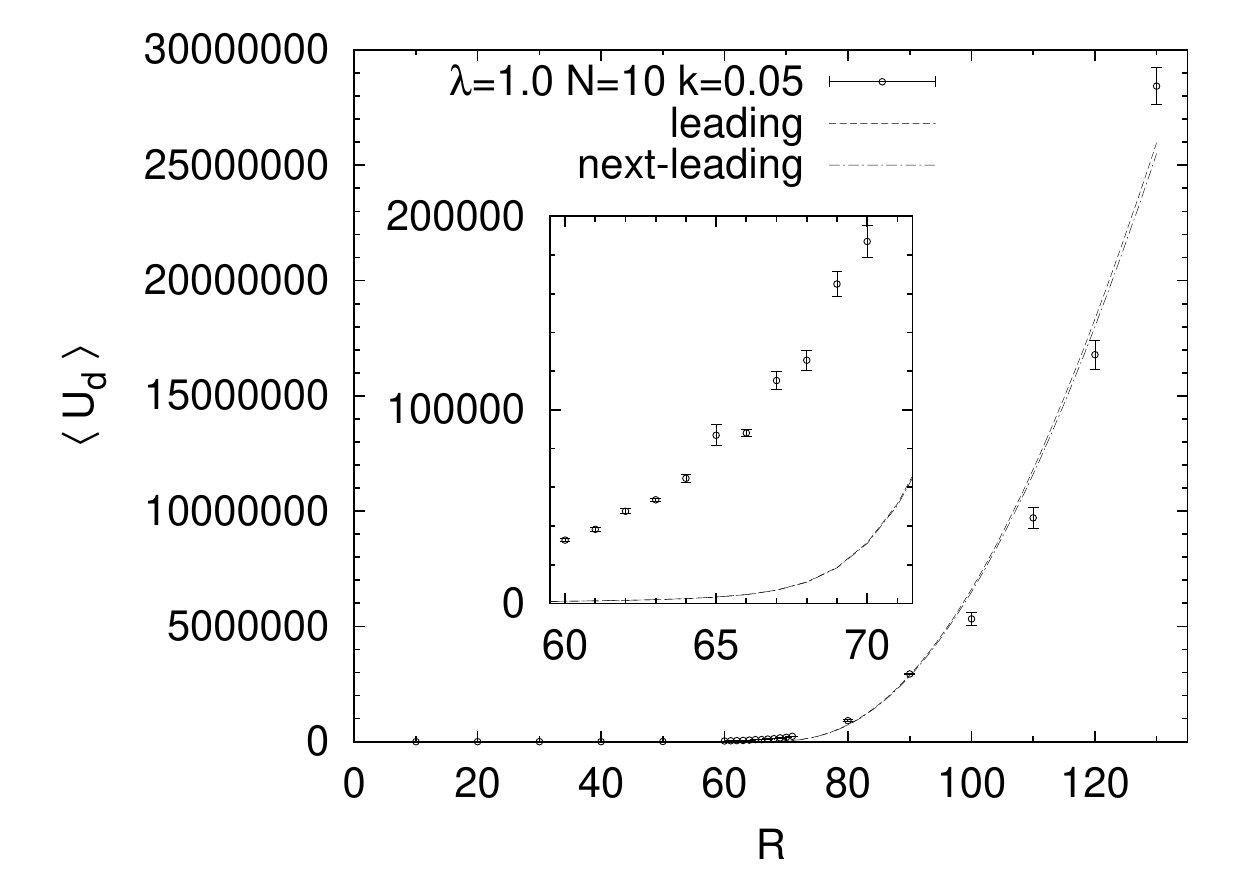}\\ 
\includegraphics[width=47.0mm,clip]{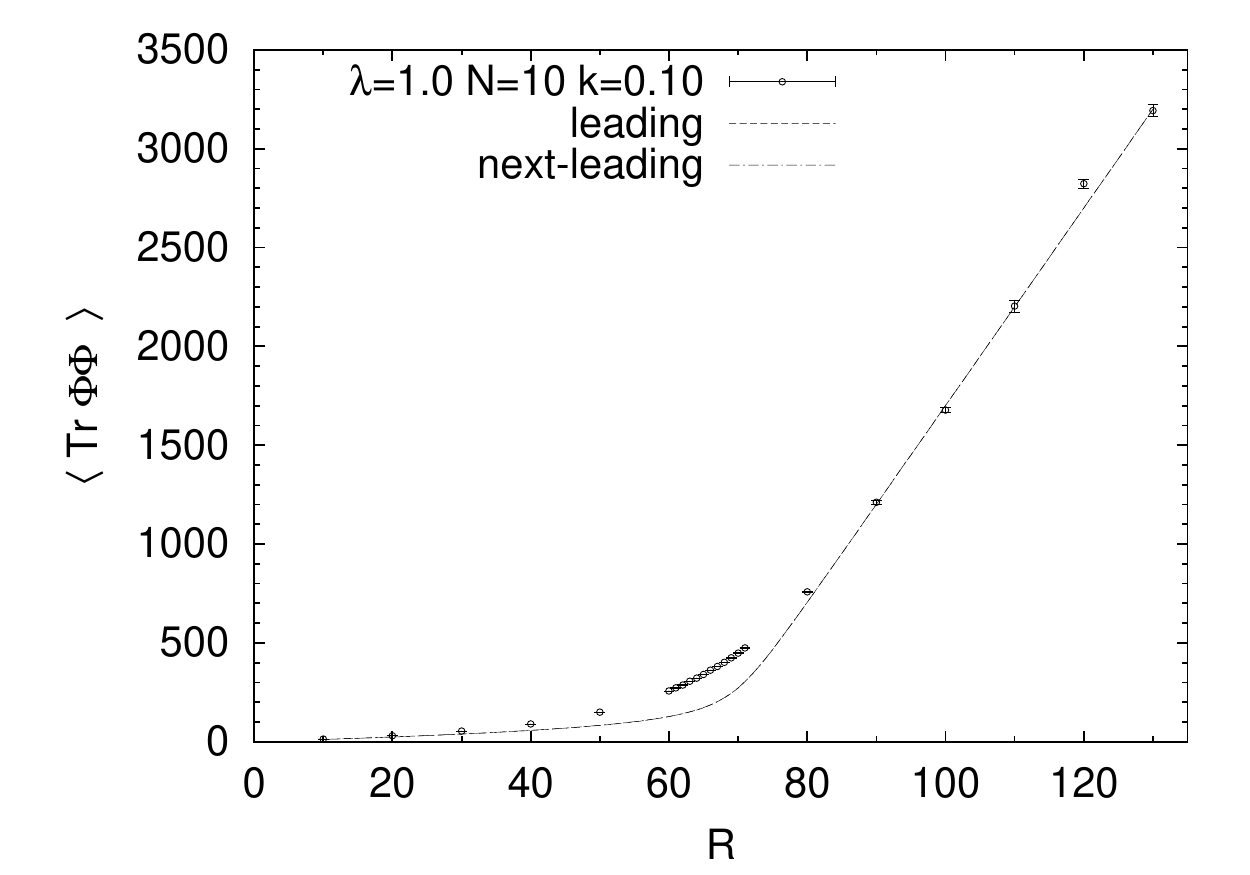}
\includegraphics[width=47.0mm,clip]{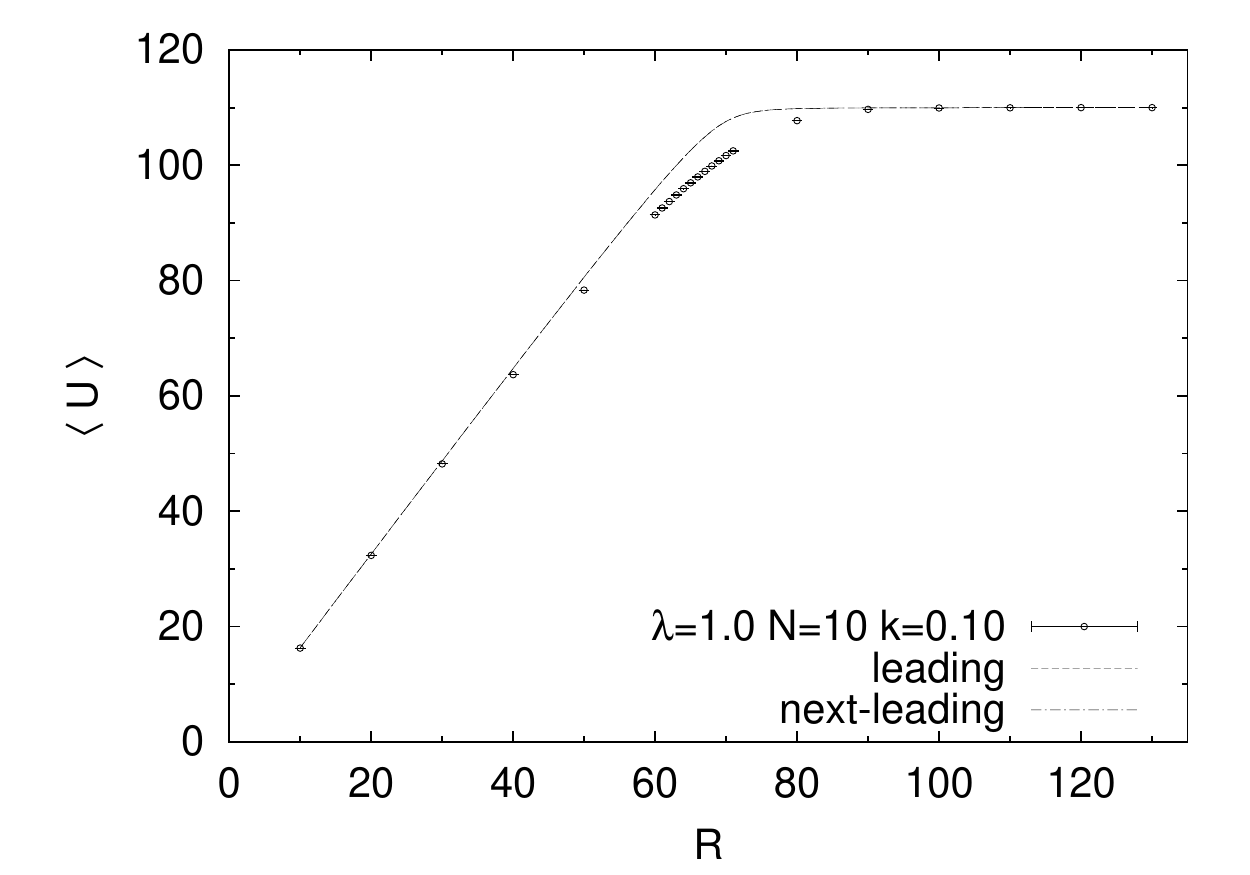} 
\includegraphics[width=47.0mm,clip]{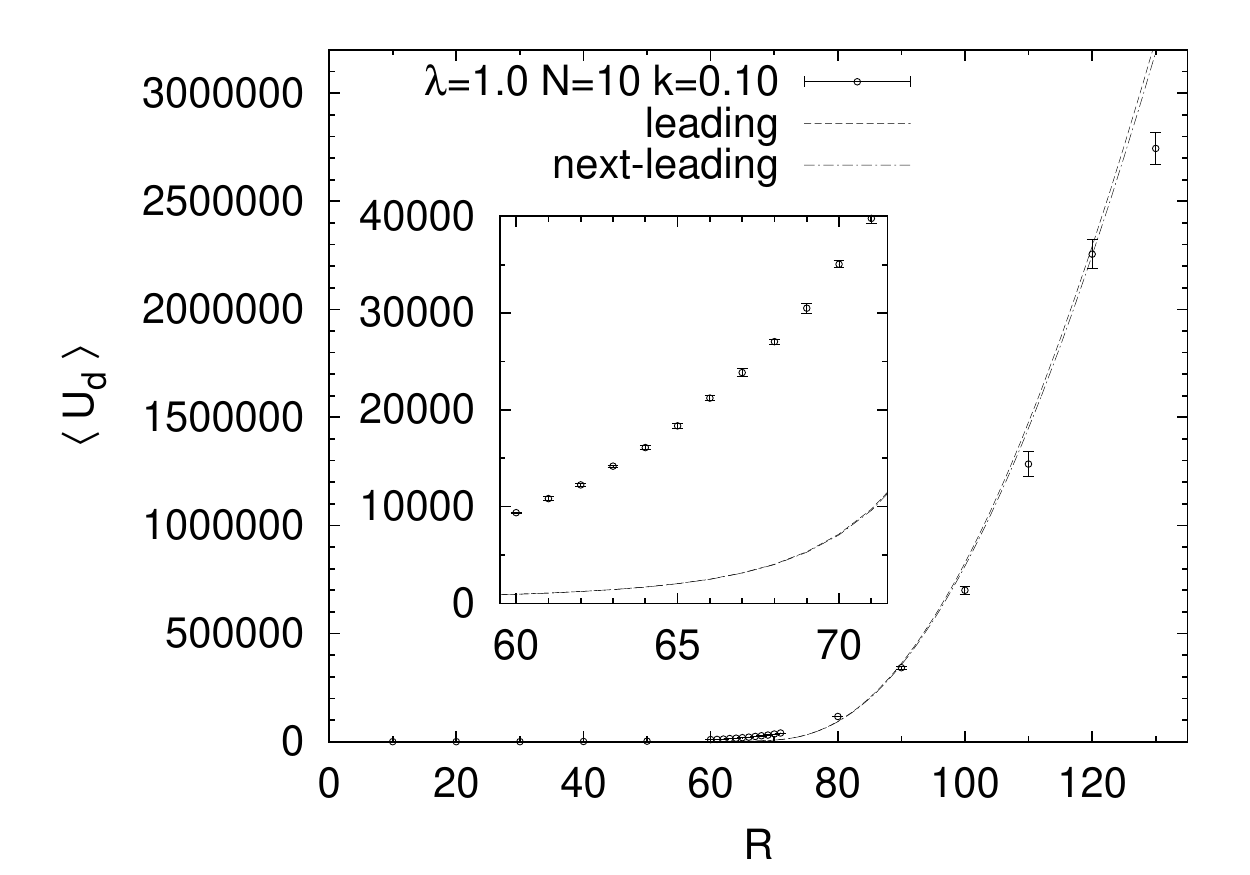}\\
\includegraphics[width=47.0mm,clip]{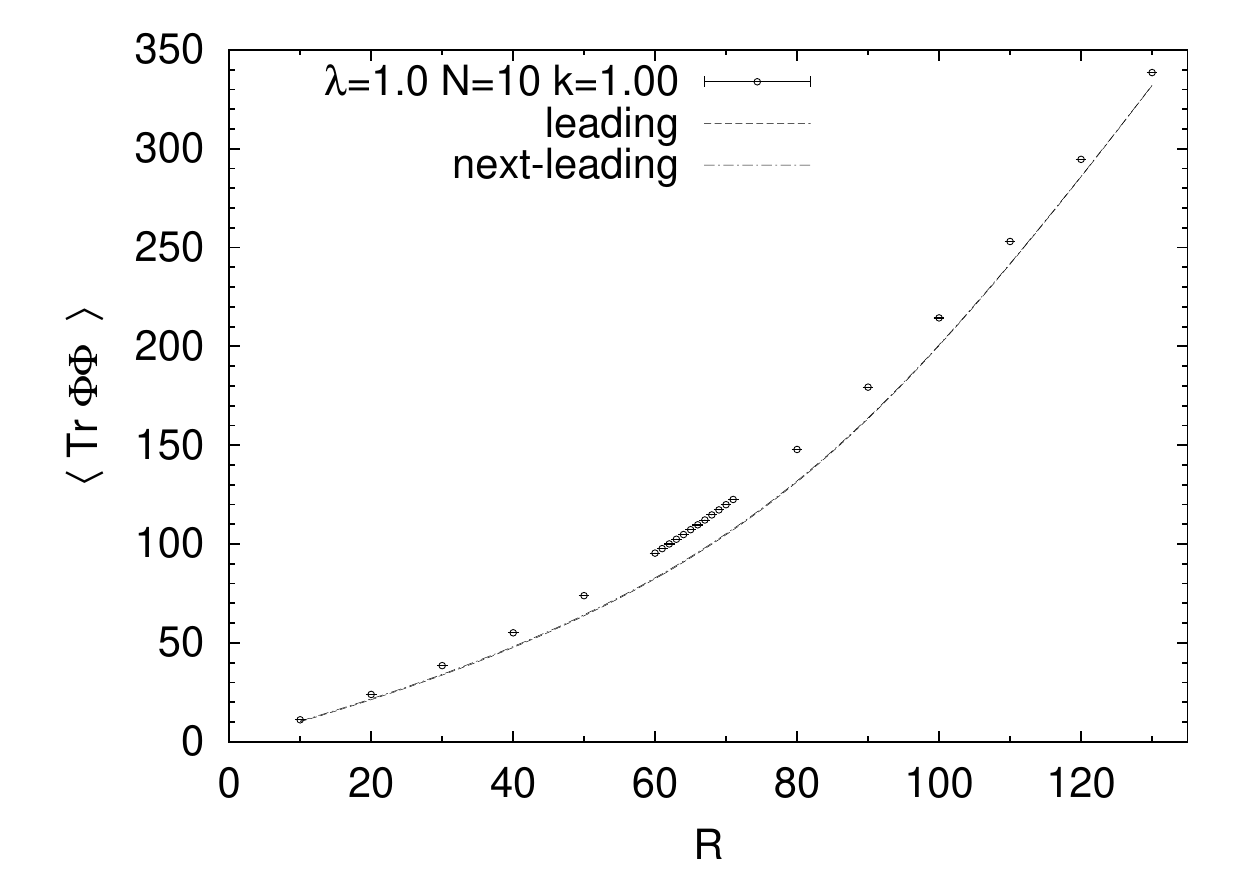} 
\includegraphics[width=47.0mm,clip]{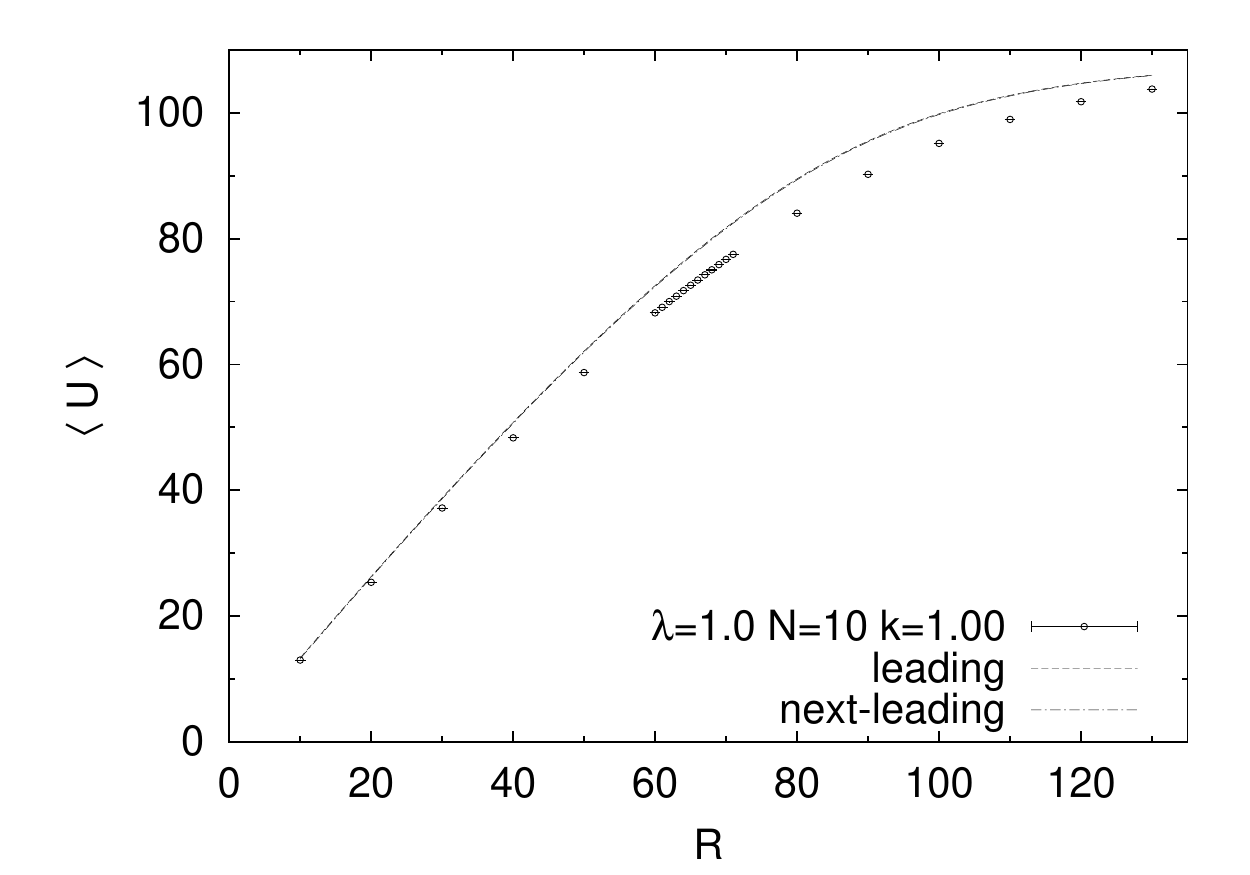}
\includegraphics[width=47.0mm,clip]{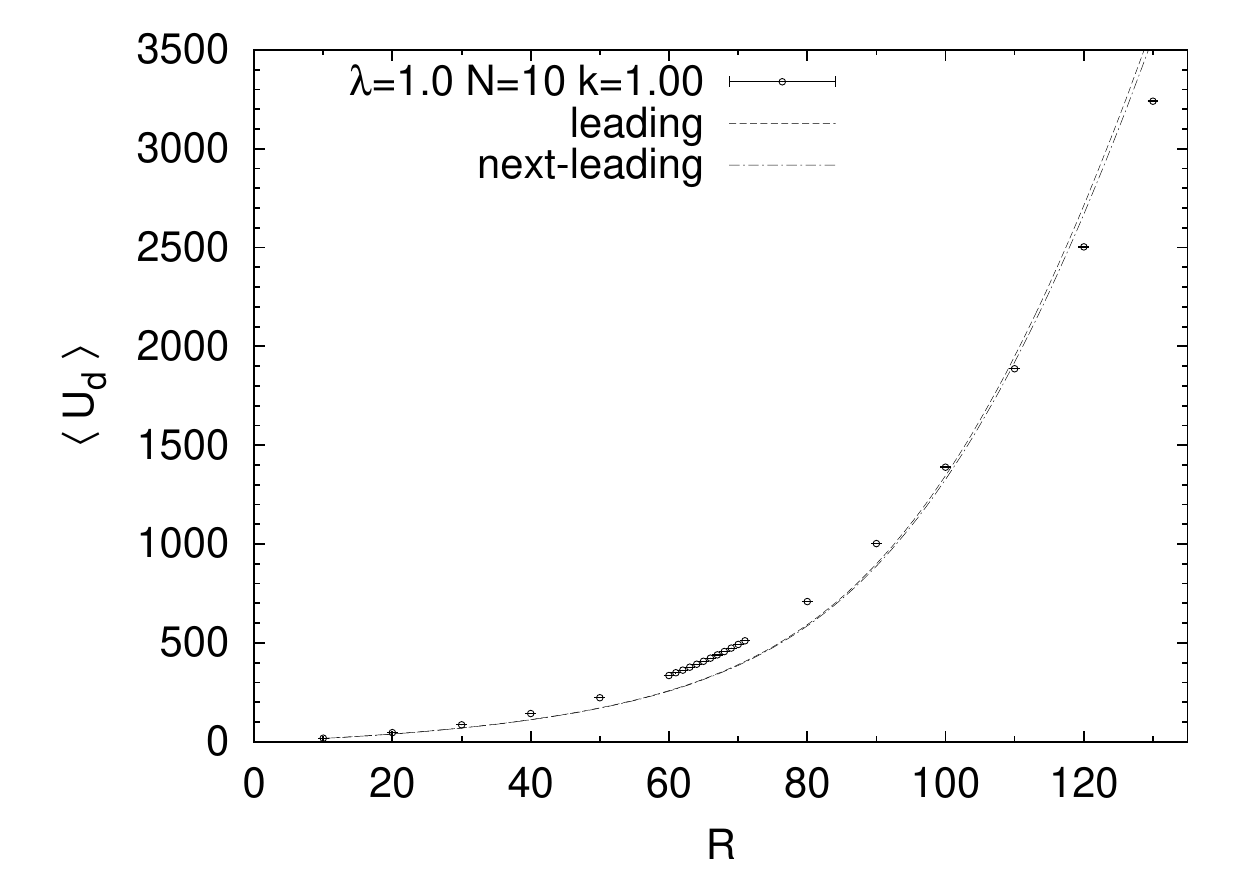}
\end{center}   
\caption{  
$\lambda=1$, 
$N=10$ and 
$k=0.05$ (top three),  
$k=0.10$ (middle three) 
and $k=1.00$ (bottom three) against $R$. 
The same notations are used, 
and some small windows are put for the same purposes as 
in Figures \ref{FigN05}.
For the clarity of the small $R$ regions that are unclear in some of the figures,
we provide Figure \ref{FigN10a}.
}
\label{FigN10} 
\end{figure*}   
%===================== 
\begin{figure*}[h!] 
\begin{center} 
\includegraphics[width=47.0mm,clip]{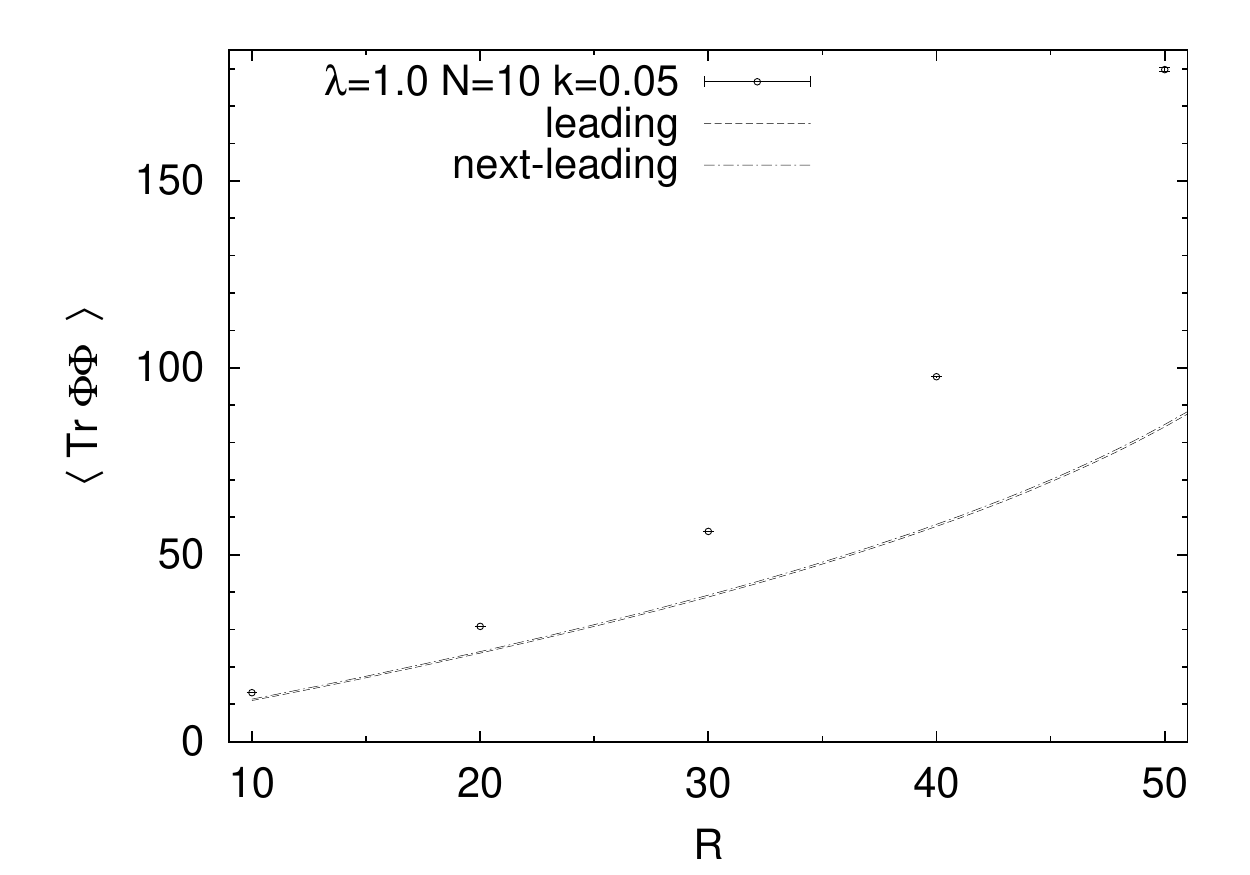}
\includegraphics[width=47.0mm,clip]{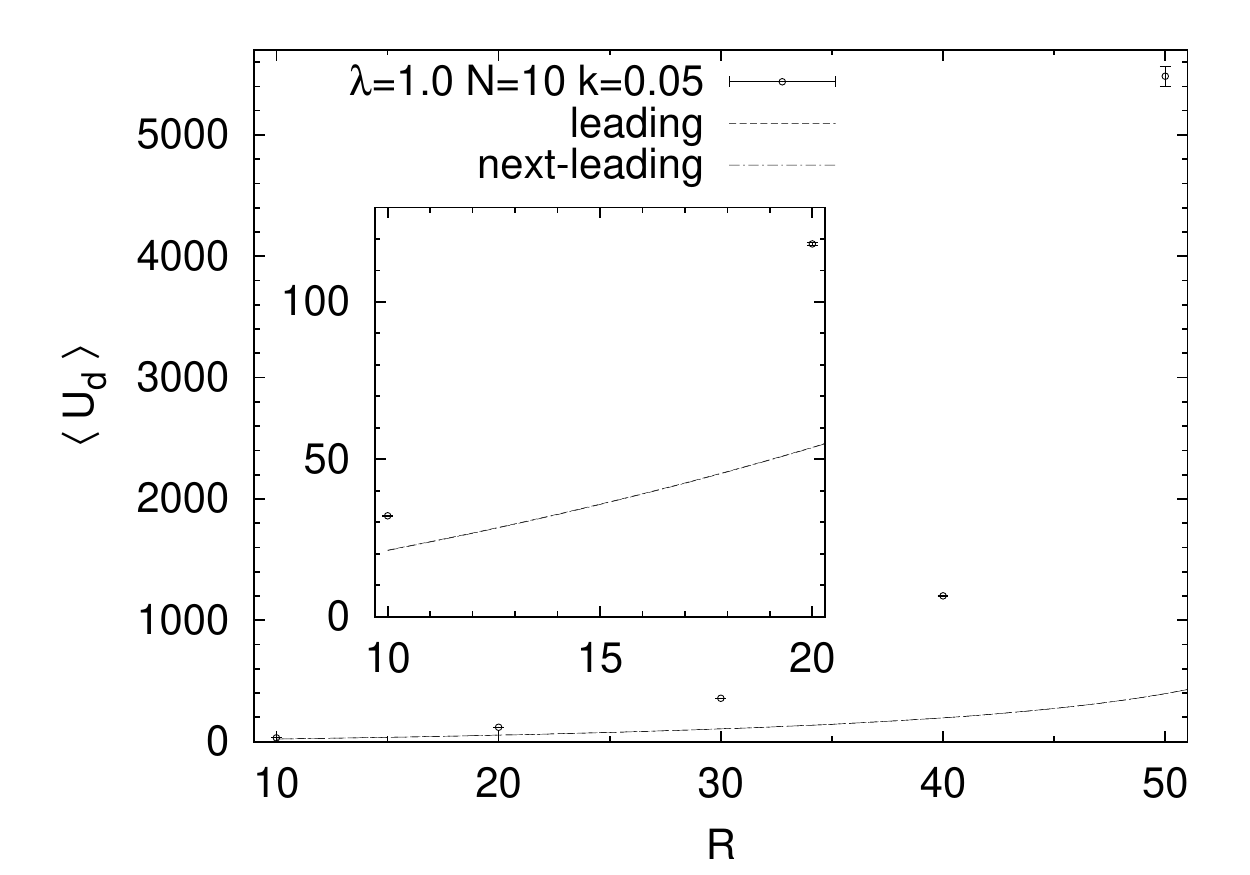}\\  
\includegraphics[width=47.0mm,clip]{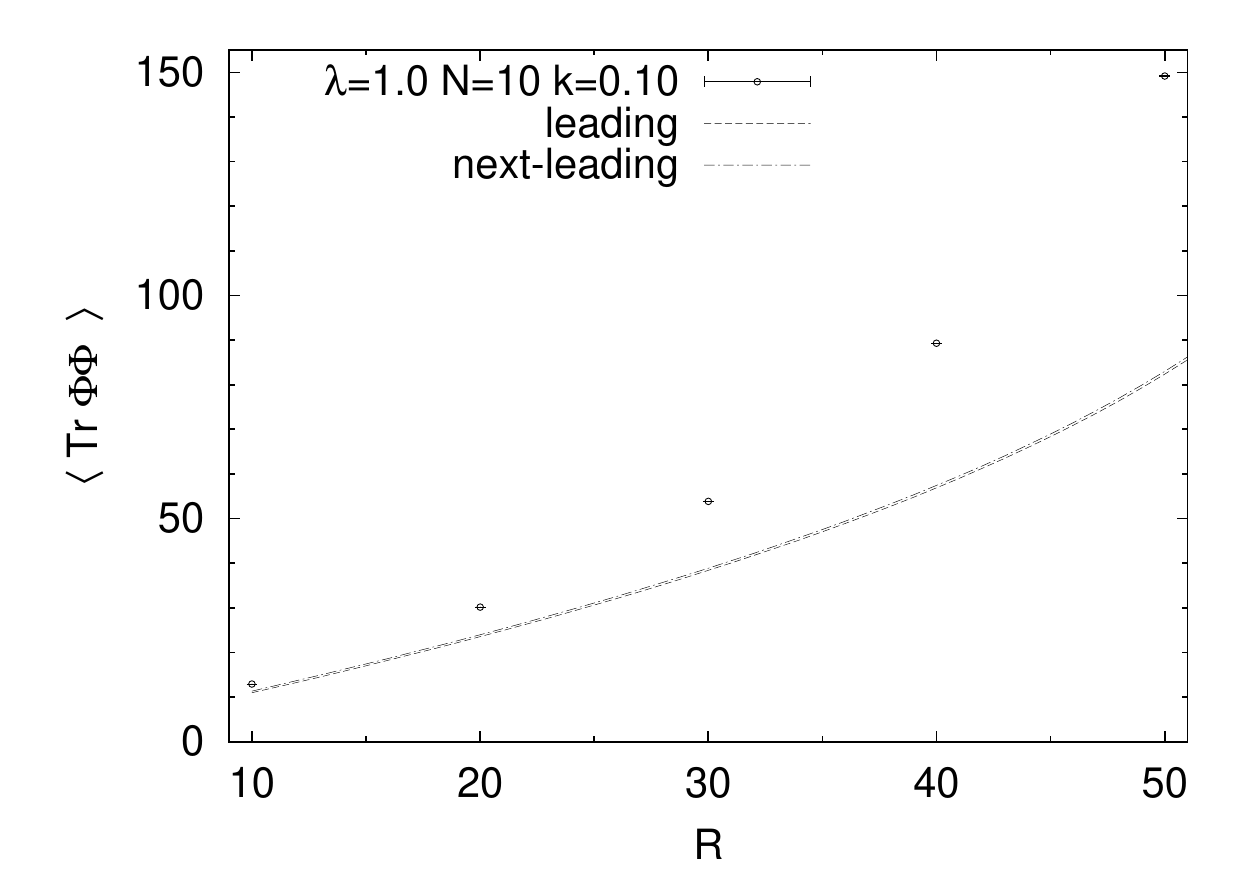}
\includegraphics[width=47.0mm,clip]{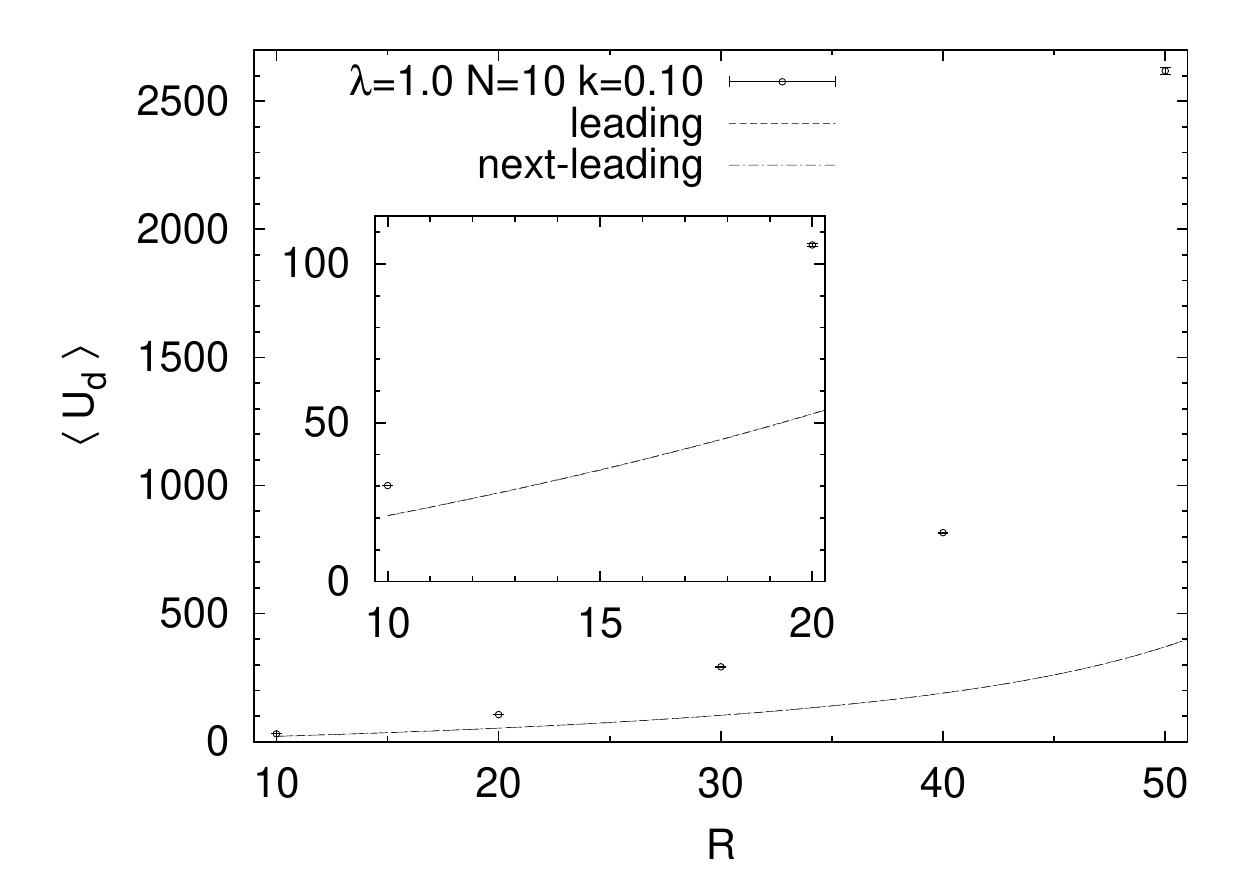}
%\\
%\includegraphics[width=47.0mm,clip]{figures/N10k1o00trpp.pdf} 
%\includegraphics[width=47.0mm,clip]{figures/N10k1o00U.pdf}
%\includegraphics[width=47.0mm,clip]{figures/N10k1o00Ud.pdf}
\end{center}    
\caption{ 
Magnification of the small $R$ regions that are unclear in Figure \ref{FigN10}.
}
\label{FigN10a} 
\end{figure*}  
%=====================
%%%%%%%%%%%%%%%%%%%%%%%%%%%%
%%%%%%%%%%%%%%%%%%%%%%%%%%%%%%%
%%%%%%%%%%%%%%%%%%%%%%%%%%%%%%%

In Section~\ref{sec:fextend},
we have computed the leading order approximation of $f_{N,R,\lambda,\lambda_d}(t)$ defined in \eq{eq:flamd}
in a perturbative method,
and have obtained the result \eq{eq:fleadwithh} along with \eq{eq:exph} and so on.
We have also derived the next-leading order correction \eq{eq:final4} with \eq{eq:finalexps4}, 
the details of which are given in \ref{app:fnext}.
With those results, we can numerically calculate the expectation values (e.v.) of the observables given in 
\eq{eq:obsbyf} by the expressions on the righthand sides.  
However, note that the above approximations of $f_{N,R,\lambda,\lambda_d}(t)$ 
are based on taking the perturbative expansion of $S_{eff}$ given in \eq{eq:seff}
up to the second order in $t$. 
Therefore they seem to require the implicit assumption of small values of $t$, 
and may not generally be trusted for the computation of \eq{eq:obsbyf},
because $t$ is finally assigned with $r^6$, and $r$ is integrated over zero to infinity.

In view of the question above,
it would be interesting to compute our model without any adoption of approximation methods.
More specifically, in this section we compute the e.v. of the observables, 
\eq{eq:evO} with ${\cal O}(\phi)=\hbox{Tr}(\phi^t \phi),U(\phi),U_d(\phi)$,  
by the Monte Carlo simulations, and compare them with 
the analytical results obtained by numerically integrating the righthand sides of \eq{eq:obsbyf},
where we put our perturbative results for $f_{N,R,\lambda,\lambda_d}(t)$. 
Note that, in our strategy of the approximations, 
$R$ is not an expansion parameter, only $t$ is, and therefore it is meaningful 
to compare the results in the full range of $R$.  

The results of the comparison are summarized in Figures~\ref{FigN05},~\ref{FigN05a},~\ref{FigN10} and \ref{FigN10a}.
The points, each with an error bar (though it's very small), represent the Monte Carlo results. 
For the information about the parameters taken in the Monte Carlo simulations, refer to the captions. 
In particular, we take $\lambda=1$ in all the computations, because
$\lambda$ in the model \eq{eq:parfun} can be scaled out by a scale transformation\footnote{The change of the overall factor of the partition function caused by the transformation is irrelevant in the Monte Carlo simulation.}
$\phi\rightarrow \lambda^{-1/6}\phi$ so that it is absorbed into $k$ as $k/\lambda^{1/3}$.  
The dotted and chained lines represent the values of the e.v. of the observables 
in the leading and the next-leading orders by adopting $f_{N,R,\lambda,\lambda_d}(t)^{leading/next-leading}$ 
to \eq{eq:obsbyf}, respectively.
In some figures, it is difficult to distinguish these two lines, almost overlapping with each other. 
Therefore, we put some small windows within the figures, 
where these two lines can be distinguished, clearly showing which of them 
exists above/below the other. 
In fact, the relative locations (namely, above/below) of the two lines remain unchanged for each observable 
throughout the parameters $N,R$ and $k$ in the figures.
%---
In the small windows within the figures of $\langle U_d \rangle$, one can more clearly see 
the results in the vicinity of the transition region around $R=R_c$. 
In addition, for the clarity of the small $R$ regions of some of the figures, 
we provide Figures~\ref{FigN05a} and \ref{FigN10a}, where one can in particular find
that the analytical and Monte Carlo results approach each other as $R$ becomes smaller.

An important thing that can be observed in the figures is that, for each $N$, 
there exists a region of $R$ that separates the smaller and larger regions of $R$ with
different qualitative behaviors of the observables.
This is more clearly seen for larger $N$ and smaller $k$.
The transition region we observe indeed exists around the value $R_c=(N+1)(N+2)/2$,
which was obtained as the critical point 
from the saddle point analysis in Section~\ref{sec:saddle} (See Figure~\ref{fig:obs}.).
  
It is an important  physical question whether this transition of behavior is a phase-transition or just a crossover in the thermodynamic limit $N\rightarrow \infty$.
However, we cannot currently answer this question for certain with the Monte Carlo results presently available, 
and this would require larger scale Monte Carlo simulations. 
It seems also difficult to answer this question by our perturbative analytical methods because of the following
reason.
In the figures, 
we can find good agreement between the perturbative computations and the Monte Carlo results  
in the regions away from the transition region. This would support the validity of our perturbative calculations
in those outside regions. 
On the other hand, we can observe 
that there exist some deviations between the perturbative computations
and the Monte Carlo results in the transition region.
The deviations appear in such a way that the Monte Carlo results smoothen the transition to make it more like a crossover.
Therefore, it seems that the analytical expressions we have obtained as approximations do not seem to be
reliable in the transition region. 
 
We can further discuss this complication from another view point as follows. 
Let us look at the figures more closely. Then we
can find that, as to the numerical relations among $\langle \textrm{Tr} \phi^t\phi \rangle$, $\langle U \rangle$,
and $\langle U_d \rangle$ in the leading and the next-leading orders, 
the following hold:
\s[
\langle \textrm{Tr}\,\phi^t\phi\rangle
\,:& 
\textrm{
including next-leading $>$ leading},\\
\langle U \rangle 
\,:& 
\textrm{
including next-leading $<$ leading},\\
\langle U_d \rangle\,:&
\textrm{
including next-leading $<$ leading},
\s]
for all $R$.
Therefore, while the next-leading order corrections indeed improve the approximations so that they approach 
the Monte Carlo results in the outside regions and this is also so for $\langle \textrm{Tr}\,\phi^t\phi\rangle$ 
and $\langle U\rangle$
in the transition region, the last inequality about $\langle U_d\rangle$ is in the opposite direction. 
This suggests that our perturbative treatment seems to have some difficulties in correctly taking 
into account some configurations that mainly contribute to $U_d$
in the transition region. 
It would be an interesting future problem to identify these configurations.
 
Let us briefly explain our actual Monte Carlo simulations. 
We have performed Monte Carlo simulations 
with the standard Metropolis update method for the model (\ref{eq:parfun})
by using KEKCC, the cluster system of KEK.
For each calculation shown in the figures, we performed 2 billion sweeps, 
where the time taken for this was generally about 7 and 23 hours with $R=10$ and $130$, respectively.
 We stored the data of the observables once per 400 sweeps, and computed their mean values 
 and the $1\sigma$-errors by the Jackknife resampling method.
In each calculation we always set the acceptance rate to be around 60$\%$. 
However, to realize this 60$\%$, we had to tune the step sizes in our Metropolis method to quite small values,
especially in the region $R\gtrsim N^2/2$.

Let us further comment on the last peculiar nature we encountered in the simulations.
We have performed the simulations for $k=1,0.1$ and $0.05$ with $N=10$ and $k=1.0$ and
$0.1$ with $N=5$, respectively, as shown in the figures. 
As suggested by the results in the figures, 
the transition could be sharpened, if we performed simulations with
smaller values of $k$ than those in the figures.  
However, when we tried to do so, we encountered a serious difficulty
in particular in the region $R\gtrsim N^2/2$.  
It was that the Metropolis step sizes must be tuned to very small values to keep the reasonable 
acceptance rate like 60\%.  
Then, the performance of the simulations became so slow that we could not find the timing
when the system had reached thermodynamic equilibriums: 
The system always looked like being in the middle stage of 
moving very slowly toward thermodynamic equilibriums, 
at least during one week of continuous running or so.    
Therefore we took relatively large values of $k$ as those in the figures to avoid the serious difficulty
that makes the simulations unreliable.

Finally, let us qualitatively explain why the analytical results computed by the perturbative method 
and the Monte Carlo results agree with each other 
outside the transition region of $R$.
Let us start with a qualitative estimation of the effective action \eq{eq:seff} 
with respect to the orders in $R$ and $t$. One can obtain
\[
S_{eff}(P)\sim (1+b_0\, t) P^2 + b_1\,t^2 P^4+\cdots,
\label{eq:seffestimate}
\]
where we have ignored all the index structures of $P^i_{abc}$ for notational simplicity, and 
the dominant $R$-dependencies of the coefficients are given by
\s[
b_0\sim \frac{c_0}{R^2}, \ b_1 \sim \frac{c_1}{R^5}.
\s]
Here the dominant $R$-dependence of $b_0$ is obtained by putting the eigenvalue 
$\lambda_{ev}\sim R$ into \eq{eq:pquad} and recalling $\gamma_3\sim 1/R^3$, as defined 
in \eq{eq:defofgamma}; The estimation of $b_1$ is also straightforward and is given in \ref{sec:estimation}.
Then, normalizing the quadratic part by rescaling $P\rightarrow P/\sqrt{1+b_0 t}$ in \eq{eq:seffestimate},
one finds that the actual quartic coupling of $S_{eff}$ is estimated as 
\[
\frac{b_1\,t^2}{(1+b_0 t)^2}\sim \frac{c_1 t^2}{R(R^2+c_0 t)^2},
\label{eq:expansionpara}
\]
which is different from the naive value $b_1\, t^2$.
When $R$ is small, $r^2={\rm Tr}(\phi^t \phi)$ is dominated by small values as shown in the Monte Carlo 
computations above. This means that, 
since $t\sim r^6$ in the usage \eq{eq:zwithf}, \eq{eq:expansionpara} is also dominated by small values, and therefor the quadratic term $S_{eff}^{(2)}$ will give good analytic estimations. 
On the other hand, when $R$ is large, $t\sim r^6$ is dominated by large values
as shown in the Monte Carlo results.
Nonetheless what is remarkable is that \eq{eq:expansionpara} is always suppressed by $1/R$ irrespective of the 
values of $t$. 
Therefore the leading order term $S_{eff}^{(2)}$ will again give good estimations. 
In the middle region, however, \eq{eq:expansionpara} 
could generally become large, and higher order and/or non-perturbative corrections may
substantially contribute, as suggested by the deviations between the Monte Carlo and analytical results. 

%%%%%%%%%%%%%%%%%%%%%%%%
%%%%%%%%%%%%%%%%%%%%%%%%
%%%%%%%%%%%%%%%%%%%%%%%%%

\section{Topological structure of configurations}

\label{sec:persistent}
 In this section, we will explain our observation on the topological structure of the 
configurations generated by the Monte Carlo simulations. 
Topology of a value of the matrix $\phi_a^i$ can be analyzed by persistent homology \cite{carlsson_topology_2009},
which is a modern technique of the topological data analysis
(See \ref{app:persistent} for a brief introduction of persistent homology.).
More specifically, we performed the Monte Carlo simulations for $N=4$ and $R=10,15,20,25$
with $\lambda=1,\ k=0.01$,
and, for each case, uniformly took 100 samples of the values of $\phi_a^i$ 
during a large number of sequential updates of order $10^8$ 
after thermodynamic equilibriums were seen to be reached. 
Then, the samples were analyzed in terms of persistent homology.
The analysis shows that the favored topology of the configurations is $S^1$ 
for $R=10,15$, but gradually changes to higher dimensional cycles, 
when $R$ is increased.
We will first explain the background motivation for this analysis,
and will then show the results. 
 
 One of the present authors and his collaborators have been studying a tensor model 
 in the canonical formalism, which we call canonical tensor model \cite{Sasakura:2011sq,Sasakura:2012fb}, 
 as a model of quantum gravity. 
 In \cite{Obster:2017dhx}, it has been shown that the exact wave function of the tensor model 
 has peaks at the configurations that are invariant under Lie groups. 
 This phenomenon, which we call symmetry highlighting phenomenon, potentially has
 an important physical significance, since this phenomenon would imply the dominance of spacetimes symmetric under Lie groups through the correspondence between 
 tensors and spaces developed in \cite{Kawano:2018pip}. 
 This symmetry highlighting phenomenon has first been shown for a toy wave function \cite{Obster:2017pdq}, which slightly simplifies the wave function of the canonical tensor model. 
 The toy wave function is given by\footnote{This wave function has actually the same form as the spherical
 $p$-spin model \cite{pspin,pedestrians} except for the following differences:
 The coupling constants of the former are pure imaginary, while they are real for the latter;
 There is a spherical constraint $\phi_a \phi_a=const.$ in the latter, while there is none in the former. 
 In addition, $R\rightarrow 0$ is taken in the study of the latter in the replica trick, 
 but we have to take $R\sim N^2/2$ for the consistency of the tensor model.
 }
 \[
 \tilde \varphi(P)=\int_{\mathbb{R}^N} \prod_{a=1}^N d\phi_a \exp \left( I P_{abc}\phi_a \phi_b \phi_c +\left( I \kappa -\epsilon\right) \phi_a \phi_a\right),
 \label{eq:wavefun}
 \]
 where $I$ denotes the imaginary unit, and $\epsilon$ is a small positive regularization parameter to assure the convergence of the integral. The symmetry highlighting phenomenon is that 
 the wave function has large peaks at $P_{abc}$ that are invariant under 
 Lie groups:  $P_{abc}=h_{a}^{a'}h_{b}^{b'}h_{c}^{c'}P_{a'b'c'}$ under ${}^\forall h \in H$, where
 $H$ is a representation of a Lie group.  The phenomenon can qualitatively be understood 
 by the following rough argument: If $P_{abc}$ is invariant under a Lie group, 
 the integration over $\phi_a$ in \eq{eq:wavefun} will contribute coherently along the 
 gauge orbit $h_{a}^{a'} \phi_{a'} \ (^\forall h\in H)$,
 but, if this is not so, the contributions tend to cancel among themselves due to the phase oscillations
 of the integrand and the wave function takes relatively small values. 
 In \cite{Obster:2017pdq}, some tractable simple cases have explicitly been studied, and 
the presence of the phenomenon has indeed be shown.
 
 Other than the simple case studies, the peak structure of the toy wave function 
 and that of the tensor model are largely unknown. 
 One reason is that the number of independent components of  $P_{abc}$, 
 which is about $\sim N^3/6$, 
 is so large that it is practically not possible to 
 go over the whole configuration space of $P_{abc}$. 
 Rather, we will be able to obtain rough knowledge by integrating over $P_{abc}$:
 \s[
 &\int_{-\infty}^\infty \prod_{a\leq b\leq c=1}^N dP_{abc} \ \tilde \varphi(P)^R \exp\left(-\alpha 
 P_{abc}P_{abc} \right) 
 ={\rm const.}\, Z_{N,R}\left( \frac{1}{4\alpha},k \right),
 \label{eq:integrated}
 \s]
 where we have denoted $k=-I \kappa+\epsilon$, and we have considered 
 an arbitrary power $R$ of the wave function, 
 because the actual wave function of the tensor model has the corresponding power\footnote{The corresponding power may rather be $R=R_c=(N+1)(N+2)/2$ than what is in the text, 
 because one of the integration variables is fixed in the toy wave function compared to the actual wave function of the tensor model. This slight difference is not important at the present stage of study, but may become so
 in the future.} 
 which specifically takes $R=(N+2)(N+3)/2$ \cite{Obster:2017dhx,Narain:2014cya,Sasakura:2013wza}
 (See \ref{Chap:Introtoctm}.).
 In \eq{eq:integrated}, we see that the integration of a power of the wave function over $P_{abc}$ 
 with a Gaussian weight produces the matrix model \eq{eq:parfun}.
 If the integration is dominated by such peaks associated to Lie groups, 
 we can expect that the $N$-dimensional vectors, 
 $\phi_a^i\ (i=1,2,\ldots,R)$, in the matrix model tend to exist along some gauge orbits
 in the vector space.
 
In general, there exist a number of peaks (or rather ridges) associated with 
various Lie group symmetries and gauge orbits for the wave function \eq{eq:wavefun}. 
As argued and shown explicitly in \cite{Obster:2017pdq,Obster:2017dhx}, 
peaks associated with lower dimensional Lie group symmetries generally 
exist more abundantly than those with higher dimensional symmetries,
because the number of symmetry conditions that must be satisfied by $P_{abc}$ is smaller 
for the former than for the latter. On the contrary, 
the peaks of the latter are generally higher than those of the former,
because the gauge orbits of the latter have larger dimensions and provide more coherent contributions
than the former. Therefore, there are competitions 
between height and abundancy, and it is generally a subtle 
question which of lower or higher dimensional Lie group symmetries
is probabilistically favored in a given case.

Let us discuss this question in view of the correspondence to the matrix model \eq{eq:integrated},
especially by considering changing the value of $R$.
As $R$ is the power of the wave function as in \eq{eq:integrated},  larger $R$ will enhance 
higher peaks compared to the lower ones. 
Therefore we expect that, for larger $R$, the contributions from 
peaks with higher dimensional symmetries will be more dominant. 
Otherwise, peaks with lower dimensional Lie group symmetries will dominate
because of their abundant existence. 
In addition, when $R$ is very small, non-symmetric configurations
will dominate, since they exist most abundantly.

In the Monte Carlo simulation of our model, 
if a symmetric peak dominates as explained above, the $N$-dimensional vectors $\phi_a^i\ (i=1,2,\ldots,R)$
will be randomly distributed along the associated gauge orbit. 
A caution here is that the gauge orbit can take any $O(N)$-transformed location in the $N$-dimensional 
vector space 
due to the $O(N)$ symmetry of the model.  Therefore we should not simply plot all the samples of $\phi_a^i$ 
generated by the Monte Carlo simulations in the $N$-dimensional vector space, 
since this will merely provide a trivial
$O(N)$-invariant spherical distribution of points without any characteristics of the Lie-group symmetry
associated to the dominant peak. Rather, we have to analyze each sample of $\phi_a^i$ generated by the Monte Carlo simulations to extract its characteristics, and then pile up all the extractions over all the samples 
to find allover characteristic properties.
 
 \begin{figure*} 
\begin{center}
\includegraphics[width=4cm]{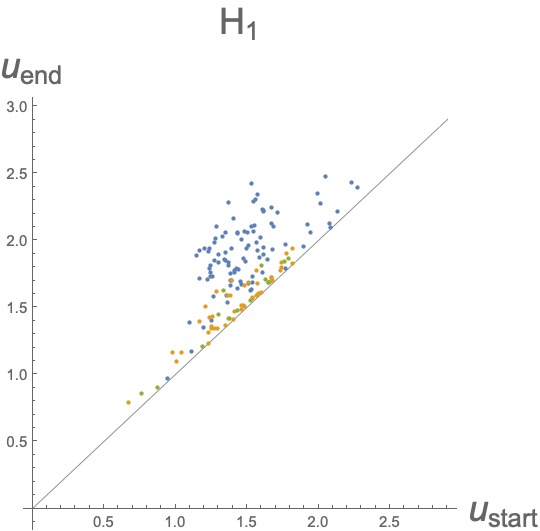}
\hfil
\includegraphics[width=4cm]{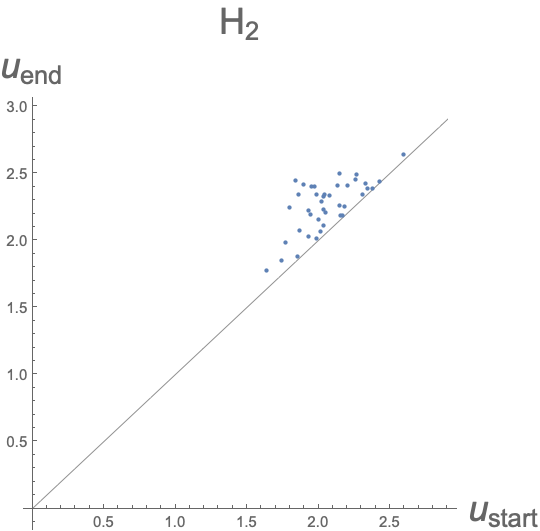}
\hfil
\includegraphics[width=4cm]{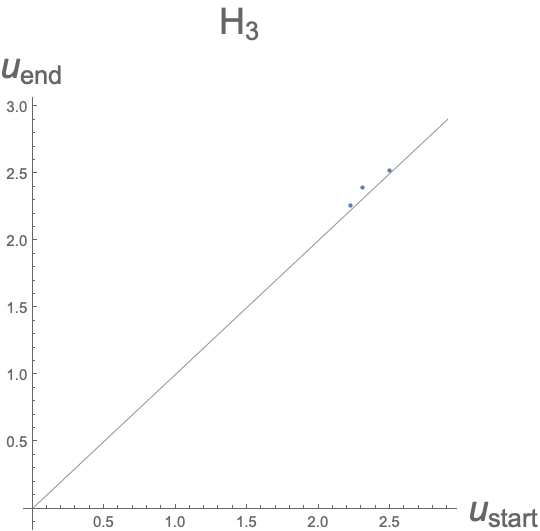}\\
\vspace{2mm}
\includegraphics[width=4cm]{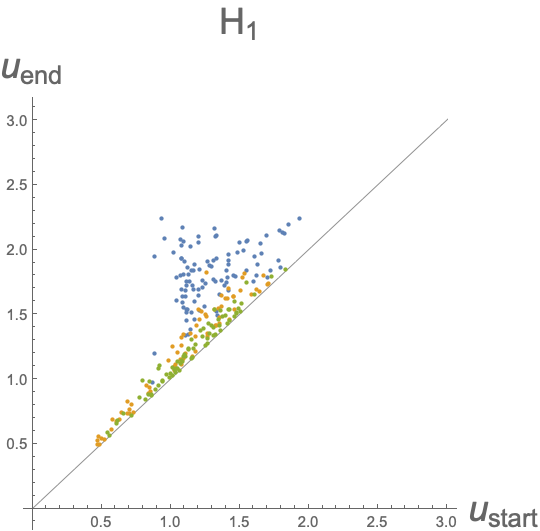}
\hfil
\includegraphics[width=4cm]{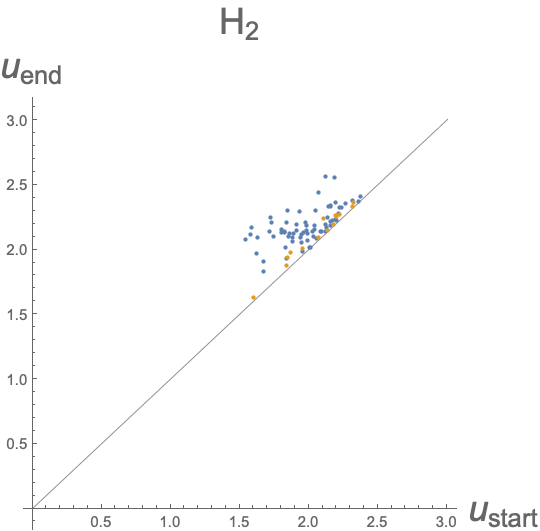}
\hfil
\includegraphics[width=4cm]{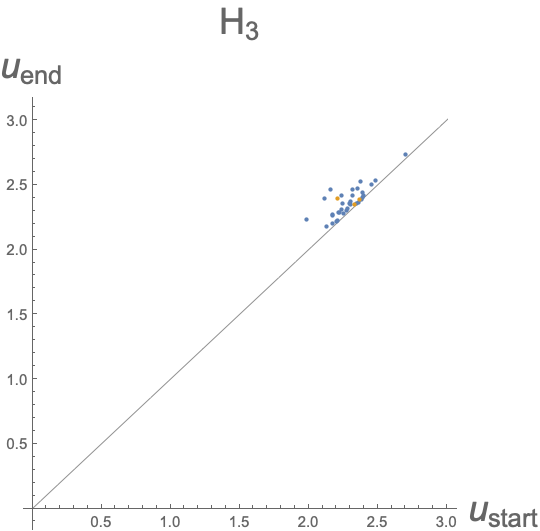}\\
\vspace{2mm}
\includegraphics[width=4cm]{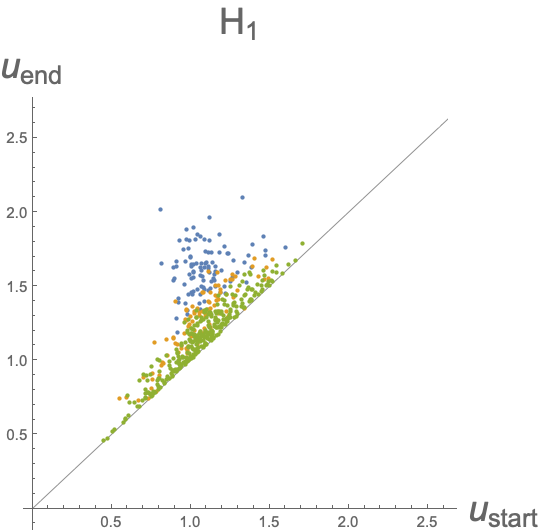}
\hfil
\includegraphics[width=4cm]{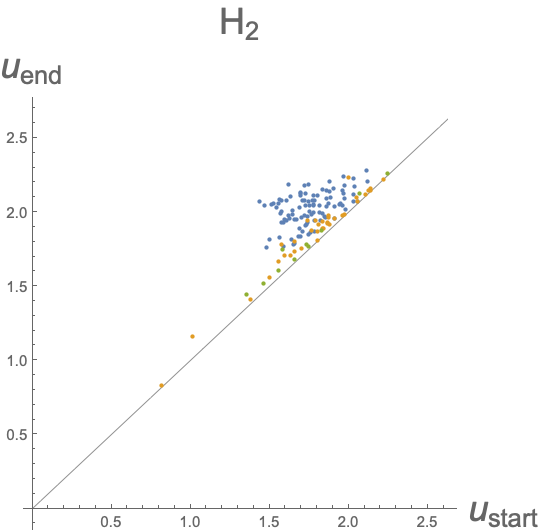}
\hfil
\includegraphics[width=4cm]{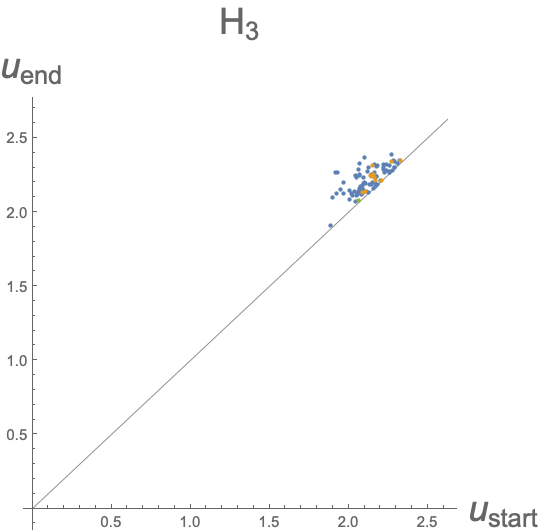}\\
\vspace{2mm}
\includegraphics[width=4cm]{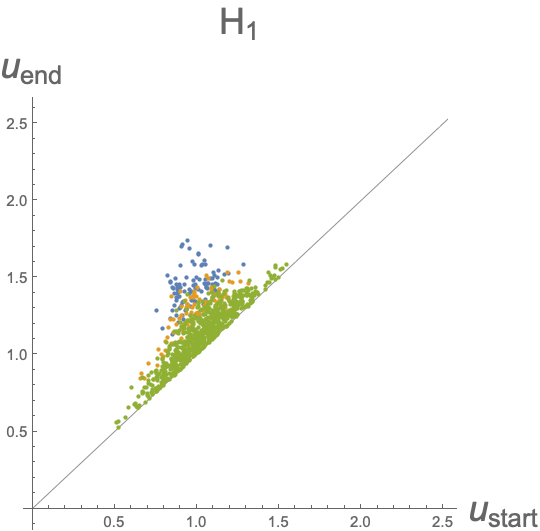}
\hfil
\includegraphics[width=4cm]{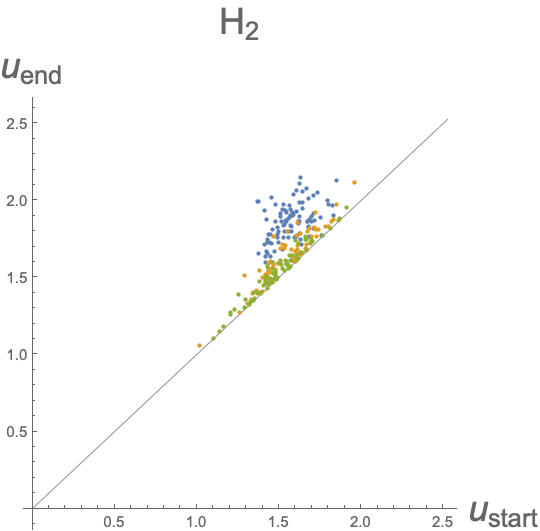}
\hfil
\includegraphics[width=4cm]{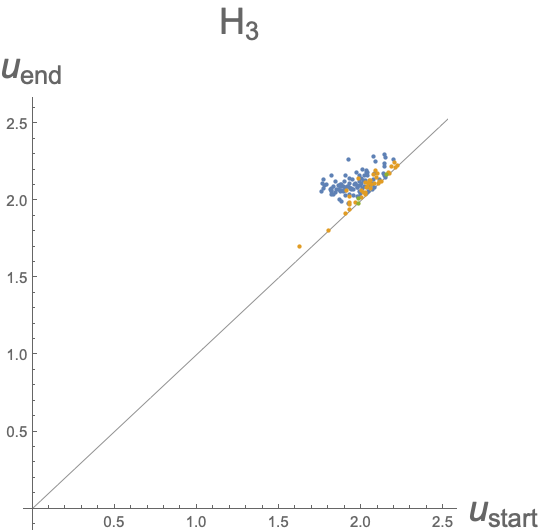}
\caption{Persistent diagrams obtained from the Monte Carlo data of $N=4,\ k=0.01,\ \lambda=1$ with $R=10,15,20,25$ (from the top to the bottom). 
To avoid the dependence of the initial values, 10 independent Monte Carlo sequences were run, 
and the sampling were performed uniformly from the sequence of updates of $\sim 10^9$ 
after thermodynamic equilibriums were seen. 
100 configurations of $\phi_a^i$ were uniformly sampled and the persistent homologies were
analyzed (one- to three-dimensional homologies from the left to the right).
The results of 100 samples are plotted on the same persistent diagrams. 
Blue dots represent the longest-life elements in each 
dimensional persistent homology group of each data, 
the yellow ones the second, and the green ones all after the second.
The dots away from the diagonal line represent long-life persistent homology 
group elements, which are considered to be characteristics of a data,
while those near the diagonal line are regarded as ``noises". 
The highest blue dots, namely those with the largest $u_{\rm end}$ 
that represent the largest structure, move from $H_1$ to $H_2$ and then $H_3$ with the increase of $R$.}
\label{fig:persistent}
\end{center}
\end{figure*}

Topological structure of each sample of $\phi_a^i$ can be analyzed by using the technique of 
persistent homology \cite{carlsson_topology_2009}.
This is a modern applied mathematical technique of the topological data analysis, and 
can extract homology groups of a data (See \ref{app:persistent}.).  
Here an input data should be a set of points with relative distances. 
We used an open-source c++ program that is called Ripser\footnote{This open-source software can be downloaded from https://github.com/Ripser/ripser.} for the analysis and plotted the output with Mathematica.  For a configuration $\phi_a^i$,
we consider the replica number $i=1,2,\ldots,R$ to represent the label of ``points" of a data set, and define the distances between two points $i$ and $j$ as 
\[
d(i,j):=\arccos\left( \frac{\phi_a^i\phi_a^j}{\sqrt{  \phi_a^i \phi_a^i \phi_b^j \phi_b^j} }\right).
\label{eq:distance}
\]  
The gist of this definition is that the $N$-dimensional vectors $\phi_a^i\ (i=1,2,\ldots,R)$ 
are projected onto the unit sphere $S^{N-1}$, 
and the geodesic distances along the sphere are regarded as the distances.
In particular, this definition is suited for detecting a gauge orbit $h_{a}^b \phi_b\ (h\in H)$, since 
it is projected on the sphere irrespective of the size of the vector $\phi_a$.
     
We want to see the phenomenon explained above in the actual Monte Carlo simulations.
For this initial study, choosing small $N$ would be preferred for simper analysis, 
because then there exist a small number of possibilities of gauge orbits with small dimensions, 
and also thermodynamic equilibriums can easily be reached due to the small number 
of degrees of freedom. 
Note however that this trades off the clarity of the homology structure
being detected by persistent homology. This is because, for small $N$,  
the values of $R$ at which the phenomenon appears
are also rather small in the order of $R\sim N^2/2$, as we will see in the analysis below.
The homology cycles formed by small numbers (namely $R$) of points
necessarily become obscure, especially higher dimensional cycles are difficult to be 
clearly detected.
 
For the actual simulation, we considered $N=4$. 
In $N=4$, as explicitly solved in \cite{Obster:2017pdq}, there exist only two possibilities of
 Lie group symmetries, $SO(2)$ and $SO(3)$, and the gauge orbits are $S^1$ and $S^2$,
 respectively.
 In fact, the ridges of the $P_{abc}$ with these symmetries reach the origin $P_{abc}=0$, and 
 therefore we should also add the trivial possibility of $S^3$ with the $SO(4)$ symmetry,
 which is the symmetry of $P_{abc}=0$ and is maximal.
 Figure~\ref{fig:persistent} shows the persistent diagrams obtained from the Monte Carlo 
 simulations with $R=10,15,20,25$.
 Statistically speaking, one can observe that, starting from $S^1$ at $R=10,15$, 
 higher dimensional cycles gradually appear and become the largest structure when $R$ is increased,
 while lower dimensional cycles gradually take smaller values of $u$.   

%%%%%%%%%%%%%%%%%%%%%%%%%%%%%%%%%%
%%%%%%%%%%%%%%%%%%%%%%%%%%%%%%%%%%%
%%%%%%%%%%%%%%%%%%%%%%%%%%%%%%%%%%
\section{Summary and future prospects}
In this paper, we studied a matrix model containing non-pairwise index
contractions \cite{Lionni:2019rty},
which has a motivation from a tensor model of quantum gravity \cite{Sasakura:2011sq,Sasakura:2012fb}.
More specifically, it has $\phi_a^i\ (a=1,2,\ldots, N, \ i=1,2,\ldots,R)$
as its degrees of freedom, where the lower indices are pairwise contracted, but the latter 
are not always done so.  
This matrix model has the same form as what appears in the replica trick of the spherical 
$p$-spin model for spin glasses \cite{pspin,pedestrians}, 
though the variable and parameter ranges of our interest are different.
We performed Monte Carlo simulations with the Metropolis update method,
and compared the results with some analytical computations in the leading order,
mostly based on the previous treatment in \cite{Lionni:2019rty}.
They are in good agreement 
outside the transition region, that is located around $R\sim N^2/2$.
In the transition region,  however, there exist 
deviations between the simulations and the analytical results, 
and the deviations cannot be corrected well, even if
the next-leading order contributions are included.
It has not been determined whether the transition is a phase transition or a crossover
in the thermodynamic limit $N\rightarrow \infty$,
because of the limited range of the parameters like $N\lesssim 10$
available in our Monte Carlo simulation. 
Our Monte Carlo simulation tended to slow down especially at $R\gtrsim N^2/2$ with large $\lambda/k^3$,
suspecting that the system gets glassy nature in the region, 
but no conclusive argument has been made for this aspect.
We also studied the topological characteristics of the configurations $\phi_a^i$ generated 
in the Monte Carlo simulations by using the modern technique called persistent homology
\cite{carlsson_topology_2009} in topological data analysis. 
This technique extracts the homology structure of a data, 
which is a configuration of $\phi_a^i$ in our case.
We observed that, in the vicinity of the transition region,
the homology structure of the configurations gradually changes from $S^1$
to higher-dimensional cycles with the increase of $R$. 

A particularly interesting result of this paper is that there seems to exist a transition
region around $R\sim N^2/2$. Intriguingly, this value of $R$ coincides 
with what is required by the consistency of the tensor model (namely, the hermiticity of
the hamiltonian constraint. See \ref{Chap:Introtoctm}.) \cite{Obster:2017dhx,Narain:2014cya,Sasakura:2013wza}.
Moreover, our model seems to have the most interesting properties in this region, but
they are not well understood:
There are some deviations between the simulation and the analytical results
in this region, but the reason is not clear;
The transition of the homological structure of the dominant configurations in this vicinity
is peculiar but not well understood;
Whether the transition is a phase transition or a crossover in the thermodynamic limit $N\rightarrow \infty$
is not determined.
For the better understanding in the future, it seems necessary to treat larger $N$ cases 
with large $\lambda/k^3$ by employing more efficient methods of Monte Carlo simulations and  
finding more powerful analytical methods. 
   
\vspace{1cm}
\section*{Acknowledgements}
%%%%%%%%%%%%%%%%%%%%%%%%%%%%%%%%%%%%%%%%%%%%%%%
%\centerline{\bf Acknowledgements} 
The Monte Carlo simulations in this study were performed using KEKCC, the cluster system of KEK.
The work of N.S. is supported in part by JSPS KAKENHI Grant No.19K03825. 
 %%%%%%%%%%%%%%%%%%%%%%%%%%%%%%%%%%%%%%%%%%%%%%%%

\appendix
\def\thesection{Appendix \Alph{section}}

\section*{Appendices}

%================================================================================================================== 
\section{The motivation for the model (\ref{eq:parfun}) from the viewpoint of the canonical tensor model}   
\label{Chap:Introtoctm}
%==================================================================================================================
In this appendix, we explain the motivation for the model (\ref{eq:parfun})
by summarizing the developments of the canonical tensor model so far.  
\newline

An interesting direction in the attempt to formulate quantum gravity 
is to pursue the existence of the spacetime as an emergent phenomenon. 
%---
In this context, the matrix models have achieved great success in the description 
of the two-dimensional quantum gravity \cite{DiFrancesco:1993cyw}. 
%--- 
To extend the success to higher dimensions, 
 the tensor 
models \cite{Ambjorn:1990ge,Sasakura:1990fs,Godfrey:1990dt,Gurau:2009tw} 
were proposed
as generalization of the matrix models.
%---
Though various interesting results have been obtained, 
emergence of spacetimes has not been achieved so far by these tensor models,
suffering from the dominance of singular configurations instead of 
macroscopic space-like ones. 
Moreover, in these tensor models under the interpretation employed in 
these original papers \cite{Ambjorn:1990ge,Sasakura:1990fs,Godfrey:1990dt,Gurau:2009tw}, the numbers of the indices of the tensorial dynamical variables 
have direct connections to the dimensions of the spaces supposed to be emergent.
This is a drawback from the viewpoint of quantum gravity, 
since the number of dimensions should also be dynamically determined rather than predetermined
as an input.
 
This drawback may be overcome by introducing a new interpretation of the tensor models.
In fact, tensors themselves have rich structures enough to describe spaces. 
This can be seen by supposing that a three-index tensor, say $C_{ab}{}^c$, define the structure constants 
of an algebra of functions, say $f_a$, over a space by $f_af_b = C_{ab}{}^c f_c$. 
It is known that spaces can be defined by the algebras of functions on them satisfying certain properties (see
for example \cite{Nestruev}.).
Though such algebras  for usual spaces are commutative and associative, 
we would be able to suppose that more general $C_{ab}{}^c$, which 
defines noncommutative or/and nonassociative algebras, would define ``fuzzy spaces''.
This idea is explored to propose a new interpretation of three-index tensor models 
as models of dynamical fuzzy spaces \cite{Sasakura:2006pq}.
Under this new interpretation,
a three-index tensor model can in principle
generate various dimensional spaces, not being restricted to 
certain dimensions \cite{Sasakura:2006pq}. One can also see the emergence of Euclidean general relativity on them \cite{Sasakura:2007sv,Sasakura:2007ud,Sasakura:2008pe,Sasakura:2009hs}.

However, these favorable results were obtained only by doing some fine-tunings to the models \cite{Sasakura:2007sv,Sasakura:2007ud,Sasakura:2008pe,Sasakura:2009hs}. Moreover, time is missing
in these tensor models. 
In fact, time might play essentially important roles in the mechanism of emergence of spacetime,
as the success of the causal dynamical triangulation \cite{Loll:2019rdj} shows over the dynamical triangulation.
Generally speaking, introducing time requires delicate treatment in quantum gravity, since such models must
respect the spacetime diffeomorphism invariance on emergent spacetimes.
To introduce time with this delicate requirement, 
a tensor model, which we call the canonical tensor model (CTM), has been formulated 
as a constrained system  in the Hamilton formalism \cite{Sasakura:2011sq,Sasakura:2012fb}. 
The model is composed of a number of first-class constraints analogous to those 
in the ADM formalism of general relativity.
Therefore, CTM embodies an analogue of the spacetime diffeomorphism as its gauge symmetry.
CTM is supposed to describe time-evolutions of fuzzy spaces. 

The classical aspects of CTM have been compared with GR in a formal continuum limit with $N\rightarrow
\infty$:
The constraint algebra of CTM has been shown to agree with that of the ADM formalism of GR 
\cite{Sasakura:2015pxa}; the classical equation of motion of CTM 
has been shown to agree with that of GR in the Hamilton-Jacobi formalism with a certain Hamilton's principal function \cite{Chen:2016ate}. 
Simple cases with large but finite $N$ have also been 
studied, and mutual agreement has been obtained \cite{Kawano:2018pip}. 

The canonical quantization of CTM is straightforward \cite{Sasakura:2013wza}. The classical constraints are 
transformed to the quantized ones with one additional term coming from operator ordering.
The algebra essentially remains the same form, especially implying the closure of the commutation algebra
among the quantized constraints. This assures the consistency of the physical state conditions given by
\[
\hat{\cal H}_a|\Psi_{phys}\rangle=\hat{\cal J}_{ab}|\Psi_{phys}\rangle=0,
\label{eq:physstate}
\]
where $\hat{\cal H}_a$ and $\hat{\cal J}_{ab}$ denote the quantized constraints of CTM,
respectively corresponding to the Hamiltonian and momentum constraints of ADM.
By taking a certain representation, these equations become a system of partial differential equations for
physical wave functions.  At first sight they looked too complicated to solve, but it has turned out that
there exist various explicit exact solutions to these equations \cite{Narain:2014cya}. In particular, a systematic solution is given in the $P$-representation by \cite{Narain:2014cya,Obster:2017dhx}
\[
\Psi(P)_{phys}=\varphi(P)^{\lambda_H/2}, 
\label{eq:psiphysbyvar}
\]
where $\lambda_H=(N+2)(N+3)/2$, and 
\[
\varphi(P)=\int_{\mathbb{R}^{N+1}} d\tilde\phi \prod_{a=1}^N d\phi_a\, e^{I(P\phi^3 -\phi^2 \tilde \phi+\frac{4}{27\lambda}\tilde \phi^3)}
\label{eq:rawwave}
\]
with $P\phi^3\equiv P_{abc}\phi_a \phi_b \phi_c$, $\phi^2\equiv\phi_a\phi_a$, and $\lambda$
is an arbitrary real parameter\footnote{The $N=1$ case of CTM produces the mini-superspace approximation 
of GR with the parameter $\lambda$ as the cosmological constant \cite{Sasakura:2014gia}.}, which can be normalized as $0,\pm 1$.
The above value of $\lambda_H$ is uniquely determined by the hermiticity of
the ``Hamilotonian constraints'' $\hat{\cal H}_a$ of CTM.

In principle, emergence of spacetime can be analyzed by investigating the properties of the wave 
function $\Psi(P)_{phys}$. The ideal situation would be that,
 in the configuration space, the quantum probability density $|\Psi(P)_{phys}|^2$ form some ridges
that can be considered to be trajectories of classical time-evolutions of spaces. 
In fact, it has been qualitatively argued and actually been found in some tractable simple cases that such ridges 
exist at $P_{abc}$'s that
are invariant under Lie-groups\footnote{On the ridges, the Lie-group symmetry associated with $P_{abc}$ has 
a definite signature like $SO(n)$, but one can find that the Lie-group symmetry 
associated with the whole terms of the exponent of \eq{eq:rawwave} has an indefinite signature like $SO(n,1)$ \cite{Obster:2017dhx}.}. 
Since symmetries are ubiquitous in the actual spacetimes\footnote{For instance, Lorentz symmetry, de Sitter symmetry, etc.}, this result is encouraging in the pursuit for spacetime emergence.
However, further analysis of $\Psi(P)_{phys}$ is not easy, and most of the properties of the 
wave function are still unknown. Therefore, it is presently not possible to discuss emergent spacetimes 
in CTM.

Yet, aiming for better understanding of $\Psi(P)_{phys}$, we consider in this paper a model which can be 
obtained after two simplifications.
One is to integrate the quantum probability density over the configuration space:
\[
\int_{\mathbb{R}^{\# P}} dP \, e^{-\alpha P^2} |\Psi(P)_{phys}|^2,
\label{eq:intpsi2}
\] 
where $dP\equiv \prod_{a\leq b\leq c}^N dP_{abc}$, $P^2\equiv P_{abc}P_{abc}$ and $\alpha$ is an
arbitrary positive number.
The other is to fix (or ignore) the integration variable $\tilde \phi$ of $\varphi(P)$ in \eq{eq:rawwave} 
so that  $\tilde \varphi(P)$ in \eq{eq:wavefun} is used as $\varphi(P)$ in \eq{eq:psiphysbyvar}.
The wave function after this replacement is still interesting enough, because $\tilde \varphi(P)$ has the similar connection between Lie-group symmetries and peaks \cite{Obster:2017pdq}.  
Then, by putting $\tilde \varphi(P)$ into \eq{eq:psiphysbyvar} and performing the Gaussian integration over $P$ in \eq{eq:intpsi2}, one obtains 
\[
\int_{\mathbb{R}^{\# P}} dP e^{-\alpha P^2} \tilde\varphi(P)^{\lambda_H}=const. Z_{N,\lambda_H} \left(\frac{1}{4 \alpha},k\right),
\]
where we have used the reality of the wave function $\tilde\varphi(P)$
for the real value of $k=-I \kappa+\epsilon$. Here we have actually obtained the partition function 
$Z$ in \eq{eq:parfun} of the matrix model with 
the particular choice $R=\lambda_H\ (=(N+2)(N+3)/2)$.
Since $\lambda_H\sim R_c\ (=(N+1)(N+2)/2)$, 
CTM seems to be located in the transition region of the matrix model. 

\section{$R=2$ case}
\label{app:Req2}
In this appendix, we consider the partition function for $R=2$,
and see that the partition function is finite even for $k=0$.  
This is not trivial, because, for general $R>1$,
the solution to $U(\phi)=0$ is non-empty (see below), 
and therefore there is a potential risk of runaway
behavior, $\phi^2\rightarrow \infty$ with $U(\phi)=0$.

Let us first see that $U(\phi)=0$ is non-empty for general $R>1$. Since 
\[
U(\phi)=\left( \sum_{i=1}^R \phi^i_a \phi^i_b \phi^i_c \right)  \left( \sum_{j=1}^R \phi^j_a \phi^j_b \phi^j_c \right),  
\]
$U(\phi)\geq 0 $ holds, and $U(\phi)$ vanishes iff
\[
\sum_{i=1}^R \phi^i_a \phi^i_b \phi^i_c=0.
\label{eq:u0cond}
\]
An obvious set of solutions are given by $\phi_a^{2i-1}=-\phi_a^{2i},\ (i=1,2,\lfloor R/2 \rfloor)$
with $\phi^R_a=0$ if $R$=odd.

For $R=2$, one can see that this is the only solution as follows. 
By contracting the indices $b$ and $c$ in \eq{eq:u0cond}, we obtain
\[
\phi^1_a  \phi^1_b \phi^1_b+\phi^2_a  \phi^2_b \phi^2_b=0.
\]
This implies that $\phi^1$ and $\phi^2$ are linearly dependent, and putting this back to 
\eq{eq:u0cond}, we obtain $\phi_a^1=-\phi^2_a$.

In the $R=2$ case, since $U(\phi)$ depends only on the relative directions and the sizes of $\phi_a^1$ and $\phi_a^2$, the partition function for $k=0$ can be written as 
\s[
&Z_{N,R=2}(\lambda,k=0)=\hbox{Vol}(S^{N-1})\hbox{Vol}(S^{N-2})
\int_0^\infty dr_1 dr_2\ r_1^{N-1} r_2^{N-1} \int_0^\pi d\theta \sin^{N-2} (\theta) \ e^{-\lambda U(\theta)},
\label{eq:zr2}
\s]
where
\[
U(r_1,r_2,\theta)=r_1^6+r_2^6+2 r_1^3 r_2^3 \cos^3 (\theta)
\]
with $r_1$ and $r_2$ being the sizes, and $\theta$ being the relative angle. As shown above,
the only case with $U=0$ is given by $r_1=r_2,\ \theta=\pi$. 
Therefore let us perform the reparameterization,
\[
r_1=r,\ r_2=r(1+x),\ \theta=\pi-y.
\]
Then the integral of \eq{eq:zr2} at large $r$ can be approximated by expanding 
the integrand for small $x,y$, and we obtain
\[
 \sim \int dr dx dy \ r^{2N-1} y^{N-2} e^{-3 \lambda r^6 (3 x^2+y^2)}
\sim \int dr \ r^{-N-1}.
\label{eq:zbehav}
\]
This shows that the partition function is convergent for $R=2$ and $k=0$. 

There are two things which can be learned from this simplest case. 
One is that, if $R$ is small enough,
 the partition function is convergent even for $k=0$. Another thing is that the large-$r$ 
 asymptotic behavior
of the integrand is much slower than the leading order result,
 \[
 \int dr r^{NR-1} f^{leading}_{N,R}(\lambda r^6) \sim \int dr\, r^{2N-1-N(N+1)(N+2)/2}.
  \]
  Therefore the asymptotic behavior derived from the leading order result  cannot be
  correct down to $R=2$.

\section{Derivation of \eq{eq:connecttilde}}
\label{app:derivationofconnect}
In this section, we derive \eq{eq:connecttilde}. This was previously derived in \cite{Lionni:2019rty}, but this is repeated here to make the present paper self-contained.

For $m=$odd,  the equation holds, because the both sides vanish. 

Let us assume $m=2p$ with a positive integer $p$.  Let us start with the following equation:
\s[
&\frac{1}{\int_{S^{NR-1}} d \tilde \phi} 
\int_{S^{NR-1}} d \tilde \phi \, \tilde\phi_{a_1}^{i_1}\tilde \phi_{a_2}^{i_2}\cdots \tilde \phi_{a_{2p}}^{i_{2p}} 
=
\frac{1}{\int_{{\mathbb R}^{NR}} d  \phi\  e^{-\beta {\rm Tr}\phi^t \phi}} 
\int_{{\mathbb R}^{NR}} d \phi  \ 
\frac{\phi_{a_1}^{i_1}\phi_{a_2}^{i_2}\cdots  \phi_{a_{2p}}^{i_{2p}}}{({\rm Tr}\phi^t \phi)^p}
 e^{-\beta\, {\rm Tr}\phi^t \phi},
 \label{eq:toconnect}
 \s]
 where $\beta$ is an arbitrary positive constant.
 This can easily be proven by reparameterizing $\phi_a^i$ with the radial and the angular 
 variables as 
 $\phi_a^i=r \tilde \phi_a^i$ on the righthand side, and observing that the integrations over $r$ decouples from the angular part and cancels between the numerator and the denominator.  

Let us consider the numerator on the righthand side of \eq{eq:toconnect},
\[
A(\beta):=\int_{{\mathbb R}^{NR}} d \phi  \ 
\frac{\phi_{a_1}^{i_1}\phi_{a_2}^{i_2}\cdots  \phi_{a_{2p}}^{i_{2p}}}{({\rm Tr}\phi^t \phi)^p}
 e^{-\beta {\rm Tr}\phi^t \phi}.
 \label{eq:abeta}
\]
Taking the $p$-th derivative of $A(\beta)$ with respect to $\beta$ 
cancels the $({\rm Tr}\phi^t \phi)^p$ in the denominator of the inetegrand. 
On the other hand, by performing the rescaling $\phi_{a}^i\rightarrow \beta^{-1/2} \phi_a^i$, 
it is obvious that $A(\beta)$ has the dependence $\beta^{-NR/2}$ on $\beta$. 
Therefore, by performing the $p$-th derivative of the both sides of \eq{eq:abeta}, 
we obtain the relation,
\[ 
\frac{\Gamma\left(\frac{NR}{2}+p\right)}{\Gamma\left(\frac{NR}{2}\right)} \beta^{-p} A(\beta)=
\int_{{\mathbb R}^{NR}} d \phi  \ 
\phi_{a_1}^{i_1}\phi_{a_2}^{i_2}\cdots  \phi_{a_{2p}}^{i_{2p}} e^{-\beta\, {\rm Tr}\phi^t \phi}.
\]   
By solving for $A(\beta)$ and putting it into \eq{eq:toconnect}, we obtain \eq{eq:connecttilde}. 

%%%%%%%%%%%%%%%%%%%%%%%%%%%%%%%%%
%%%%%%%%%%%%%%%%%%%%%%%%%%%%%%%%%%%%%%%%%%%%%
%%%%%%%%%%%%%%%%%%%%%%%%%%%%%%%%%%%%%%%

\section{Computation of $f_{N,R,\Lambda_{ij}}(t)$ in the next-leading order}
\label{app:fnext}
\begin{figure*}
\begin{center}
\includegraphics[width=\textwidth]{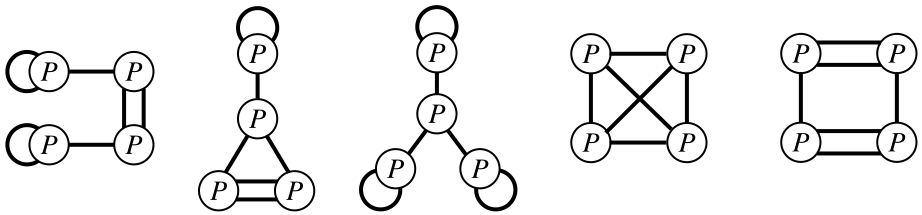}
\caption{
The Feynman diagrams for the forth order interaction term, $\langle (P\phi^3)^4 \rangle^c_\phi$.
}
\label{fig:quad}
\end{center}
\end{figure*}

In this appendix, we will compute the fourth-order term 
 $\langle (P\tilde \phi^3)^4 \rangle_{\tilde \phi}^c$ in \eq{eq:seff}.
 From the definition \eq{eq:defofcum} of cumulants 
 and the formula \eq{eq:connecttilde}, we obtain
 \s[
 \langle (P\tilde \phi^3)^4 \rangle_{\tphi}^c&=\langle (P\tilde \phi^3)^4 \rangle_{\tphi}
 -3(\langle (P\tilde \phi^3)^2 \rangle_{\tphi})^2 \\
&=\gamma_6 (2\beta)^{6}  \langle (P\phi^3)^4 \rangle_\phi-3 \left(\gamma_3 (2\beta)^{3} \langle (P\phi^3)^2 \rangle_\phi\right)^2 \\
 &=\gamma_6 (2\beta)^6
 \langle (P\phi^3)^4 \rangle^c_\phi-3 (2\beta)^6\left(
\gamma_3^2-\gamma_6
 \right) \left(\langle (P\phi^3)^2 \rangle^c_\phi\right)^2.
 \label{eq:pphi4c}  
 \s]  
The $\langle (P\phi^3)^2 \rangle^c_\phi$ in the last term has already been computed in Section~\ref{sec:fextend}.
As for $\langle (P\phi^3)^4 \rangle^c_\phi$,
Wick contractions \eq{eq:wick} give the five connected diagrams in Figure~\ref{fig:quad}.
By counting the number of ways to connect the legs, one can find that the degeneracies 
are given by $2^3 \cdot 3^5$,  $2^4 \cdot 3^5$, $2^3\cdot 3^4$, $2^4\cdot 3^4$, 
and $2^3\cdot 3^5$, respectively,
from the left to the right diagrams\footnote{A non-trivial check of these numbers is to see whether the sum of them 
agrees with $12!/(2^6 \cdot 6!)-3\cdot 15^2$, where the former number counts
all the possibilities of connecting $12$ legs of the four vertices, and the latter
is the subtraction of the disconnected diagrams among them. } in Figure~\ref{fig:quad}.
These Feynman diagrams represent the ways of the index contractions of $P_{abc}^i$'s in
 the fourth order interaction term.
For example, the leftmost diagram gives
\s[
&\langle (P\phi^3)^4 \rangle^{c,{\rm leftmost}}_\phi=
\frac{ 2^3 3^5}{(2\beta)^6}
 \sum_{i_1,i_2,i_3,i_4,j=1}^R
 \tilde \Lambda_{i_1j}\tilde \Lambda_{i_2j}\tilde \Lambda_{i_3j}\tilde \Lambda_{i_4j}
P^{i_1}_{aab}P^{i_2}_{bcd}P^{i_3}_{cde}P^{i_4}_{eff},
\label{eq:intleftmost}
 \s]
where the numerator of the numerical factor is the degeneracy, and the denominator 
comes from the factor of the Wick contraction \eq{eq:wick}.
 It is also straightforward to write down the explicit expressions for all the other diagrams in Figure~\ref{fig:quad}.
 
 Now let us suppose we have obtained the explicit expressions of $S^{(4)}_{eff}(P)$
 by the above procedure.
 An immediate difficulty of this forth order term is that $S^{(4)}_{eff}(P)$ has the negative 
 all over sign due to $I^4$ as in \eq{eq:seff},
 and therefore the system with $S_{eff}(P)=S^{(2)}_{eff}(P)+S^{(4)}_{eff}(P)$ is not stable. 
 This may be changed if we include the next order term $S^{(6)}_{eff}(P)$,
 which has a positive coefficient, but the computation will become more complicated than 
 $S^{(4)}_{eff}(P)$ and will not be performed in this paper. 
To treat this situation in a consistent manner, we only take the first correction coming from $S^{(4)}_{eff}(P)$ as  $e^{-S^{(4)}_{eff}(P)}\sim 1-S^{(4)}_{eff}(P)$ rather than the full exponential form. 
 Note that this can consistently be understood as taking the first correction coming from the full expression of the interactions,
 $e^{-S^{(4)}_{eff}(P)-S_{eff}^{(6)}(P)-\cdots}=1-S^{(4)}_{eff}(P)-S_{eff}^{(6)}(P)+S^{(4)}_{eff}(P)^2/2-\cdots$, in the order of $P$.
 Then $f_{N,R,\Lambda_{ij}}(t)$ with this first correction of the quartic order 
 can be obtained by computing
 \s[
 f^{(4)}_{N,R,\Lambda_{ij}}(t)&=const. \int dP \, e^{-S_{eff}^{(2)}(P)}(1- S_{eff}^{(4)}(P))\\
 &=f^{(2)}_{N,R,\Lambda_{ij}}(t) 
 \left(1 - \langle S_{eff}^{(4)}(P) \rangle_P \right),
 \label{eq:f4estimation}
 \s]
 where $f^{(2)}_{N,R,\Lambda_{ij}}(t)$ is given in \eq{eq:expf2},
 the allover constant has been determined by requiring $f^{(4)}_{N,R,\Lambda_{ij}}(0)=1$,
 and $\langle \cdot \rangle_P$  is defined by 
 \[
 \langle {\cal O}(P) \rangle_P:=\frac{1}{\int dP e^{-S_{eff}^{(2)}(P) } }\int dP
\,  {\cal O}(P)\,   e^{-S_{eff}^{(2)}(P) }
 \]
 with the quadratic action $S_{eff}^{(2)}(P)$ given in \eq{eq:seff2}. 

 \begin{figure*}
 \begin{center}
\includegraphics[width=10cm]{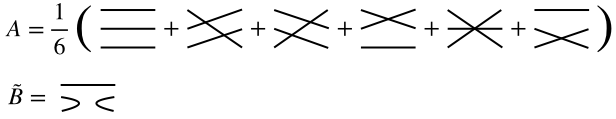}
 \caption{Graphical representations of the matrices $A$ and $\tilde B$.}
 \label{fig:AandB}
 \end{center}
 \end{figure*}
 
The computation of $f^{(4)}_{N,R,\Lambda_{ij}}(t)$ in \eq{eq:f4estimation} has
now been reduced to that of $\langle S_{eff}^{(4)}(P) \rangle_P$.
 This can be computed by the Wick theorem,
using the Wick contraction for $P_{abc}^i$ determined from $S_{eff}^{(2)}(P)$.
The Wick contraction of $P_{abc}^i$ can be 
obtained by taking the inverse of the coefficient matrix in the quadratic 
action $S_{eff}^{(2)}(P)$ of $P_{abc}^i$.
Since $S_{eff}^{(2)}(P)$ has the form of the direct product with respect to the 
upper and lower indices, they can be treated separately. 

Let us first treat the lower indices by starting with \eq{eq:pquad}. 
 Let us introduce the following matrices (See Figure~\ref{fig:AandB}):
 \[
 A_{abc,def}&=\frac{1}{6}\sum_{\sigma}
 \delta_{a\sigma_d}\delta_{b\sigma_e}\delta_{c\sigma_f},\\
 B_{abc,def}&=\frac{3}{N+2}\left(A \tilde B A\right)_{abc,def}, \\
 \tilde B_{abc,def}&=\delta_{ad}\delta_{bc}\delta_{ef},
 \]
 where the summation over $\sigma$ denotes the sum over all the permutations of $d,e,f$, and
 the product of two matrices, say $X$ and $Y$, is defined by
 \[
 (XY)_{abc,def}=\sum_{g,h,i=1}^N X_{abc,ghi}Y_{ghi,def}. 
 \]
 Note that $A$ acts as the identity on a symmetric tensor, namely, $(AP)_{abc}=P_{abc}$.  
One can easily check the following properties:
\[
A^2=A,\ AB=BA=B,\ B^2=B.
\]
Furthermore, one can check that $A-B$ and $B$ give the projectors to the 
tensor and vector parts of $P_{abc}$, respectively, as
\[
(A-B)P=P^T,\ BP=P^V.
\]
Therefore, by using these matrices, \eq{eq:quadev} can be rewritten in the form,
\[
P\left( c_{T,\lambda_{ev}} (A-B) +c_{V,\lambda_{ev}} B\right) P,
\label{eq:PmP}
\]
where $c_{T,\lambda_{ev}}$ and $c_{V,\lambda_{ev}}$ are the coefficients associated 
to the tensor and vector parts of $P_{abc}$,
namely,
\s[
c_{T,\lambda_{ev}}&=1+12\gamma_3 \lambda_{ev}\,  t, \\
c_{V,\lambda_{ev}}&=1+6(N+4)\gamma_3\lambda_{ev}\,  t.
\s]
The inverse of the matrix in \eq{eq:PmP} is given by
\s[
(c_{T,\lambda_{ev}} (A-B) +c_{V,\lambda_{ev}} B)^{-1}
&=\frac{1}{c_{T,\lambda_{ev}}} (A-B)+\frac{1}{c_{V,\lambda_{ev}}} B  \\
&=\frac{1}{c_{T,\lambda_{ev}}} A+\left(\frac{1}{c_{V,\lambda_{ev}}}-\frac{1}{c_{T,\lambda_{ev}}} \right) B.
\label{eq:propP}
\s]
For later convenience, let us define
\s[
a_{\lambda_{ev}}&:=\frac{1}{c_{T,\lambda_{ev}}}, \\
b_{\lambda_{ev}}&:=\frac{1}{c_{V,\lambda_{ev}}}-\frac{1}{c_{T,\lambda_{ev}}}.
\s]

Let us next take into account the upper indices. 
To derive the above result we started with the expression \eq{eq:pquad} with an eigenvalue 
$\lambda_{ev}$ of the matrix $\Lambda_{ij}$. 
Therefore it is the result for the corresponding eigenspace.
Considering the projection to each eigenspace,
the final form of the Wick contraction for $P_{abc}^i$ is obtained as 
\[
\langle P_{abc}^i P_{def}^j\rangle_P=
\sum_{\lambda_{ev}} \frac{1}{2} M_{\lambda_{ev}}^{ij}
\left( a_{\lambda_{ev}} A+b_{\lambda_{ev}} B \right)_{abc,def}, 
\label{eq:fullpropP}
\]
where $M_{\lambda_{ev}}^{ij}$ denotes the projector to the eigenspace of
the eigenvalue $\lambda_{ev}$ of $\Lambda_{ij}$.

\begin{figure}
\begin{center}
\includegraphics[width=8cm]{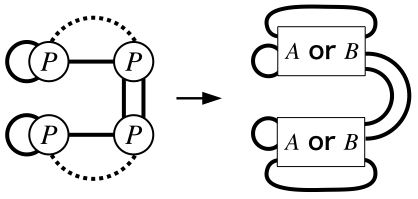}
\caption{An example of Wick contraction of $P_{abc}^i$'s. This can be computed by 
inserting the matrices $A$ and $B$ at the locations of contractions.}
\label{fig:exAB}
\end{center}
\end{figure}

Let us next start computing $\langle S^{(4)}_{eff}(P)\rangle_P$ by using the Wick contraction
\eq{eq:fullpropP}.
Let us first compute the factors coming from the projectors.
Let us restrict ourselves to the 
case $\Lambda_{ij}=\lambda+\lambda_d \delta_{ij}$, which is of our interest as 
explained in Section~\ref{sec:fextend}. 
As given there, the
eigenvector for $\lambda_{ev}=\lambda R+\lambda_d$ is $(1,1,\cdots,1)$, and 
those for $\lambda_{ev}=\lambda_d$ are the vectors transverse to that.
Therefore the projectors are $M_{\lambda R+\lambda_d}^{ij}=1/R$ for $\lambda_{ev}=\lambda R+\lambda_d$,
and $M_{\lambda_d}^{ij}=\delta_{ij}-1/R$ for $\lambda_{ev}=\lambda_d$,
respectively.
For later usage, let us compute the projectors sandwiched between $\tilde \Lambda_{ij}$:
\s[
&(\tilde \Lambda M_{\lambda_{ev}} \tilde \Lambda)_{ij}:=\sum_{k,l=1}^R \tilde \Lambda_{ik}M_{\lambda_{ev}}^{kl} \tilde \Lambda_{lj}=
\left \{
\begin{array}{ll}
\lambda+\frac{\lambda_d}{R}, & \hbox{for }\lambda_{ev}=\lambda R+\lambda_d,\\
\lambda_d\delta_{ij}-\frac{\lambda_d}{R}, & \hbox{for }\lambda_{ev}=\lambda_d,
\end{array}
\right.
\label{eq:projtilde}
\s]
where we have used the following explicit solution to \eq{eq:defoftildelam} for 
$\Lambda_{ij}=\lambda+\lambda_d \delta_{ij}$:
\[
\tilde \Lambda_{ij}=p \delta_{ij}+q \hbox{ with } p^2=\lambda_d,\ 2pq+Rq^2=\lambda.
\label{eq:explicitlamtilde}
\]

As for the factor coming from the $\tilde \Lambda_{ij}$ in \eq{eq:intleftmost}, the Wick contractions \eq{eq:fullpropP} of $P^i_{abc}$ will generate the factor  
$\sum_j (\tilde \Lambda M_{\lambda_{ev}} \tilde \Lambda)_{jj}
(\tilde \Lambda M_{\lambda_{ev}'} \tilde \Lambda)_{jj}$.
Thus, using \eq{eq:projtilde}, we obtain the following results for each case:
\s[
&\sum_{j=1}^R (\tilde \Lambda M_{\lambda_{ev}} \tilde \Lambda)_{jj}
(\tilde \Lambda M_{\lambda_{ev}'} \tilde \Lambda)_{jj}
=\left\{
\begin{array}{cl}
R\left( \lambda+\frac{\lambda_d}{R} \right)^2 ,&\hbox{for }
\lambda_{ev}=\lambda_{ev}'=\lambda R+\lambda_d,\\
R\left( \frac{(R-1)\lambda_d}{R}\right)^2 ,&\hbox{for }
\lambda_{ev}=\lambda_{ev}'=\lambda_d,\\
R\left( \lambda+\frac{\lambda_d}{R} \right)\frac{(R-1)\lambda_d}{R},
&\hbox{otherwise}.
\end{array}
\right.
\label{eq:lamfor4}
\s]
It is easy to see that all the other diagrams in Figure~\ref{fig:quad}
have the same factor. 

As for the Wick contractions \eq{eq:fullpropP} 
of $\left(\langle (P\phi^3)^2 \rangle^c_\phi\right)^2$
in \eq{eq:pphi4c}, there exist two cases.
One is to take the contractions within each $\langle (P\phi^3)^2 \rangle^c_\phi$,
or otherwise. The former case can be called the disconnected case,
and the latter the connected case, based on their diagrammatic characters.
From \eq{eq:cum2} and \eq{eq:fullpropP}, 
the factors coming from the $\Lambda_{ij}$'s are obtained as  
\[
\sum_{i,j=1}^R(\Lambda M_{\lambda_{ev}})_{ii}(\Lambda M_{\lambda'_{ev}})_{jj}
&=\lambda_{ev} \lambda_{ev}' \hbox{Tr}(M_{\lambda_{ev}})\hbox{Tr}(M_{\lambda'_{ev}}) 
\\ &
=\left\{
\begin{array}{cl}
(\lambda R+\lambda_d)^2, &\hbox{for }
\lambda_{ev}=\lambda_{ev}'=\lambda R+\lambda_d,\\
(R-1)^2\lambda_d^2 ,&\hbox{for }
\lambda_{ev}=\lambda_{ev}'=\lambda_d,\\
(\lambda R+\lambda_d)(R-1)\lambda_d,
&\hbox{otherwise},
\end{array}
\right.
\label{eq:s4dis}\\
\sum_{i,j=1}^R(\Lambda M_{\lambda_{ev}})_{ij}(\Lambda M_{\lambda'_{ev}})_{ji}&=
 \lambda_{ev}^2 \hbox{Tr}(M_{\lambda_{ev}})\delta_{\lambda_{ev}\lambda_{ev}'}
 \nonumber\\
 &=\left\{
\begin{array}{cl}
(\lambda R+\lambda_d)^2, &\hbox{for }
\lambda_{ev}=\lambda_{ev}'=\lambda R+\lambda_d,\\
(R-1)\lambda_d^2 ,&\hbox{for }
\lambda_{ev}=\lambda_{ev}'=\lambda_d,\\
0,
&\hbox{otherwise}.
\end{array}
\right.
 \label{eq:s4con}
\]
for the disconnected and the connected cases, respectively.

The last ingredient for the computation of $S^{(4)}_{eff}(P)$ is to 
take into account the lower index part of the Wick contraction \eq{eq:fullpropP}.
As can be seen in \eq{eq:fullpropP}, in general, this is to insert $x A +y A \tilde B A$ with
some $x,y$ at the location of the Wick contraction (See Figure~\ref{fig:exAB} for an example).  
Diagrammatically, this is to insert the diagrams in Figure~\ref{fig:AandB} at the location
of the Wick contraction. This insertion generates many diagrams with a number of loops.
The number of loops gives the degeneracy of each diagram in powers of $N$. 
The summation over all the diagrams is too many to do so by hand, so we performed this task 
by using Mathematica.  We have obtained
\s[
G_1(x,y)&=6 N (2 + N) (4 + N) (225 x^2 + 90 N x^2 + 9 N^2 x^2 + 456 x y \\&+ 
   174 N x y + 18 N^2 x y + 232 y^2 + 84 N y^2 + 8 N^2 y^2)
   \label{eq:count4}
\s]
for $\langle (P\phi^3)^4 \rangle^c_\phi$.
As for  $\left(\langle (P\phi^3)^2 \rangle^c_\phi\right)^2$, we have obtained
\[
G_2(x,y)=&N^2 (2 + N)^2 (4 + N)^2 (x + y)^2, 
\label{eq:G2}\\
G_3(x,y)=&2 N (2 + N) (4 + N) (15 x^2 + 24 x y  + 6 N x y + 8 y^2 + 6 N y^2 + 
   N^2 y^2),
 \label{eq:G3}
\]
respectively, for the disconnected and connected cases.

Let us combine all the results above. By using \eq{eq:fullpropP}, \eq{eq:lamfor4}
and \eq{eq:count4} and summing over all the possibilities of the eigenspaces of $\Lambda_{ij}$,
we obtain
\[
 \langle \langle (P\phi^3)^4 \rangle^c_\phi\rangle_P=\frac{R}{2^2 (2\beta)^6} G_1 (x_1,y_1),
\label{eq:finalg1}
\]
where 
\[
x_1&= \left(\lambda+\frac{\lambda_d}{R} \right) a_{\lambda R+\lambda_d}+
\frac{(R-1)\lambda_d}{R} a_{\lambda_d},\\
y_1&=\frac{3}{N+2} \left( \left(\lambda+\frac{\lambda_d}{R} \right) b_{\lambda R+\lambda_d}+
\frac{(R-1)\lambda_d}{R} b_{\lambda_d}\right).
\]
For the disconnected case of 
$\langle \left(\langle (P\phi^3)^2 \rangle^c_\phi\right)^2\rangle_P$, 
from \eq{eq:fullpropP}, \eq{eq:s4dis} and \eq{eq:G2} we obtain
\[
 \langle \left(\langle (P\phi^3)^2 \rangle^c_\phi\right)^2 \rangle_P^{discon}=
\frac{1}{2^2 (2\beta)^6} G_2(x_2,y_2),
\label{eq:finalg2}
\]
where 
\[
x_2&= (\lambda R+\lambda_d) a_{\lambda R+\lambda_d}+
(R-1) \lambda_d a_{\lambda_d},\\
y_2&=\frac{3}{N+2} \left( (\lambda R+\lambda_d) b_{\lambda R+\lambda_d}+
(R-1) \lambda_d b_{\lambda_d}\right).
\]
For the connected part of $\langle \left(\langle (P\phi^3)^2 \rangle^c_\phi\right)^2\rangle_P$, 
from \eq{eq:fullpropP}, \eq{eq:s4con} and \eq{eq:G3} we obtain
\s[
  &\langle \left(\langle (P\phi^3)^2 \rangle^c_\phi\right)^2 \rangle_P^{con}
 =\frac{1}{2^2(2\beta)^6}\\&\cdot \left(
  (\lambda R+\lambda_d)^2 G_3(x_3,y_3)+
   (R-1) \lambda_d^2 G_3(x_4,y_4) \right),
   \label{eq:finalg3}
\s]
where 
\s[
(x_3,y_3)&=\left(a_{\lambda R+\lambda_d},\frac{3b_{\lambda R+\lambda_d}}{N+2}\right), \\
(x_4,y_4)&=\left(a_{\lambda_d},\frac{3b_{\lambda_d}}{N+2}\right).
\label{eq:x34}
\s]
By putting \eq{eq:finalg1}, \eq{eq:finalg2} and \eq{eq:finalg3} into \eq{eq:pphi4c}, we obtain
the final result given in \eq{eq:final4} and \eq{eq:finalexps4}.

%%%%%%%%%%%%%%%%%%%%%%%
%%%%%%%%%%%%%%%%%%%%%%%%%
%%%%%%%%%%%%%%%%%%%%%%%%%%%    
 \section{Estimation of the $R$-dependence of $b_1$}
\label{sec:estimation}
As for $b_1$, there are two kinds of contributions as given in \eq{eq:pphi4c}.
The dominant $R$-dependence of the former term of \eq{eq:pphi4c} can be estimated by considering 
the expression \eq{eq:intleftmost}, because the contractions 
of the upper indices among $P^i_{abc}$ are the same for all the diagrams in Figure~\ref{fig:quad}.
It is sufficient to consider $\lambda_d=0$, since this is of our interest, 
and the mode of $P^i_{abc}$ contributing in this case 
with the appropriate normalization has the form $\frac{1}{\sqrt{R}}(1,1,\ldots,1) P_{abc}$, 
which has the form of the eigenvector of the matrix $\Lambda_{ij}=\lambda$ for the eigenvalue $\lambda R$. 
Since $\tilde \Lambda_{ij}\sim 1/\sqrt{R}$ (see \eq{eq:explicitlamtilde}) and there are five free
summations over the upper indices in \eq{eq:intleftmost}, the former term of \eq{eq:pphi4c}
is estimated as
\[
b_1^{former}\sim \gamma_6  R^5 \sqrt{R}^{-4} \sqrt{R}^{-4} \sim R^{-5}.
\]

As for the latter term of  \eq{eq:pphi4c}, $\langle (P\phi^3)^2 \rangle^c_\phi \sim R P^2$ by a
similar argument as above. Therefore,
\[
b_1^{latter}\sim (\gamma_3^2-\gamma_6)R^2\sim R^{-5},
\]
where it is important to note that the contents of the parentheses give $R^{-7}$ due to the cancellation
of the leading $R^{-6}$ terms of both. 

The above two estimations conclude $b_1\sim R^{-5}$.

\section{Brief introduction of persistent homology}
\label{app:persistent}
In this appendix, we give a brief introduction of persistent homology for this paper
to be self-contained. More details can be found for instance in \cite{carlsson_topology_2009}.

Persistent homology is a notion that characterizes the topological aspects of a data
in terms of homology. 
A data  to be analyzed should be a set of points with relative distances.
From a data, a stream (or a filtration) of simplicial complexes parameterized by a scale parameter, say $u$,
is constructed. The details of the construction of a stream will be given at the end 
of this appendix.
Once a stream is constructed, the homology groups of the simplicial complexes at each value 
of $u$ are computed. By increasing the value of $u$ from zero, a homology group element 
will appear at a certain value of $u$, say $u_{\rm start}$, and will continue to exist until it disappears 
at another value, say $u_{\rm end}$. If the life period $u_{\rm end}-u_{\rm start}$ is large, 
one may regard the element 
as a persistent homology group element, which has a long life. 
The collection of persistent homology group elements 
characterizes the topological property of a data set.
There will also be a number of short-life elements, but they are often regarded as ``noises",
which are not robust against small perturbations of the data.
Roughly speaking, the scale $u$ parameterizes the sizes of topological structure 
of interest, and persistent homology characterizes a data with multi-scale homology groups.

\begin{figure*}
\begin{center}
\includegraphics[width=5cm]{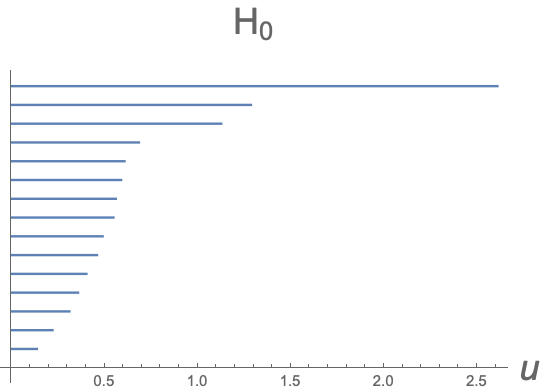}
\hfil
\includegraphics[width=5cm]{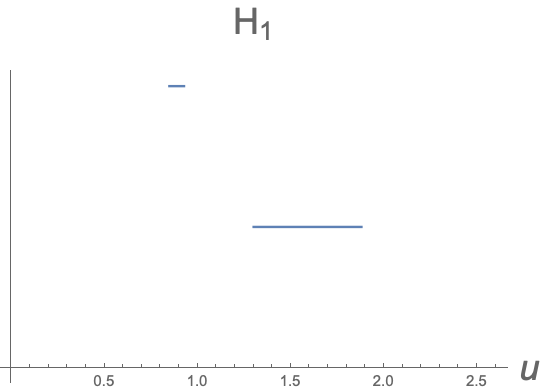}
\caption{An example of barcode diagrams. The example is obtained by analyzing a configuration $\phi_a^i\ (i=1,2,\ldots,R)$ generated by the actual Monte Carlo simulation for $N=4$, $R=15$, and $k=0.01$. The presence of a long-life one-dimensional homology group element can be observed
in this particular data. }
\label{fig:exh0h1}
\end{center}
\end{figure*}

\begin{figure}
\begin{center}
\includegraphics[width=5cm]{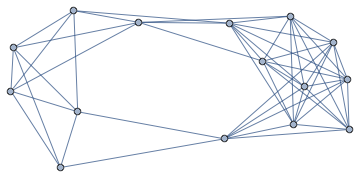}
\caption{A graph made by connecting the points with relative distances smaller than $1.5$.
The data is the same one used for Figure~\ref{fig:exh0h1}.
This choice of the distance cut-off is made because a one-dimensional cycle is expected to 
exist at this scale, seeing the right of Figure~\ref{fig:exh0h1}.  }
\label{fig:exh1graph}
\end{center}
\end{figure}

There are two kinds of diagrams that are convenient for visualizing the persistent homology
of a data.
One is called barcode diagram, where each horizontal line segment $[u_{start},u_{end}]$
represents a homology group element that exists during the period.  An example of barcode diagrams
constructed from a sample of $\phi_a^i$ 
generated in the Monte Carlo simulation is shown in Figure~\ref{fig:exh0h1}.
How to construct a set of points with relative distances from $\phi_a^i$ is given in Section~\ref{sec:persistent}.
The left figure shows the barcode diagram for the 0-dimensional homology,
and the right that for the 1-dimensional homology. 
The left diagram indicates that the initially separated points form one connected component 
over the scale $u\sim 1.4$.
The right figure shows that there exists a one-dimensional 
cycle which has the size of $u\sim 1.8$, while there is a small ``noise" around $u\sim 0.9$.
In Figure~\ref{fig:exh1graph}, we provide the graph constructed by connecting the points with relative distances $u<1.5$, using the same $\phi_a^i$ for Figure~\ref{fig:exh0h1}. 
One can actually see the presence of a one-dimensional cycle
consistent with the barcode diagram.

\begin{figure}
\begin{center}
\includegraphics[width=5cm]{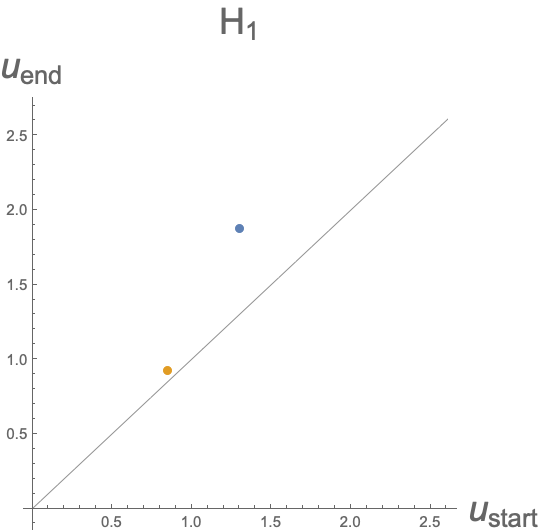}
\caption{The persistent diagram corresponding to the right barcode diagram in Figure~\ref{fig:exh0h1}. A line segment in a barcode diagram is represented by a dot
in a persistent diagram. The longest one is represented by a blue dot, and the second by a yellow one.
The colors are also used in the persistent diagrams in Section~\ref{sec:persistent}.}
\label{fig:exh1per}
\end{center}
\end{figure}

The other kind of diagram is called persistent diagram. An element that
is represented by a line segment $[u_{\rm start},u_{\rm end}]$ in
a barcode diagram is represented by a dot
located at the two-dimensional coordinate $(u_{\rm start},u_{\rm end})$
in a persistent diagram. Since there are multiple elements in general, and 
$u_{\rm start}<u_{\rm end}$, 
a persistent diagram consists of a number of dots in the region over the diagonal line. 
The long-life elements are represented by the dots that exist away from the diagonal line, and 
those in the vicinity of the diagonal line are regarded as ``noises".
An example of a persistent diagram is given in Figure~\ref{fig:exh1per},
which corresponds to the right barcode diagram in Figure~\ref{fig:exh0h1}.
What is convenient in a persistent diagram is that one can easily superimpose 
persistent diagrams from multiple data. If there is a common characteristics
through multiple data, one can find it as a characteristic pattern in a superimposed persistent diagram.
Therefore, we use persistent diagrams to find statistically favored structure common in 
the configurations generated by the Monte Carlo simulation.

Let us finally explain the actual construction of a stream of simplicial complexes parameterized by 
$u$. In fact, there exist various streams depending on purposes in the literature,
but let us restrict ourselves to the Vietoris-Rips stream, which is used in the open-source c++ code called
Ripser. 
For a given data that stores distances between points, 
the Vietoris-Rips stream ${\rm VR}(V,u)$ is defined as follows:
\begin{itemize}
\item The vertex set is given by the point set $V$ of a data.
\item For vertices $i$ and $j$ with distance $d(i,j)$, 
the edge $[i,j]$ are included in ${\rm VR}(V,u)$, if and only if $d(i,j)\leq u$.
\item A higher-dimensional simplex is included in ${\rm VR}(V,u)$, if and only if all of its edges are.
\end{itemize}

From the above definition, there is an obvious property, ${\rm VR}(V,u) \subset {\rm VR}(V,u')$
for $u<u'$. An important fact is that this induces a map: ${\rm H}_k({\rm VR}(V,u)) 
\rightarrow {\rm H}_k({\rm VR}(V,u'))$ for $u<u'$.
Therefore, the development of each homology group element under the change of the
value of $u$ can be followed, and its life is characterized by the two endpoint values of $u$. 
This makes barcode and persistent diagrams convenient ways for the description.

%%%%%%%%%%%%%%%%%%%%%%%%%%%%%%%%
%%%%%%%%%%%%%%%%%%%%%%%%%%%%%%%%
%%%%%%%%%%%%%%%%%%%%%%%%%%%%%%%%

\end{document}